\documentclass[12pt]{article}
\usepackage{amsmath}
\usepackage{amssymb}
\usepackage{amsthm}
\usepackage[breaklinks,linktocpage=true]{hyperref} 
\begin{document}

\vspace*{0.5in}

\begin{center}

{\large\bf Analogues of Mathai-Quillen forms in sheaf cohomology}

{\large\bf and applications to topological field theory}

\vspace*{0.2in}

Richard S. Garavuso$^1$, Eric Sharpe$^2$

\vspace*{0.2in}

\begin{tabular}{cc}
{\begin{tabular}{l}
$^1$ Harish-Chandra Research Institute\\
Chhatnag Road\\
Jhunsi, Uttar Pradesh 211019\\
India\\
\end{tabular}
} &
{\begin{tabular}{l}
$^2$ Department of Physics\\
Robeson Hall, 0435\\
Virginia Tech\\
Blacksburg, VA  24061, USA\\
\end{tabular}
}
\end{tabular}

{\tt garavuso@hri.res.in}, {\tt ersharpe@vt.edu}

$\,$

\end{center}

We construct sheaf-cohomological analogues of Mathai-Quillen forms,
that is, holomorphic bundle-valued differential forms whose cohomology
classes are independent of certain deformations, and which are believed
to possess Thom-like properties.  Ordinary Mathai-Quillen forms are
special cases of these constructions, as we discuss.  These 
sheaf-theoretic variations arise physically in A/2 and B/2 model
pseudo-topological field theories, and we comment on their
origin and role.

\begin{flushleft}
October 2013
\end{flushleft}

\newpage

\tableofcontents

\newpage

\section{Introduction}

The Mathai-Quillen formalism \cite{MathaiQuillen:Superconnections}
provides mathematics with a by-now well-known realization of
Thom classes, which is to say, it allows for integrals of differential forms
over total spaces of vector bundles to be reduced to integrals over base 
spaces.  Given a complex manifold $Y = \{s = 0 \} \subset M$ 
for $s$ a section of
a holomorphic vector bundle ${\cal G} \rightarrow M$, say, a typical application of
a Mathai-Quillen form 
$ U({\cal G},\nabla) $ 
is to use its pullback 
$ s^* U({\cal G},\nabla) $ 
to relate an integral of a differential form over 
$ Y $ 
to an integral over 
$ M $. 
Specifically, if 
$ \omega_{MQ} \propto s^* U({\cal G},\nabla) $, 
then
\begin{displaymath}
\int_Y \alpha \: \propto \:
\int_M \tilde{\alpha} \wedge \omega_{MQ}
\end{displaymath}
for a differential form $\alpha$ and a suitable pullback
$\tilde{\alpha}$.  In physics, Mathai-Quillen forms play a further role in
understanding topological field theories, see {\it e.g.}
\cite{Kalkman:BRST,Blau:The-Mathai-Quillen,Wu:On-the-Mathai-Quillen,CordesMooreRamgoolam:Lectures,Wu:Mathai-Quillen}.

In this paper, we propose six analogues of Mathai-Quillen forms
for sheaf cohomology valued
in locally-free sheaves, for various circumstances.
Specifically, 
given a complex manifold $Y$ as above,
with bundle ${\cal E}'$ and
isomorphism $\det {\cal E}'^* \cong K_Y$ (or $\det {\cal E}' \cong K_Y$)
so that integrals of the form
\begin{displaymath}
\int_Y {\cal O}_1 \wedge \cdots \wedge {\cal O}_n,
\end{displaymath}
for 
\begin{displaymath}
{\cal O}_i \: \in \: H^{\bullet}\left(Y, \wedge^{\bullet} {\cal E}'^* \right)
\mbox{ or }
H^{\bullet}\left(Y, \wedge^{\bullet} {\cal E}' \right),
\end{displaymath}
are well-defined, we propose sheaf cohomology classes
$\omega$ such that for 
systematically-defined cohomology classes
$\tilde{\cal O}_i$ lifting the ${\cal O}_i$,
\begin{displaymath}
\int_Y {\cal O}_1 \wedge \cdots \wedge {\cal O}_n
\: \propto \:
\int_Z \tilde{\cal O}_1 \wedge \cdots \wedge \tilde{\cal O}_n
\wedge \omega .
\end{displaymath}
Here, $Z = M$ or $ Z $ is the total space $ X $ of a holomorphic vector bundle over $ M $.

Our constructions rest on a representation of sheaf cohomology valued in 
locally-free sheaves as $\overline{\partial}$-cohomology classes of
bundle-valued differential forms, and on an analogous representation of
hypercohomology valued in a complex of locally-free sheaves.

In the special case that ${\cal E}'$ is the tangent bundle to $Y$, so that
\begin{displaymath}
H^{\bullet}\left(Y, \wedge^{\bullet} {\cal E}'^* \right)
\: = \: H^{\bullet, \bullet}(Y),
\end{displaymath}
one of our six constructions will specialize to ordinary Mathai-Quillen
forms describing $Y$ as the zero locus of a section of ${\cal G}$.

We demonstrate that our bundle-valued forms define suitable elements of
cohomology, and have some of the same cohomological-invariance properties
of ordinary Mathai-Quillen forms.  However, the demonstration that
\begin{displaymath}
\int_Y {\cal O}_1 \wedge \cdots \wedge {\cal O}_n
\: \propto \:
\int_Z \tilde{\cal O}_1 \wedge \cdots \wedge \tilde{\cal O}_n
\wedge \omega .
\end{displaymath}
is left for later work:  it is a consequence of the physical origin
of these analogues (specifically, it is a mathematical prediction of
the renormalization group), 
but we do not offer a rigorous mathematical demonstration.

Our construction is motivated by physics.  In much the same way that
ordinary Mathai-Quillen forms enter ordinary topological field theories,
the analogues we propose enter heterotic analogues of topological field
theories, the A/2 and B/2 models,
which is where they were first observed.  (See for example 
\cite{katz-s,ade,s-b,McOristMelnikov:Summing,dgks1,dgks2,mss}
for further information on the A/2 and B/2 models.)  Part of the purpose of
this paper is to try to extract precise mathematical predictions about
analogues of Mathai-Quillen forms from those heterotic constructions, and
we devote the latter part of the paper to explaining physical origins and
applications.  

In particular, this paper was originally motivated by the
desire to understand claims made in \cite{McOristMelnikov:Summing}, regarding
the dependence of correlation functions in the A/2 model on certain complex
and bundle moduli.  Our original hope was to establish their claims in  
a clean mathematical setting.  Although we did succeeed in establishing
some of their claims, others are left for future work.

We begin in section~\ref{Review-MQ-formalism} 
by briefly reviewing pertinent aspects of the
Mathai-Quillen formalism.  In section~\ref{Deformed-Mathai-Quillen} we
discuss various sheaf-theoretic generalizations.  The basic format is
that we take a bundle on $Y$ which can be realized as {\it e.g.} a kernel
or cokernel on a larger space including $Y$, and lift the sheaf
cohomology computation to that larger space.  We give representatives
of sheaf cohomology, $\overline{\partial}$-closed bundle-valued
differential forms, whose cohomology classes are independent of
certain deformations.  In particular, in this section we observe
how ordinary Mathai-Quillen forms can be understood as special cases of
the kernel construction, and also discuss how the results of
\cite{McOristMelnikov:Summing} can be understood in this context.  

Finally in section~\ref{sect:apps:tft}
we discuss how these analogues of Mathai-Quillen forms arise in
heterotic versions of topological field theories.  
Corresponding to each analogue of a Mathai-Quillen form discussed
earlier, we give a Landau-Ginzburg model that renormalization group
flows to a nonlinear sigma model.  The analogues of Mathai-Quillen forms
provide a mathematical mechanism for understanding how correlation functions
in the (UV) Landau-Ginzburg theories can match those of (IR) nonlinear
sigma models.  See 
\cite{GuffinSharpe:A-twisted,GuffinSharpe:A-twistedheterotic} 
for a discussion of the pertinent Landau-Ginzburg models, whose descriptions
are further elaborated upon here.

This paper could be viewed as a step in the further development of
Landau-Ginzburg models over nontrivial spaces, further developing results
in {\it e.g.} 
\cite{GuffinSharpe:A-twisted,GuffinSharpe:A-twistedheterotic,AndoSharpe,clarke1,bmp1}.

A mathematician solely interested in analogues of Mathai-Quillen forms
can safely skip section~\ref{sect:apps:tft}, as it is intended to give
context to physicists reading this article.

\section{Brief review of the Mathai-Quillen formalism}
\label{Review-MQ-formalism}

Consider a complex vector bundle
$ {\cal G} \overset{\pi}{\longrightarrow} M $ with standard fiber
$ V $, over a complex manifold $M$.
Suppose that $ {\cal G} $ has fiber metric $ (\cdot,\cdot)_{\cal G} $ 
and compatible connection $ \nabla $.
Under these circumstances, the Mathai-Quillen formalism
\cite{MathaiQuillen:Superconnections,
Kalkman:BRST,
Blau:The-Mathai-Quillen,
Wu:On-the-Mathai-Quillen,
CordesMooreRamgoolam:Lectures,
Wu:Mathai-Quillen,BerlinGetzlerVergne:Heat}
provides an explicit representative $ U({\cal G},\nabla) $ 
of the Thom class of $ {\cal G} $.
Furthermore, the pullback $ \mathfrak{s}^* U({\cal G},\nabla) $ of $ U({\cal G},\nabla) $ 
by any 
section $ \mathfrak{s}: M \rightarrow {\cal G} $ of $ {\cal G} $ 
is a representative of the 
top Chern class of $ {\cal G} $.
Let us review the formalism in more detail.
%%%%%%%%%%%%%%%%%%%%%%%%%%%%%%%%%%%%%%%%%%%%%%%%%%%%%%%%%%%%%%%%%
\subsection{Conventions}

Our conventions for $ M $, $ {\cal G} $, and the dual $ {\cal G}^* $ 
of $ {\cal G} $ are as follows.
The exterior derivatives on $ M $ and $ {\cal G} $ 
are respectively denoted by $ d $ and $ d^{\cal G} $.
We choose local coordinates $ \phi^I $ on $ M $.
The connection on $ {\cal G} $ is then given by $ \nabla = d \phi^I \nabla_I $.
In terms of this connection, the curvature 2-form on $ {\cal G} $ 
is given by $ \mathcal{R} = \nabla^2 $.
We choose a local oriented orthonormal frame $ \{ e_A \} $ for $ {\cal G} $ 
and let $ \{ f^A \} $ be the dual coframe.
The section $ s $ may thus be expressed as $ s = s^A e_A $.
Similarly, we write $ \rho = \rho_A f^A $, 
where the $ \rho_A $ are anticommuting orthonormal coordinates on
$ {\cal G}^{*} $.
The dual pairing on $ {\cal G} $ is denoted by 
$ \langle \cdot, \cdot \rangle_{\cal G} $.
Finally, the metric on $ {\cal G}^* $ is denoted by 
$ (\cdot,\cdot)_{{\cal G}^*} $.

Now, consider the pullback bundle
$ \pi^* {\cal G} \rightarrow {\cal G} $.
This bundle has fiber metric
\begin{displaymath}
 \pi^* (\cdot,\cdot)_{\cal G} \: \equiv \: (\cdot,\cdot)_{\pi^* {\cal G}} ,
\end{displaymath}
compatible connection
$ \pi^* \nabla \equiv \widetilde{\nabla} $,
curvature 2-form
$ \pi^* \mathcal{R} \equiv \widetilde{\mathcal{R}} $,
local oriented orthonormal frame
$ \{ \pi^* e_A \} \equiv \{ \tilde{e}_A \} $, and
tautological section
$ \tilde{x} = \tilde{x}^A \tilde{e}_A $.
The dual bundle
$ (\pi^* {\cal G})^* \rightarrow {\cal G} $
has coframe
$ \{ (\pi^*)^* f^A \} \equiv \{ \tilde{f}^A \} $
and metric
$ (\cdot,\cdot)_{(\pi^* {\cal G})^*} $.
We write
$ \tilde{\rho} = \tilde{\rho}_A \tilde{f}^A $,
where the
$ \tilde{\rho}_A \equiv (\pi^*)^* \rho_A $
are anticommuting orthonormal coordinates on $ (\pi^* {\cal G})^* $.
The dual pairing on $ \pi^* {\cal G} $ is denoted by 
$ \langle \cdot, \cdot \rangle_{\pi^* {\cal G}} $.

%%%%%%%%%%%%%%%%%%%%%%%%%%%%%%%%%%%%%%%%%%%%%%%%%%%%%%%%%%%%%%%%%
\subsection{Definition}

The Mathai-Quillen form $ U({\cal G},\nabla) $ is defined to be proportional to
%%%%%%%%%%%%%%%%%
\begin{equation}
\label{MQ-form}
u({\cal G},\nabla)
   =  \int d \tilde{\rho} \,
     \exp
     \left(
          - \tilde{\mathcal{A}}
     \right), 
\end{equation}
%%%%%%%%%%%%%%%%%
where
%%%%%%%%%%%%%%%%%
\begin{equation}
\tilde{\mathcal{A}} 
   = \frac{1}{2} 
     \Bigl( 
            \tilde{x}, \tilde{x} 
     \Bigr)_{\pi^* {\cal G}} 
   + \left\langle 
            \widetilde{\nabla} \tilde{x}, \tilde{\rho} 
     \right\rangle_{\pi^* {\cal G}}
   + \frac{1}{2} 
     \left(
            \tilde{\rho}, \widetilde{\mathcal{R}} \tilde{\rho} 
     \right)_{(\pi^* {\cal G})^{*}}.                
\end{equation}

By construction,
\begin{equation*}
\tilde{\mathcal{A}} 
   \in \overset{}{\underset{i}{\oplus}} \Omega^i
       \left(
              {\cal G}, 
              \wedge^i 
              \left(
                     \pi^* {\cal G} 
              \right)^{*} 
       \right),                  
\end{equation*}
hence
\begin{equation*}
\exp
\left(
     - \tilde{\mathcal{A}}
\right)
   \in \overset{}{\underset{i}{\oplus}} \Omega^i
       \left(
              {\cal G}, 
              \wedge^i 
              \left(
                     \pi^* {\cal G} 
              \right)^*
       \right) ,
\end{equation*}
hence $ u({\cal G},\nabla) \in \Omega^r({\cal G}) $.
Moreover, the Mathai-Quillen form is closed:
\begin{displaymath}
d^{\cal G} u({\cal G},\nabla) \: = \: 0.
\end{displaymath}

To show that it is closed, first note that
since $ \widetilde{\nabla} $ is compatible with the metric 
$ \left( \cdot, \cdot \right)_{\pi^* {\cal G}} $, it follows that
%%%%%%%%%%%%%%%%%
\begin{equation*}
d^{\cal G} \int d \tilde{\rho} \, 
\tilde{\alpha}
   = \int d \tilde{\rho} \,
     \widetilde{\nabla} 
     \tilde{\alpha} \, ,           
\end{equation*}
%%%%%%%%%%%%%%%%%
where
$
\tilde{\alpha}
   \in \Omega
       \left(
              {\cal G}, 
              \wedge
              \left( 
                     \pi^* {\cal G}
              \right)^*
       \right)
$.
Furthermore,
%%%%%%%%%%%%%%%%%
\begin{align}
\left( 
       \widetilde{\nabla}
     + \tilde{x}_A
       \frac{\partial}{\partial \tilde{\rho}_A}   
\right)
\tilde{\mathcal{A}}
  &= \left(
            \widetilde{\nabla} \tilde{x}, \tilde{x}
     \right)_{\pi^* {\cal G}}  
   + \left\langle
            \widetilde{\mathcal{R}} \tilde{x}, \tilde{\rho}
     \right\rangle_{\pi^* {\cal G}}
   - \frac{1}{2}
     \left(
            \tilde{\rho}, \widetilde{\nabla} \widetilde{\mathcal{R}} \tilde{\rho} 
     \right)_{\left(\pi^* {\cal G}\right)^*}
\nonumber
\\[1ex]
  &\phantom{=}
   - \left(
            \widetilde{\nabla} \tilde{x}, \tilde{x}
     \right)_{\pi^* {\cal G}}  
   - \left\langle
            \widetilde{\mathcal{R}} \tilde{x}, \tilde{\rho}
     \right\rangle_{\pi^* {\cal G}} ,
\nonumber
\\[2ex]
  &= 0 \, ,                            
\end{align}
%%%%%%%%%%%%%%%%%
where we have used the Bianchi identity
$ \widetilde{\nabla} \, \widetilde{\mathcal{R}} = 0 $.
From these results, we obtain
%%%%%%%%%%%%%%%%%
\begin{align*}
d^{\cal G} u({\cal G},\nabla)
  &=  
     d^{\cal G} \int d \tilde{\rho} \, 
     \exp 
     \left( 
          - \tilde{\mathcal{A}} 
     \right) ,
\\[1ex]
  &=  \int d \tilde{\rho} \,
     \widetilde{\nabla} 
     \exp
     \left(
          - \tilde{\mathcal{A}}
     \right) ,
\\[1ex]
  &=  \int d \tilde{\rho} \,
     \left(
            \widetilde{\nabla}
          + \tilde{x}_A
            \frac{\partial}{\partial \tilde{\rho}_A}  
     \right)
     \exp
     \left(
          - \tilde{\mathcal{A}}
     \right) ,
\\[1ex]
  &=  \int d \tilde{\rho} \,
     \left[
          - \left(  
                   \widetilde{\nabla}
                 + \tilde{x}_A
                   \frac{\partial}{\partial \tilde{\rho}_A}
            \right)
            \tilde{\mathcal{A}}         
     \right]
     \exp
     \left(
          - \tilde{\mathcal{A}}
     \right) ,
\\[1ex]
   &= 0 \, .                      
\end{align*}
%%%%%%%%%%%%%%%%%
Here, the third equality holds because
$ 
\tilde{x}_A 
\left( 
       \partial / \partial \tilde{\rho}_A 
\right)
e^{
  - \tilde{\mathcal{A}}
  }
$
contributes nothing to the Grassmann integral.

The Mathai-Quillen form has additional properties, most importantly that
\begin{displaymath}
\int_V U({\cal G},\nabla) \: = \: 1.
\end{displaymath} 
This implies that the Mathai-Quillen form is a representative
of the Thom class\footnote{
A representative of the Thom class of $ {\cal G} $ is
a $ d^{{\cal G}} $-closed differential form 
$ U({\cal G}) \in \Omega^r({\cal G}) $
such that $ \int_V U({\cal G}) = 1 $, for $r = {\rm rk}\, {\mathcal G}$.} 
of $ {\cal G} $.
This fact plays an important role in physics; for example, as a Thom class
it provides an understanding of renormalization group flow in 
Landau-Ginzburg models 
\cite{GuffinSharpe:A-twisted,GuffinSharpe:A-twistedheterotic,AndoSharpe}.  
However, we will not establish analogues of this in
this paper, we merely propose that they exist,
and as the property is standard, we omit its derivation here.

\subsection{Pullbacks of Mathai-Quillen forms}

What will be more relevant for this paper is 
the pullback of the Mathai-Quillen form $ U({\cal G},\nabla) $ by any section $ \mathfrak{s} $ of
$ {\cal G} $.
We shall write
%%%%%%%%%%%%%%%%%
\begin{equation}
\label{MQ-Euler}
\mathfrak{s}^* u({\cal G},\nabla)
   = 
     \int d \rho \,
     \exp 
     \left(
          - {\cal A}     
     \right), 
\end{equation}
%%%%%%%%%%%%%%%%%
where 
%%%%%%%%%%%%%%%%%
\begin{equation}
\label{S}
{\cal A} = \frac{1}{2} 
    \left(
           \mathfrak{s}, \mathfrak{s}
    \right)_{\cal G} 
  + \left\langle 
            \nabla \mathfrak{s}, \rho 
    \right\rangle_{\cal G}
  + \frac{1}{2} 
    \left(
           \rho, \mathcal{R} \rho
    \right)_{{\cal G}^{*}}.
\end{equation}
%%%%%%%%%%%%%%%%%
%%%%%%%%%%%%%%%%%%%%%%%%%%%%%%%%%%%%
The form $ \mathfrak{s}^* u({\cal G},\nabla) $ satisfies
\begin{enumerate}

\item[(i)]
$ \mathfrak{s}^* u({\cal G},\nabla) \in \Omega^{ {\rm rk}\, {\cal G}}(M) \, , $

\item[(ii)]
$ d \mathfrak{s}^* u({\cal G},\nabla) = 0 \, . $

\end{enumerate}

We see these properties as follows.
The first follows immediately from the fact that, by construction,
%%%%%%%%%%%%%%%%%
\begin{equation*}
{\cal A} \in \underset{i}{\oplus} \Omega^i
      \left(
             {\cal G}, 
             \wedge^i {\cal G}^{*} 
      \right). 
\end{equation*}
%%%%%%%%%%%%%%%%%
The proof of the second property is similar
to that of the argument that $u({\cal G}, \nabla)$ is $d^{\cal G}$-closed
in the last section, and uses the results
%%%%%%%%%%%%%%%%%
\begin{equation*}
d \int d \rho \, 
\alpha
   = \int d \rho \,
     \nabla 
     \alpha \, ,           
\end{equation*}
%%%%%%%%%%%%%%%%%
where
$
\alpha
   \in \Omega
       \left(
              {\cal G}, 
              \wedge {\cal G}^*
       \right)
$, and
%%%%%%%%%%%%%%%%%
\begin{equation}
\label{nabla-S-equation}
\left(
       \nabla 
     + \mathfrak{s}_A \frac{\partial}{\partial \rho_A} 
\right) {\cal A}
   = 0 \, . 
\end{equation}
%%%%%%%%%%%%%%%%%

For our purposes, the most important mathematical property of
the pullback  $ \mathfrak{s}^* u({\cal G},\nabla) $ is that the
$ d $-cohomology class of 
$ \mathfrak{s}^* u({\cal G},\nabla) $ is independent of the section $ s $.
We can see this as follows.
Let $ \mathfrak{s}_{\tau} = \mathfrak{s} + \tau \mathfrak{s}^{\prime} $ 
be an affine one-parameter family of sections of $ {\cal G} $ and let
%%%%%%%%%%%%%%%%%
\begin{equation*}
{\cal A}_{\tau}
   =
     \frac{1}{2} 
     \left(
            \mathfrak{s}_{\tau}, \mathfrak{s}_{\tau}
     \right)_{\cal G} 
   + \left\langle
            \nabla \mathfrak{s}_{\tau}, \rho
     \right\rangle_{\cal G}
   + \frac{1}{2} 
     \left(
            \rho, \mathcal{R} \rho
     \right)_{{\cal G}^{*}}.
\end{equation*}
%%%%%%%%%%%%%%%%%
Then
%%%%%%%%%%%%%%%%%
\begin{align*}
\frac{d}{d \tau} 
\mathfrak{s}^*_{\tau} u({\cal G},\nabla)
  &=  
     \frac{d}{d \tau}
     \int d \rho \,     
     \exp 
     \left(
          - {\cal A}_{\tau}
     \right) ,
\\[1ex]          
  &=
   - 
     \int d \rho \,
     \Bigl[
            \left(
                   \mathfrak{s}^{\prime}, \mathfrak{s}_{\tau}
            \right)_{\cal G}  
          + \left\langle
                   \nabla \mathfrak{s}^{\prime}, \rho
            \right\rangle_{\cal G}   
     \Bigr]          
     \exp 
     \left(
          - {\cal A}_{\tau}
     \right) ,            
\nonumber
\\[1ex]
  &=
   -   
     \int d \rho \,
     \left\{
            \left[
                   \nabla 
                 + \left(
                          \mathfrak{s}_{\tau}
                   \right)_A       
                   \frac{\partial}{\partial \rho_A}
            \right]       
            \left\langle
                   \mathfrak{s}^{\prime}, \rho
            \right\rangle_{\cal G}
     \right\}
     \exp 
     \left(
          - {\cal A}_{\tau} 
     \right) ,
\nonumber
\\[1ex]
  &=
   -   
     \int d \rho \,
     \left[
            \nabla 
          + \left(
                   \mathfrak{s}_{\tau}
            \right)_A       
            \frac{\partial}{\partial \rho_A}
     \right]
     \Bigl[       
            \left\langle
                   \mathfrak{s}^{\prime}, \rho
            \right\rangle_{\cal G} \,
            \exp 
            \left(
                 - {\cal A}_{\tau} 
            \right)
     \Bigr] ,    
\nonumber
\\[1ex]
  &=
   -  
     d
     \int d \rho \,   
     \left\langle
            \mathfrak{s}^{\prime}, \rho
     \right\rangle_{\cal G} \,          
     \exp 
     \left(
          - {\cal A}_{\tau} 
     \right).                         
\end{align*}
%%%%%%%%%%%%%%%%%
It follows that
%%%%%%%%%%%%%%%%%
\begin{equation*}
\mathfrak{s}^*_{\tau_2} u({\cal G},\nabla) - \mathfrak{s}^*_{\tau_1} u({\cal G},\nabla)
   =
   -   
     d \int^{\tau_2}_{\tau_1} d \tau
     \int d \rho \,   
     \left\langle
            \mathfrak{s}^{\prime}, \rho
     \right\rangle_{\cal G} \,                 
     \exp 
     \left(
          - {\cal A}_{\tau} 
     \right).
\end{equation*}
%%%%%%%%%%%%%%%%%          
Thus, for arbitrary sections $ \mathfrak{s}_{\tau_1} $ and $ \mathfrak{s}_{\tau_2} $ of $ {\cal G} \, , $ 
the $ d $-closed forms
$ \mathfrak{s}^*_{\tau_1} u({\cal G},\nabla) $ and
$ \mathfrak{s}^*_{\tau_2} u({\cal G},\nabla) $ differ by a $ d $-exact form 
and hence are cohomologous.

Mathematically,
the form $ \mathfrak{s}^* U({\cal G},\nabla) $ is cohomologous to the differential form
\begin{equation*}
    \frac{1}{(2 \pi)^\frac{{\rm rk}\, {\cal G}}{2}} \int d \rho \,
     \exp \left[
                 \frac{1}{2} 
                 \left(
                        \rho, \mathcal{R} \rho
                 \right)_{{\cal G}^{*}}
          \right]
    = \mathrm{Det} \left( 
                           \frac{\mathcal{R}}{2 \pi}
                    \right)
\end{equation*}
%%%%%%%%%%%%%%%%%
and hence is a representative of the top Chern class of $ {\cal G} $.
We can recover this simply by choosing $ \mathfrak{s} $ to be the zero section.

As an aside, if
$ \mathfrak{s} $ intersects the zero section of $ {\cal G} $ transversely, then
$ \mathfrak{s}^* U({\cal G},\nabla) $ is Poincar\'{e} dual to $ \mathfrak{s}^{-1}(0) $, i.e.
%%%%%%%%%%%%%%%%%
\begin{equation}
\int_{\mathfrak{s}^{-1}(0)}
\alpha
   =
\int_M 
\tilde{\alpha}
\wedge 
\mathfrak{s}^* U({\cal G},\nabla) \, , 
\end{equation}
%%%%%%%%%%%%%%%%%
where $ \tilde{\alpha} \in \Omega^{{\rm dim}\, M - {\rm rk}\, {\cal G}}(M) $ is 
$ d $-closed.
Also, when $ {\rm dim}\, M = {\rm rk}\, {\cal G} $, 
integrating $ \mathfrak{s}^* U({\cal G},\nabla) $ over $ M $ 
yields a representation of the integral of the
top Chern class of ${\cal G}$.

\section{Sheaf-cohomological Mathai-Quillen analogues}
\label{Deformed-Mathai-Quillen}

In this section, we will propose six analogues of Mathai-Quillen forms
for sheaf cohomology.  The general prototype is as follows.
We propose analogues of Mathai-Quillen forms whose insertions relate
integrals of cup products of sheaf cohomology classes
\begin{displaymath}
H^{\bullet}(N, \wedge^{\bullet} {\cal E}^*)
\mbox{ or }
H^{\bullet}(N, \wedge^{\bullet} {\cal E}) \, ,
\end{displaymath}
(or suitable hypercohomology classes, depending upon the analogue,)
where ${\cal E}$ is a locally-free sheaf on $N$,
to corresponding integrals of cup products of sheaf cohomology classes
\begin{displaymath}
H^{\bullet}(Y, \wedge^{\bullet} {\cal E}'^* )
\mbox{ or }
H^{\bullet}(Y, \wedge^{\bullet} {\cal E}' )
\end{displaymath}
over $Y \subset N$, where ${\cal E}'$ is a locally-free sheaf on $ Y $
constructed in part from the data in the analogue of the Mathai-Quillen form.

The first analogue we will discuss, the first kernel construction, will
specialize to (pullbacks of) ordinary Mathai-Quillen forms in 
the case that ${\cal E}' = TY$, as we shall discuss.

\subsection{Kernels}
\label{sect:kernels}

\subsubsection{First kernel construction}

Suppose that one is interested in computing integrals over
some space
$Y \equiv \{ s=0 \} \subset M$ ($s \in \Gamma({\cal G})$) of
sheaf cohomology classes
\begin{displaymath}
{\cal O} \: \in \: H^{\bullet}(Y, \wedge^{\bullet} {\cal E}'^* ) ,
\end{displaymath}
where ${\cal E}'$ is a holomorphic vector bundle on $Y$.
Such integrals will be of the form
\begin{displaymath}
\int_Y {\cal O}_1 \wedge \cdots \wedge {\cal O}_n 
\end{displaymath}
and will be well-defined if $\det {\cal E}'^* \cong K_Y$, and one picks
a particular isomorphism.

We propose that if ${\cal E}'$ is given as the restriction to $Y$ of
the kernel of a smooth surjective map
$\tilde{F}: 
{\cal F}_1 \rightarrow {\cal F}_2$ (${\cal F}_1$, ${\cal F}_2$
holomorphic vector bundles on $M$), whose restriction to $Y$ is
holomorphic, 
then at least for some ${\cal O}$'s it is possible to 
write the integral in the different form
\begin{displaymath}
\int_Y {\cal O}_1 \wedge \cdots \wedge {\cal O}_n \: \propto \:
\int_M \tilde{\cal O}_1 \wedge \cdots \wedge \tilde{\cal O}_n \wedge \omega_{K1} ,
\end{displaymath}
where $\tilde{\cal O}_i$ (when it exists) is an element of
$H^{\bullet}(M, \wedge^{\bullet} {\cal F}_1^*)$ `lifting' ${\cal O}_i$,
in a manner we shall describe shortly, and 
$\omega_{K1}$ is a $\overline{\partial}$-closed
analogue of a Mathai-Quillen form,
\begin{equation}  \label{eq:MQ:kernel:defn}
\omega_{K1} \: \in \: H^{g}\left(M, \wedge^{f_2} {\cal F}_1^*
\otimes  \det {\cal G}^* \otimes \det {\cal F}_2
\right)
\end{equation}
($f_i = {\rm rk}\, {\cal F}_i$, $g = {\rm rk}\, {\cal G}$), 
where one has an isomorphism
\begin{displaymath}
K_M \: \cong \: \det {\cal F}_1^* \otimes \det {\cal F}_2
\otimes \det {\cal G}^*
\end{displaymath}
that restricts to the isomorphism
$\det {\cal E}'^* \cong K_Y$ appearing above.

As a consistency check, notice that
\begin{displaymath}
\tilde{\cal O}_1 \wedge \cdots \wedge \tilde{\cal O}_n \: \in \:
H^{{\rm dim}\, M \: - \: g}\left(M, \wedge^{f_1 - f_2} {\cal F}_1^*
\right)
\end{displaymath}
so that the cohomology class of $\omega_{K1}$ is correct for the integrand
of $M$ to be a top-form.

We propose an expression for $\omega_{K1}$ below, and check its properties.

First, let us describe the relation between 
\begin{displaymath}
{\cal O} \: \in \: H^{\bullet}\left(Y, \wedge^{\bullet} {\cal E}'^* \right) 
\mbox{ and }
\tilde{\cal O} \: \in \: H^{\bullet}\left( M, \wedge^{\bullet} {\cal F}_1^*
\right).
\end{displaymath}
Let $i: Y \hookrightarrow M$ denote inclusion.  Then, given
$\tilde{\cal O}$, we have
\begin{displaymath}
i^* \tilde{\cal O} \: \in \: H^{\bullet}\left(Y, \wedge^{\bullet}
{\cal F}_1^* |_Y \right).
\end{displaymath}
Next, dualizing the short exact sequence
\begin{displaymath}
0 \: \longrightarrow \: {\cal E}' \: \longrightarrow \:
{\cal F}_1 |_Y \: \longrightarrow \: {\cal F}_2 |_Y \: \longrightarrow \: 0
\end{displaymath}
to
\begin{displaymath}
0 \: \longrightarrow \: {\cal F}_2^* |_Y \: \longrightarrow \: 
{\cal F}_1^* |_Y \: \longrightarrow \: {\cal E}'^* \: \longrightarrow \: 0,
\end{displaymath}
we see that there is a surjective map 
\begin{displaymath}
\wedge^{\bullet} {\cal F}_1^*|_Y
\: \longrightarrow \: \wedge^{\bullet} {\cal E}'^*
\end{displaymath}
which induces
\begin{displaymath}
j_*: \: H^{\bullet}\left(Y, \wedge^{\bullet} {\cal F}_1^* |_Y \right) \:
\longrightarrow \: H^{\bullet}\left(Y, \wedge^{\bullet} {\cal E}'^* \right).
\end{displaymath}
Hence, a pair ${\cal O}$, $\tilde{\cal O}$, when they exist, are related as
\begin{displaymath}
{\cal O} \: = \: j_* i^* \tilde{\cal O} .
\end{displaymath}

Our proposal for $\omega_{K1}$ 
is given by
the Grassmann integral
\begin{displaymath}
\omega_{K1} \: = \: \int \prod d \lambda^{\overline{x}} d \chi^r \, 
\exp( - {\cal A}_{K1} ) ,
\end{displaymath}
where
\begin{equation}   \label{eq:MQ:kernels}
{\cal A}_{K1} \: = \: 
h^{x \overline{x}} s_x \overline{s}_{\overline{x}}
\: + \: \chi^{\overline{\imath}} \lambda^{\overline{x}} \,
\overline{D}_{\overline{\imath}} \overline{s}_{\overline{x}}
\: + \: \chi^{r} \lambda^{\gamma} \tilde{F}_{r \gamma} \: + \:
F_{\overline{\imath} 
r \overline{x} \gamma} \chi^{\overline{\imath}} 
\chi^{r} \lambda^{\overline{x}} \lambda^{\gamma} ,
\end{equation}
where $x$ indexes local coordinates along the fibers of ${\cal G}$,
$\gamma$ indexes local coordinates along the fibers of
${\cal F}_1$, $r$ indexes local coordinates (denoted $p$) along
the fibers of ${\cal F}_2^*$, and $i$ indexes local
coordinates on $M$.
The curvature term
\begin{displaymath}
F_{\overline{\imath} 
r \overline{x} \gamma} \chi^{\overline{\imath}} 
\chi^{r} \lambda^{\overline{x}} \lambda^{\gamma}
\end{displaymath}
is defined by the condition\footnote{
This constraint is imposed physically by supersymmetry.  Mathematically,
it can be shown that one can always find $F_{\overline{\imath} 
r \overline{x} \gamma}$ satisfying this
condition \cite{donagipriv}.  For example, if ${\cal G}$ is a line
bundle, then this curvature term is the coboundary of
\begin{displaymath}
\tilde{F}|_Y \: \in \: H^0\left(Y, {\cal F}_1^*|_Y \otimes {\cal F}_2|_Y \right)
\end{displaymath}
in the long exact sequence derived from tensoring
\begin{displaymath}
0 \: \longrightarrow \: {\cal O}(-Y) \: \longrightarrow \:
{\cal O} \: \longrightarrow \: {\cal O}_Y \: \longrightarrow \: 0
\end{displaymath}
by ${\cal F}_1^* \otimes {\cal F}_2$.  In this special case, the constraint
above is merely the specification of the coboundary map.  For another
special case,
when one specializes to ordinary Mathai-Quillen
forms, $\tilde{F}_{i x} = D_i s_x$, and the constraint becomes
\begin{displaymath}
\overline{D}_{\overline{\imath}} D_j s_x \: = \:
\left[ \overline{D}_{\overline{\imath}}, D_j \right] s_x \: = \:
R_{\overline{\imath} j x \overline{x}} h^{y \overline{x}} s_{y},
\end{displaymath}
a standard result.  As another consistency check, note that along the
locus $\{s=0\}$, the constraint~(\ref{eq:dF-stildeF}) becomes the
statement that $\tilde{F}|_Y$ is holomorphic.
}
\begin{equation}   \label{eq:dF-stildeF}
\overline{\partial}_{\overline{\imath}} \tilde{F}_{r \gamma} \: = \:
h^{x \overline{x}} s_x F_{\overline{\imath} r \gamma \overline{x}} \: = \:
- h^{x \overline{x}} s_x F_{\overline{\imath} 
r \overline{x} \gamma} 
\end{equation}
and defines an element of 
\begin{displaymath}
H^1\left(M, {\cal F}_1^* \otimes {\cal F}_2 \otimes {\cal G}^* \right).
\end{displaymath} 
(Physically, $F$ arises as part of the curvature of a holomorphic vector
bundle, hence is $\overline{\partial}$-closed by virtue of the Bianchi
identity.  Similar considerations are the reason that the curvature of a bundle
defines the Atiyah class as an element of sheaf cohomology.)

Now, let us explain some aspects of $\omega_{K1}$ in more detail.
In ${\cal A}_{K1}$, every $\lambda^{\overline{x}}$ is accompanied by a 
$\chi^{\overline{\imath}}$, so integrating out the $\lambda^{\overline{x}}$'s
should result in ${\rm rk}\, {\cal G} = g$ $\chi^{\overline{\imath}}$'s,
hence a degree $g$ cohomology class.  Similarly, each $\chi^r$ is
accompanied by a $\lambda^{\gamma}$, so integrating out the $\chi^r$'s
should result in coefficients $\wedge^{f_2} {\cal F}_1^*$.
Furthermore, the
Grassmann integral measure makes $\omega_{K1}$ couple to 
$\det {\cal G}^* \otimes \det {\cal F}_2$.
Thus, $\omega_{K1}$ is a form of the type indicated in
equation~(\ref{eq:MQ:kernel:defn}).  We shall show it is
$\overline{\partial}$-closed momentarily.

Now, we can argue formally that the analogue of a Mathai-Quillen form
defined above is $\overline{\partial}$-closed.  The central point is that 
\begin{eqnarray*}
\left( \overline{D} \: + \: h^{x \overline{x}} s_x 
\frac{\partial}{\partial \lambda^{\overline{x}}
} \right) {\cal A}_{K1} & = &
\chi^{\overline{\imath}} \chi^r \lambda^{\gamma} \left(
\overline{\partial}_{\overline{\imath}} \tilde{F}_{r \gamma}
\: + \:
h^{x \overline{x}} s_x F_{\overline{\imath} r \overline{x} \gamma}
\right),
\\
& = &
0
\end{eqnarray*}
using equation~(\ref{eq:dF-stildeF}),
where
\begin{displaymath}
\overline{D} \: = \: \chi^{\overline{\imath}} \,
\overline{\partial}_{\overline{\imath}} .
\end{displaymath}

Given the result above, it follows that
\begin{eqnarray*}
\overline{\partial} \omega_{K1} & = & 
\int \prod  d \lambda^{\overline{x}} d \chi^r
\, \overline{D} \exp( - {\cal A}_{K1} ) ,
\\
& = &
\int \prod d \lambda^{\overline{x}} d \chi^r
\left( \overline{D} \: + \: h^{x \overline{x}} s_x 
\frac{\partial}{\partial \lambda^{\overline{x}}
} \right) \exp( - {\cal A}_{K1} ) ,
\\
& = & 0 ,
\end{eqnarray*}
and so $\omega_{K1}$ defines a $\overline{\partial}$-closed form.

Next, we will argue that the cohomology class of
$\omega_{K1}$ is unchanged by antiholomorphic
deformations of the section $s$.  In other words,
consider the one-parameter family
\begin{displaymath}
{\cal A}_{K1, \tau} \: = \: 
h^{x \overline{x}} s_x ( \overline{s}_{\overline{x}} + 
\tau \overline{t}_{\overline{x}})
\: + \: \chi^{\overline{\imath}} \lambda^{\overline{x}} \,
\overline{D}_{\overline{\imath}} ( \overline{s}_{\overline{x}}
+ \tau \overline{t}_{\overline{x}} )
\: + \: \chi^{r} \lambda^{\gamma} \tilde{F}_{r \gamma} \: + \:
F_{\overline{\imath} 
r \overline{x} \gamma} \chi^{\overline{\imath}} 
\chi^{r} \lambda^{\overline{x}} \lambda^{\gamma} .
\end{displaymath}
Then,
\begin{eqnarray*}
\frac{d}{d \tau} \omega_{ K1, \tau } & = &
\frac{d}{d \tau}  \int \prod d \lambda^{\overline{x}} d \chi^r \, 
\exp( - {\cal A}_{K1, \tau} ) ,
\\
& = &
-  \int \prod d \lambda^{\overline{x}} d \chi^r
\left( h^{x \overline{x}} s_x \overline{t}_{\overline{x}}
\: + \:  \chi^{\overline{\imath}} \lambda^{\overline{x}} \,
\overline{D}_{\overline{\imath}} \overline{t}_{\overline{x}} \right)
\exp( - {\cal A}_{K1, \tau} ) ,
\\
& = &
\int \prod d \lambda^{\overline{x}} d \chi^r
\left( \overline{D} \: + \: h^{x \overline{x}} s_x 
\frac{\partial}{\partial \lambda^{\overline{x}}
} \right)
\left( - \lambda^{\overline{x}} \, \overline{t}_{\overline{x}} \right)
\exp( - {\cal A}_{K1, \tau} ) ,
\\
& = & 
\overline{\partial} 
 \int \prod d \lambda^{\overline{x}} d \chi^r
\left( - \lambda^{\overline{x}} \,\overline{t}_{\overline{x}} \right)
\exp( - {\cal A}_{K1, \tau} ) ,
\end{eqnarray*}
thus demonstrating the desired result.

As we will see later in section~\ref{sect:A2-kernels},
this analogue of a Mathai-Quillen form appears
in the A/2 model pseudo-topological
field theories \cite{GuffinSharpe:A-twistedheterotic},
where it plays a role analogous to that of the Mathai-Quillen
form in some ordinary topological field theories.  A/2 model correlation
functions amount to sheaf cohomology computations, so the fact that this
deformed Mathai-Quillen form defines an $\overline{\partial}$-cohomology
class is precisely what is needed to correlate with its physical role.

One simple special case is that in which ${\cal G}=0$, so that our
proposed analogue of a Mathai-Quillen form simply reduces to $M$ rather
than some complete intersection inside $M$.  In this case, 
${\cal A}_{K1}$ becomes simply
\begin{displaymath}
 \chi^{r} \lambda^{\gamma} \tilde{F}_{r \gamma} .
\end{displaymath}
Intuitively, its role is clear:  the Grassmann quantity $\lambda^{\gamma}$
above annihilates those $\lambda$ which are not in the kernel
of $F$, thus reducing sheaf cohomology valued in 
${\cal F}_1$ to sheaf cohomology valued in the
kernel of $\tilde{F}: {\cal F}_1 \rightarrow {\cal F}_2$.

\subsubsection{Specialization to ordinary Mathai-Quillen and its deformations}
\label{sect:specialize:22MQ}

An important special case of the construction above is to the ordinary
Mathai-Quillen form and its deformations.

First, let us outline a family of deformations of (pullbacks of) ordinary
Mathai-Quillen forms.
Consider the pullback, 
$\mathfrak{s} ^* U({\cal G}, \nabla) \propto \mathfrak{s} ^* u({\cal G}, \nabla) $,
defined earlier.  Assume that ${\cal G}$ is a holomorphic vector bundle.
Consider deforming $ \mathfrak{s}^* u({\cal G},\nabla) $ to
%%%%%%%%%%%%%%%%%
\begin{equation}
\label{omega-delta-s}
\omega_{\delta s}({\cal G},\nabla)
   = 
     \int d \rho \,
     \exp 
     \left(
          - {\cal A}_{\delta s}    
     \right),  
\end{equation}
%%%%%%%%%%%%%%%%%
\begin{equation}
{\cal A}_{\delta s}
   = {\cal A} 
   + \left\langle
            \rho^{p'} f_{p'},    
            d \phi^i
            (\delta s)_{i p}
            e^p
     \right\rangle_{{\cal G}}.
\end{equation}
%%%%%%%%%%%%%%%%%
Here, $ \cal{A} $ is given by (\ref{S^{(2,2)}_0}) and
\begin{displaymath}
(\delta s)_{i p} \: \in \: \Gamma\left( \pi^* {\cal G} \otimes \pi^* TM 
\right) .
\end{displaymath}

These deformations (and pullbacks of ordinary Mathai-Quillen forms themselves,
in the case $\delta s = 0$)
are special cases of the sheaf-cohomological analogue of the
previous section.  Specifically, this corresponds to the special
case that ${\cal F}_1 = TM$, ${\cal F}_2 = {\cal G}$, and with
map $\tilde{F}: {\cal F}_1 \rightarrow {\cal F}_2$ 
defined by
\begin{displaymath}
F_{i p} \: = \: D_i s_p \: + \: (\delta s)_{i p} ,
\end{displaymath}
where $s_p$ is a holomorphic section of ${\cal G}$.
Then, note that
\begin{displaymath}
\overline{\partial}_{\overline{\imath}} \left( D_j s_p \: + \:
(\delta s)_{j p} \right) \: = \:
\left[ \overline{D}_{\overline{\imath}}, D_j \right] s_p \: = \: 
R_{\overline{\imath} j p \overline{p} } h^{q \overline{p} } s_q
\end{displaymath}
so that the curvature term, now an element of
\begin{displaymath}
H^1\left(M, \Omega^1_M \otimes {\cal G} \otimes {\cal G}^* \right)
\end{displaymath}
is determined by the curvature of ${\cal G}$, specifically,
the Atiyah class of
${\cal G}^*$, exactly as needed to match ordinary Mathai-Quillen forms.

In terms of sheaf cohomology, this is the special case in which
${\cal E}'$ is a deformation of $TY$, with deformation determined by
$\delta s$.  If $\delta s=0$, then ${\cal E}' = TY$, and this
sheaf-cohomological analogue of a Mathai-Quillen form is relating
\begin{displaymath}
H^{\bullet}\left(M, \wedge^{\bullet} T^* M \right) \: = \:
H^{\bullet,\bullet}(M)
\end{displaymath}
to
\begin{displaymath}
H^{\bullet}\left(Y, \wedge^{\bullet} T^* Y \right) \: = \:
H^{\bullet,\bullet}(Y),
\end{displaymath}
just as expected.

Applying previous results, we know that
\begin{equation*}
\overline{\partial} \omega_{\delta s}({\cal G},\nabla)
   = 0 \, .
\end{equation*}
(If instead we deformed the original pullback of the Mathai-Quillen form
by $(\delta \overline{s})_{\overline{\imath}}$ analogously, the result would
similarly
be a $\partial$-closed differential form.)

We also know, from specializing previous results, that the cohomology
class of this
deformation of the pullback of the Mathai-Quillen form is invariant under
``antiholomorphic'' deformations of $s$, for the notion of
antiholomorphic deformation defined earlier.

One formal\footnote{
Physically, actually taking such a limit is more subtle than we have
indicated, because for example this removes the bosonic potential which adds
new scalar zero mode directions.
} consequence of the result above is that, by rescaling
$\overline{s}_{\overline{p}}$ to zero, $\omega_{\delta s}$ can be written in
the purely holomorphic
form
\begin{displaymath}
  \int d \rho \,
     \exp 
     \left(
          - \left(
\left\langle 
            \rho^{p'} f_{p'}, ( D s_p + d \phi^i (\delta s)_{i p}) e^p                   
    \right\rangle_{\cal G}
  + \frac{1}{2} 
    \left(
           \rho, \mathcal{R} \rho
    \right)_{{\cal G}^{*}}
\right) \right) ,
\end{displaymath}
which is in the same $\overline{\partial}$-cohomology class.

The $\overline{\partial}$-cohomology class of this
deformation does seem to depend upon the $(\delta s)_{i p}$, at least naively.
Let
$ 
(\delta s)^{\tau}_{i p} 
   = (\delta s)_{i p}
   + \tau (\delta s)^{\prime}_{i p}
$
and
$
{\cal A}_{\delta s, \tau}
  = {\cal A} 
  + \left\langle 
           \rho^{p'} f_{p'},
           d \phi^i
           (\delta s)^{\tau}_{i p}
           e^p
     \right\rangle_{{\cal G}}.
$
Then
\begin{align*}
\frac{d}{d \tau} 
\omega_{\delta s, \tau}({\cal G},\nabla)
  &= 
     \frac{d}{d \tau}
     \int d \rho \,
     \exp
      \left(
           - {\cal A}_{\delta s, \tau}  
     \right) ,
\\[1ex]
  &= 
   - \int d \rho \,
     \left\langle
            \rho^{p'} f_{p'},
            d \phi^i 
            (\delta s)^{\prime}_{i p}
            e^p
     \right\rangle_{{\cal G}}
     \exp
     \left(
           - {\cal A}_{\delta s, \tau}  
     \right).          
\end{align*}
It follows that
\begin{displaymath}
\omega_{\delta s, \tau_2}({\cal G},\nabla) 
- \omega_{\delta s, \tau_1}({\cal G},\nabla)
   = 
   - \int^{\tau_2}_{\tau_1} d \tau
     \int d \rho \,
     \left\langle
            \rho^{p'} f_{p'},
            d \phi^i 
            (\delta s)^{\prime}_{i p}
            e^p
     \right\rangle_{{\cal G}}     
     \exp
     \left(
          - {\cal A}_{\delta s, \tau}  
     \right) ,         
\end{displaymath}
which is at least not obviously $\overline{\partial}$-exact.
We will comment on the physical meaning of this result in
the next subsection.

\subsubsection{Application to work of Melnikov-McOrist}

In \cite{McOristMelnikov:Summing}, it was argued, based on physical
properties of gauged linear sigma models, that A/2 correlation functions
for deformations of the tangent bundle should be independent of the
deformation $\delta s$.  One of the original hopes of this work was to
see that result explicitly in a cleaner setting.

Although it is true that the deformed object
$\omega_{\delta s}$ given by (\ref{omega-delta-s}) is independent
of antiholomorphic deformations of the section $s$, as noted earlier,
we have not found a simple explanation for the claim above in this
context.

Implicitly in this paper we are discussing mathematics motivated by
Landau-Ginzburg models, which are related to the gauged linear sigma models
of \cite{McOristMelnikov:Summing} via renormalization group flow.  It is
entirely possible that their results are only visible in gauged linear
sigma models, that the renormalization group flow obscures the result
in question.

\subsubsection{Second kernel construction}

Another construction exists for kernels.
Suppose that one is interested in computing integrals
over some space $Y \equiv \{s=0\} \subset M$ ($s \in \Gamma({\cal G})$)
of sheaf cohomology classes
\begin{displaymath}
{\cal O} \: \in \: H^{\bullet}\left(Y, \wedge^{\bullet} {\cal E}' \right) ,
\end{displaymath}
where ${\cal E}'$ is a holomorphic vector bundle on $Y$.

We propose that if ${\cal E}'$ is given as the restriction to $Y$
of the kernel of a surjective holomorphic map $\tilde{F}: {\cal F}_1 \rightarrow
{\cal F}_2$ (${\cal F}_1$, ${\cal F}_2$ holomorphic vector bundles on $M$),
then at least for some ${\cal O}$'s
it is possible to write the integral in a different form
\begin{displaymath}
\int_Y {\cal O}_1 \wedge \cdots \wedge {\cal O}_n \: \propto \:
\int_X \tilde{\cal O}_1 \wedge \cdots \wedge \tilde{\cal O}_n \wedge \omega_{K2} ,
\end{displaymath}
where $X$ is the total space of $\pi: {\cal F}_2^* \rightarrow M$,
$\tilde{\cal O}_i$ (when it exists) is an element of hypercohomology 
\begin{displaymath}
{\mathbb H}^{\bullet}\left(X, \cdots \: \longrightarrow \:
\wedge^2 \pi^* {\cal F}_1 \: \longrightarrow \: \pi^* {\cal F}_1 
\: \longrightarrow
\: {\cal O}_X \right) 
\end{displaymath}
(with maps given by contraction with $p \tilde{F}$)
`lifting' ${\cal O}_i$, in a fashion we shall describe momentarily,
and $\omega_{K2}$ is 
hypercohomology class in 
\begin{displaymath}
{\mathbb H}^{g}\left(X, \pi^* \det {\cal G}^* \otimes \left(
\cdots \: \longrightarrow \:
\wedge^2 \pi^* {\cal F}_1 \: \longrightarrow \: \pi^* {\cal F}_1 
\: \longrightarrow
\: {\cal O}_X \right)
\right) ,
\end{displaymath}
where $g = {\rm rk}\, {\cal G}$
and one has an isomorphism
\begin{displaymath}
K_X \: \cong \: \pi^* \det {\cal G}^* \otimes \pi^* \det {\cal F}_1
\end{displaymath}
that restricts to the isomorphism $\det {\cal E}' \cong K_Y$ needed to
define the corresponding integral on $Y$.

We propose an expression for $\omega_{K2}$, and check its properties.

First, let us describe the relation between
\begin{displaymath}
{\cal O} \: \in \: H^{\bullet}\left(Y, \wedge^{\bullet} {\cal E}' \right)
\mbox{ and }
\tilde{\cal O} \: \in \:
{\mathbb H}^{\bullet}\left(X, \cdots \: \longrightarrow \:
\wedge^2 \pi^* {\cal F}_1 \: \longrightarrow \: \pi^* {\cal F}_1 
\: \longrightarrow
\: {\cal O}_X \right) . 
\end{displaymath}
Briefly, we can use the fact that\footnote{
This section works in the language of hypercohomology, despite this
isomorphism, because the pertinent analogue of a Mathai-Quillen form
seems most straightforwardly expressed in the language of hypercohomology.
}
\begin{displaymath}
{\mathbb H}^{\bullet}\left(X, \cdots \: \longrightarrow \:
\wedge^2 \pi^* {\cal F}_1 \: \longrightarrow \: \pi^* {\cal F}_1 
\: \longrightarrow
\: {\cal O}_X \right) \: \cong \:
H^{\bullet}\left(M, \wedge^{\bullet} {\cal E}' \right)
\end{displaymath}
(a consequence of a minor variation of an argument given in
\cite{GuffinSharpe:A-twistedheterotic}[appendix A]).
If we let $i: Y \hookrightarrow M$ denote the inclusion, then the relation
between the pair ${\cal O}$, $\tilde{\cal O}$ is simply
\begin{displaymath}
{\cal O} \: = \: i^* \tilde{\cal O},
\end{displaymath}
using the isomorphism above.

Our proposal for $\omega_{K2}$
is given by the
Grassmann integral
\begin{displaymath}
\omega_{K2} \: = \: \int \prod d \lambda^{\overline{x}} \exp(- {\cal A}_{K2}) ,
\end{displaymath}
where
\begin{displaymath}
{\cal A}_{K2} \: = \: 
h^{x \overline{x}} s_x \overline{s}_{\overline{x}}
\: + \:
h^{\gamma \overline{\gamma}} p^r \overline{p}^{\overline{r}}
\tilde{F}_{r \gamma} 
\tilde{\overline{F}}_{\overline{r} \overline{\gamma}}
\: + \:
\chi^{\overline{\imath}} \lambda^{\overline{x}} \, \overline{D}_{\overline{\imath}} 
\overline{s}_{\overline{x}} \: + \:
\chi^{\overline{r}} \theta_{\gamma} \tilde{\overline{F}}_{
\overline{r} \overline{\gamma}} h^{\gamma \overline{\gamma}} \: + \:
\chi^{\overline{\imath}} \theta_{\gamma} \,
\overline{p}^{\overline{r}} \, \overline{D}_{\overline{\imath}} \tilde{\overline{F}}_{
\overline{r} \overline{\gamma}} h^{\gamma \overline{\gamma}} ,
\end{displaymath}
where $x$ indexes local coordinates along the fibers of ${\cal G}$,
$\gamma$ indexes local coordinates along the fibers of
${\cal F}_1$, $r$ indexes local coordinates (denoted $p$) along
along the fibers of ${\cal F}_2^*$, and $i$ indexes local
coordinates on $M$.
Note that since each $\lambda^{\overline{x}}$ is
paired with a $\chi^{\overline{\imath}}$, integrating out
$\lambda^{\overline{x}}$'s results in $g$ factors of $\chi^{\overline{
\imath}}$, interpreted as a degree $g$ form, as advertised.

Now, we can argue formally that the analogue of a Mathai-Quillen form
defined above is an element of hypercohomology.  The central
point is that
\begin{eqnarray*}
\left( \overline{D} \: + \: h^{x \overline{x}} s_x \frac{\partial}{
\partial \lambda^{\overline{x}} } \right) {\cal A}_{K2} 
& = &
p^r \tilde{F}_{r \gamma} \chi^{\overline{r}} h^{\gamma \overline{\gamma}}
\tilde{ \overline{F} }_{\overline{r} \overline{\gamma} } 
\: + \:
p^r \tilde{F}_{r \gamma} \overline{p}^{\overline{r}} \chi^{\overline{\imath}} \,
\overline{\partial}_{\overline{\imath}} \left( h^{\gamma \overline{\gamma}} \,
\tilde{ \overline{F} }_{\overline{r} \overline{\gamma} } \right)
\\
& & \hspace*{2in}
\: + \:
\chi^{\overline{\jmath}} \chi^{\overline{\imath}} \theta_{\gamma} \,
\overline{p}^{\overline{r}} \, \overline{\partial}_{\overline{\jmath}}
\left( h^{\gamma \overline{\gamma} } \, \overline{D}_{\overline{\imath}}
\tilde{ \overline{F} }_{\overline{r} \overline{\gamma} } \right),
\\
& = &
- p^r \tilde{F}_{r \gamma} \frac{\partial}{\partial \theta_{\gamma} }
{\cal A}_{K2},
\end{eqnarray*}
where 
\begin{displaymath}
\overline{D} \: = \: \chi^{\overline{\imath}} \, \overline{\partial}_{
\overline{\imath}} \: + \: \chi^{\overline{r}} \, \overline{\partial}_{
\overline{r}}
\end{displaymath}
and we have used the fact that 
\begin{displaymath}
\chi^{\overline{\jmath}} \chi^{\overline{\imath}} \theta_{\gamma} \,
\overline{p}^{\overline{r}} \, \overline{\partial}_{\overline{\jmath}}
\left( h^{\gamma \overline{\gamma} } \, \overline{D}_{\overline{\imath}}
\tilde{ \overline{F} }_{\overline{r} \overline{\gamma} } \right)
\: = \: 0,
\end{displaymath}
which follows from the fact that ${\cal F}_1$, ${\cal F}_2$ are holomorphic.
Given the contribution to the coefficients from the Grassmann integral
itself, it is now straightforward to show that $\omega_{K2}$ is an
element of the aforementioned hypercohomology group.

Next, we will argue that the cohomology class of $\omega_{K2}$ is
unchanged by antiholomorphic deformations of the section $s$.
As the details are somewhat more complicated than the argument in the
previous subsection, we sketch the details here.
Consider the one-parameter family
\begin{eqnarray*}
{\cal A}_{K2, \tau} & = &
h^{x \overline{x}} s_x \left( \overline{s}_{\overline{x}}
\: + \: \tau \overline{t}_{\overline{x}} \right)
\: + \:
h^{\gamma \overline{\gamma}} p^r \overline{p}^{\overline{r}}
\tilde{F}_{r \gamma} 
\tilde{\overline{F}}_{\overline{r} \overline{\gamma}}
\: + \:
\chi^{\overline{\imath}} \lambda^{\overline{x}} \left( \overline{D}_{\overline{\imath}} 
\overline{s}_{\overline{x}} \: + \: \tau \overline{D}_{\overline{\imath}}
\overline{t}_{\overline{x}} \right)
\\
& & \hspace*{2in}
\: + \:
\chi^{\overline{r}} \theta_{\gamma} \tilde{\overline{F}}_{
\overline{r} \overline{\gamma}} h^{\gamma \overline{\gamma}} \: + \:
\chi^{\overline{\imath}} \theta_{\gamma} \,
\overline{p}^{\overline{r}} \, \overline{D}_{\overline{\imath}} \tilde{\overline{F}}_{
\overline{r} \overline{\gamma}} h^{\gamma \overline{\gamma}} ,
\end{eqnarray*}
so that
\begin{eqnarray*}
\frac{d}{d \tau} \omega_{K2, \tau} & = &
\frac{d}{d \tau} \int \prod d \lambda^{\overline{x}} \exp\left( - 
{\cal A}_{K2, \tau} \right) ,
\\
& = &
- \int \prod d \lambda^{\overline{x}} \left(
h^{x \overline{x}} s_x \overline{t}_{\overline{x}} \: + \:
\chi^{\overline{\imath}} \lambda^{\overline{x}} \, 
\overline{D}_{\overline{\imath}}
\overline{t}_{\overline{x}} \right)
\exp\left( - 
{\cal A}_{K2, \tau} \right) ,
\\
& = & 
\int \prod d \lambda^{\overline{x}} \left(
\overline{D} \: + \: h^{x \overline{x}} s_x \frac{\partial}{
\partial \lambda^{\overline{x}} } \: + \:
p^r \tilde{F}_{r \gamma} \frac{\partial}{\partial \theta_{\gamma} }
\right)
\left( - \lambda^{\overline{x}} \, \overline{t}_{\overline{x}} \right)
\exp\left( - 
{\cal A}_{K2, \tau} \right) ,
\\
& = &
\left( \overline{\partial} \: + \:
p^r \tilde{F}_{r \gamma} \frac{\partial}{\partial \theta_{\gamma} }
\right)
\int \prod d \lambda^{\overline{x}} \left( - \lambda^{\overline{x}} \, \overline{t}_{\overline{x}} \right)
\exp\left( - 
{\cal A}_{K2, \tau} \right) ,
\end{eqnarray*}
establishing the desired result.

The relevance of this construction to physics will be discussed in 
section~\ref{sect:b2-kernels}.

\subsection{Cokernels}   \label{sect:MQ:cokernels}

\subsubsection{First cokernel construction}

% From B/2 twist for cokernels

Analogous constructions exist for cokernels.
Suppose that one is interested in computing integrals over
some space
$Y \equiv \{ s=0 \} \subset M$ ($s \in \Gamma({\cal G})$) of
sheaf cohomology classes
\begin{displaymath}
{\cal O}_i \: \in \: H^{\bullet}(Y, \wedge^{\bullet} {\cal E}' ) ,
\end{displaymath}
where ${\cal E}'$ is a holomorphic vector bundle on $Y$, as before.

We propose that if ${\cal E}'$ is given as the restriction to $Y$ of
the cokernel of a smooth injective map
$\tilde{E}: 
{\cal F}_1 \rightarrow {\cal F}_2$ (${\cal F}_1$, ${\cal F}_2$
holomorphic vector bundles on $M$), whose restriction to $Y$ is
holomorphic, then at least for some ${\cal O}$'s it is possible to
write
\begin{displaymath}
\int_Y {\cal O}_1 \wedge \cdots \wedge {\cal O}_n \: \propto \:
\int_M \tilde{\cal O}_1 \wedge \cdots \wedge \tilde{\cal O}_n \wedge \omega_{CK1} ,
\end{displaymath}
where $\tilde{\cal O}_i$ (when it exists) is an element of
$H^{\bullet}(M, \wedge^{\bullet} {\cal F}_2)$ `lifting' ${\cal O}_i$,
in a fashion we shall discuss momentarily,
and
$\omega_{CK1}$ is a $\overline{\partial}$-closed
analogue of a Mathai-Quillen form,
\begin{equation}  \label{eq:MQ:cokernel1:defn}
\omega_{CK1} \: \in \: H^{g}\left(M, \wedge^{f_1} {\cal F}_2
\otimes  \det {\cal G}^* \otimes \det {\cal F}_1^*
\right)
\end{equation}
($f_i = {\rm rk}\, {\cal F}_i$, $g = {\rm rk}\, {\cal G}$),
where one has an isomorphism
\begin{displaymath}
K_M \: \cong \: \det {\cal F}_1^* \otimes \det {\cal F}_2 \otimes
\det {\cal G}^*
\end{displaymath}
that restricts to the isomorphism
$\det {\cal E}' \cong K_Y$ needed to define the corresponding integral on $Y$.

As a consistency check, notice that
\begin{displaymath}
\tilde{\cal O}_1 \wedge \cdots \wedge \tilde{\cal O}_n \: \in \:
H^{{\rm dim}\, M \: - \: g}\left( M, \wedge^{f_2 - f_1} {\cal F}_2
\right)
\end{displaymath}
so that the cohomology class of $\omega_{CK1}$ is correct for the integrand
of $M$ to be a top-form.

We propose an expression for $\omega_{CK1}$ below, and check its properties.

First, let us explain the relationship between 
\begin{displaymath}
{\cal O}  \: \in \: H^{\bullet}\left(Y, \wedge^{\bullet} {\cal E}' \right)
\mbox{ and }
\tilde{\cal O} \: \in \: H^{\bullet}\left(M, 
\wedge^{\bullet} {\cal F}_2 \right).
\end{displaymath}
Let $i: Y \hookrightarrow M$ denote inclusion, so
\begin{displaymath}
i^* \tilde{\cal O} \: \in \: H^{\bullet}\left(Y, 
\wedge^{\bullet} {\cal F}_2 |_Y \right).
\end{displaymath}
Next, from the short exact sequence
\begin{displaymath}
0 \: \longrightarrow \: {\cal F}_1|_Y \: \longrightarrow \: {\cal F}_2|_Y
\: \longrightarrow \: {\cal E}' \: \longrightarrow \: 0
\end{displaymath}
we have a surjective map 
\begin{displaymath}
\wedge^{\bullet} {\cal F}_2|_Y \: \longrightarrow \: 
\wedge^{\bullet} {\cal E}'
\end{displaymath}
which induces
\begin{displaymath}
j_*: \: H^{\bullet}\left(Y, \wedge^{\bullet} {\cal F}_2|_Y \right)
\: \longrightarrow \:
H^{\bullet}\left(Y, \wedge^{\bullet} {\cal E}' \right).
\end{displaymath}
Then, the pair ${\cal O}$, $\tilde{\cal O}$, when they exist, are related by
\begin{displaymath}
{\cal O} \: = \: j_* i^* \tilde{\cal O} .
\end{displaymath}

Our proposal for $\omega_{CK1}$
is given by
the Grassmann integral
\begin{displaymath}
\omega_{CK1} \: = \: \int \prod d \lambda^{x} d \chi^{m}
\, \exp( - {\cal A}_{CK1}) ,
\end{displaymath}
where
\begin{displaymath}
{\cal A}_{CK1} \: = \:
h_{x \overline{x}} s^x \overline{s}^{\overline{x}} \: + \:
\chi^{\overline{\imath}} \lambda^x \overline{D}_{\overline{\imath}} \overline{s}^{
\overline{x}} \, h_{x \overline{x}} \: + \:
\chi^m \theta_{\gamma}  \tilde{E}_m^{\gamma}
h_{\gamma \overline{\gamma}} \: + \:
F_{\overline{\imath} m x \overline{\gamma}} \chi^{\overline{\imath}}
\chi^m \lambda^x \theta_{\gamma} h^{\gamma \overline{\gamma}} ,
\end{displaymath}
where
$x$ indexes local coordinates along the fibers of ${\cal G}$,
$m$ indexes local coordinates (denoted $q$) along the fibers of
${\cal F}_1$, $\gamma$ indexes local coordinates
along the fibers of ${\cal F}_2$, $i$ indexes local
coordinates on $M$,
and $s^x$ now denotes a component of a holomorphic
section of ${\cal G}$ (rather than $s_x$ as was used in the discussion of
kernels, for reasons of notational sanity).
The curvature term
\begin{displaymath}
F_{\overline{\imath} m x \overline{\gamma}} \chi^{\overline{\imath}}
\chi^m \lambda^x \theta_{\gamma} h^{\gamma \overline{\gamma}}
\end{displaymath}
is fixed to solve\footnote{
Solutions exist for reasons closely analogous to those in the
analogous discussion in the kernels section.  For example, if
${\cal G}$ is a line bundle, then this curvature term is the coboundary of
\begin{displaymath}
\tilde{E}|_Y \: \in \: H^0(Y, {\cal F}_1^*|_Y \otimes {\cal F}_2|_Y)
\end{displaymath}
in the long exact sequence derived from tensoring
\begin{displaymath}
0 \: \longrightarrow \: {\cal G}^* \: \longrightarrow \: {\cal O}
\: \longrightarrow \: {\cal O}_Y \: \longrightarrow \: 0
\end{displaymath}
by ${\cal F}_1^* \otimes {\cal F}_2$.  In this special case, the
condition above amounts to the definition of the coboundary map.
} 
\begin{equation}  \label{eq:dtildeE-sF}
h_{\gamma \overline{\gamma}}
\overline{\partial}_{\overline{\imath}} \tilde{E}_m^{\gamma}
\: = \:
s^x F_{\overline{\imath} m  \overline{\gamma} x} \: = \: 
- s^x F_{\overline{\imath} m x \overline{\gamma}} 
\end{equation}
and defines an element of
\begin{displaymath}
H^1\left(M, {\cal F}_1^* \otimes {\cal F}_2 \otimes {\cal G}^* \right).
\end{displaymath}
(Physically, $F$ arises as part of the curvature of a holomorphic
vector bundle, hence is $\overline{\partial}$-closed by virtue of the
Bianchi identity, much as in the closely related discussion in the
kernels section.)

Now, let us explain some aspects of $\omega_{CK1}$ in more detail.
In ${\cal A}_{CK1}$ above, every $\lambda^x$ is accompanies by a $\chi^{
\overline{\imath}}$, so integrating out the $\lambda^x$'s should result
in ${\rm rk}\, {\cal G} = g$ of $\chi^{\overline{\imath}}$'s, hence
a degree $g$ cohomology class.  Similarly, each $\chi^m$ is 
accompanies by a $\lambda^{\overline{\gamma}}$, so integrating out the
$\chi^m$'s should result in coefficients 
$\wedge^{f_1} {\cal F}_2$.  The Grassmann integrals are responsible for
a $\det {\cal G}^* \otimes \det {\cal F}_1^*$ factor in the coefficients.
Thus, $\omega_{CK1}$ is a form of the type indicated in
equation~(\ref{eq:MQ:cokernel1:defn}).  We shall show it is
$\overline{\partial}$-closed momentarily.

Now, we can argue formally that the analogue of a Mathai-Quillen form
defined above is $\overline{\partial}$-closed.  The central point is that
\begin{eqnarray*}
\left( \overline{D} \: + \:  s^x \frac{\partial}{\partial
\lambda^x} \right) {\cal A}_{CK1} & = &
\chi^{\overline{\imath}} \chi^m \theta_{\gamma}
\left(  \overline{D}_{\overline{\imath}} 
\tilde{E}_m^{\gamma} \: + \: s^x F_{\overline{\imath} m x \overline{\gamma}}
h^{\gamma \overline{\gamma}}
\right) , \\
& = & 0 ,
\end{eqnarray*}
using equation~(\ref{eq:dtildeE-sF}), where
\begin{displaymath}
\overline{D} \: = \: \chi^{\overline{\imath}} \overline{\partial}_{
\overline{\imath}}.
\end{displaymath}
Proceeding in the same fashion as before, it is simple to show that
$\overline{\partial} \omega_{CK1} = 0$.

Next, we will argue that the cohomology class of $\omega_{CK1}$ is unchanged
by `antiholomorphic' deformations of $s$.  To that end, consider the
one-parameter family
\begin{displaymath}
{\cal A}_{CK1, \tau} \: = \:
h_{x \overline{x}} s^x \left( \overline{s}^{\overline{x}} 
\: + \: \tau \overline{t}^{\overline{x}} \right)
\: + \:
\chi^{\overline{\imath}} \lambda^x \left( \overline{D}_{\overline{\imath}} \overline{s}^{
\overline{x}} \: + \: \tau \overline{D}_{\overline{\imath}} \overline{t}^{\overline{x}}
\right) h_{x \overline{x}} \: + \:
\chi^m \theta_{\gamma}  \tilde{E}_m^{\gamma}
h_{\gamma \overline{\gamma}} \: + \:
F_{\overline{\imath} m x \overline{\gamma}} \chi^{\overline{\imath}}
\chi^m \lambda^x \theta_{\gamma} h^{\gamma \overline{\gamma}}.
\end{displaymath}
Then, 
\begin{eqnarray*}
\frac{d}{d \tau} \omega_{CK1, \tau} & = &
\frac{d}{d\tau} \int \prod d \lambda^x d \chi^m \exp\left(
- {\cal A}_{CK1, \tau} \right) ,
\\
& = & - \int \prod d \lambda^x d \chi^m \left(
h_{x \overline{x}} s^x \overline{t}^{\overline{x}} \: + \:
\chi^{\overline{\imath}} \lambda^x h_{x \overline{x}} \overline{D}_{\overline{\imath}}
\overline{t}^{\overline{x}} \right) \exp\left( - {\cal A}_{CK1, \tau} \right) ,
\\
& = & \int \prod d \lambda^x d \chi^m \left(
\overline{D} \: + \: s^x \frac{\partial}{\partial \lambda^x} \right)
\left( - h_{x \overline{x}} \lambda^x \overline{t}^{\overline{x}}
\right) \exp\left( - {\cal A}_{CK1, \tau} \right) ,
\\
& = & \overline{\partial} \int \prod d \lambda^x d \chi^m
\left( - h_{x \overline{x}} \lambda^x \overline{t}^{\overline{x}}
\right) \exp\left( - {\cal A}_{CK1, \tau} \right),
\end{eqnarray*}
from which the desired result follows.

In passing, an important special case to which this cokernels
construction is relevant is Euler sequences and generalizations describing
tangent bundles of toric varieties.  In general, the tangent bundle
of a toric variety $Z$ is the cokernel
\begin{displaymath}
0 \: \longrightarrow \: {\cal O} \otimes V \: \stackrel{\tilde{E}}{
\longrightarrow} \:
\oplus_D {\cal O}(D) \: \longrightarrow \: TZ \: \longrightarrow \: 0 ,
\end{displaymath}
where $V$ is a vector space, and the $D$'s in the middle entry are
toric divisors.  By deforming $\tilde{E}$, one can deform $TZ$ to a 
different holomorphic vector bundle on $Z$.

The physical relevance of this construction will be discussed in
section~\ref{sect:b2-cokernels}.

\subsubsection{Second cokernel construction}

% A/2 twist of cokernels

Suppose that one is interested in computing integrals over
some space
$Y \equiv \{ s=0 \} \subset M$ ($s \in \Gamma({\cal G})$) of
sheaf cohomology classes
\begin{displaymath}
{\cal O} \: \in \: H^{\bullet}(Y, \wedge^{\bullet} {\cal E}'^* ) ,
\end{displaymath}
where ${\cal E}'$ is a holomorphic vector bundle on $Y$, as before,
but the sheaf cohomology classes are valued in powers of ${\cal E}'^*$
instead of ${\cal E}'$.

We propose that if ${\cal E}'$ is given as the restriction to $Y$ of
the cokernel of a holomorphic injective map
$\tilde{E}: 
{\cal F}_1 \rightarrow {\cal F}_2$ (${\cal F}_1$, ${\cal F}_2$
holomorphic vector bundles on $M$), 
then at least for some ${\cal O}$'s it is possible to
write the integral in the different form
\begin{displaymath}
\int_Y {\cal O}_1 \wedge \cdots \wedge {\cal O}_n \: \propto \:
\int_X \tilde{\cal O}_1 \wedge \cdots \wedge \tilde{\cal O}_n \wedge \omega_{CK2} ,
\end{displaymath}
where $\tilde{\cal O}_i$ (when it exists) is an element of
hypercohomology
\begin{displaymath}
{\mathbb H}^{\bullet}
\left(X, \cdots \: \longrightarrow \: \wedge^2 \pi^* {\cal F}_2^*
\: \longrightarrow \: \pi^* {\cal F}_2^* \: \longrightarrow \: 
{\cal O}_X \right)
\end{displaymath}
(with maps given by inclusion along $q \tilde{E}$)
`lifting' ${\cal O}_i$, in a fashion we shall explain momentarily,
\begin{displaymath}
X \: = \: {\rm Tot}\, \left( 
{\cal F}_1 \: \stackrel{\pi}{\longrightarrow} \:
M \right),
\end{displaymath}
and
$\omega_{CK2}$ is a $\overline{\partial}$-closed
analogue of a Mathai-Quillen form,
\begin{equation}  \label{eq:MQ:cokernel:defn}
\omega_{CK2} \: \in \: {\mathbb H}^{g }\left(X, 
\pi^* \det {\cal G}^* \otimes \left(
\cdots \: \longrightarrow \: \wedge^* \pi^* {\cal F}_2^*
\: \longrightarrow \: \pi^* {\cal F}_2^* \: \longrightarrow \: 
{\cal O}_X \right)
\right)
\end{equation}
($g = {\rm rk}\, {\cal G}$),
where one has an isomorphism
\begin{displaymath}
K_X \: \cong \: \pi^* \det {\cal G}^* \otimes \pi^* \det {\cal F}_2^*
\end{displaymath}
that restricts to the isomorphism
$\det {\cal E}'^* \cong K_Y$ needed to define the corresponding integral on $Y$.
Note that in this section, unlike the last, $\tilde{E}$ must be holomorphic
everywhere.

We propose an expression for $\omega_{CK2}$ below, and check its properties.

First, let us discuss the relationship between
\begin{displaymath}
{\cal O} \: \in \: H^{\bullet}(Y, \wedge^{\bullet} {\cal E}'^* )
\mbox{ and }
\tilde{\cal O} \: \in \:
{\mathbb H}^{\bullet}
\left(X, \cdots \: \longrightarrow \: \wedge^2 \pi^* {\cal F}_2^*
\: \longrightarrow \: \pi^* {\cal F}_2^* \: \longrightarrow \: 
{\cal O}_X \right) .
\end{displaymath}
Briefly, we can use the isomorphism
\begin{displaymath}
{\mathbb H}^{\bullet}
\left(X, \cdots \: \longrightarrow \: \wedge^2 \pi^* {\cal F}_2^*
\: \longrightarrow \: \pi^* {\cal F}_2^* \: \longrightarrow \: 
{\cal O}_X \right)
\: \cong \:
H^{\bullet}\left(M, \wedge^{\bullet} {\cal E}'^* \right)
\end{displaymath}
discussed in \cite{GuffinSharpe:A-twistedheterotic}[appendix A].
Let $i: Y \hookrightarrow M$ denote inclusion, then the pair ${\cal O}$,
$\tilde{\cal O}$, when it exists, is related by
\begin{displaymath}
{\cal O} \: = \: i^* \tilde{\cal O}
\end{displaymath}
using the isomorphism above.

Our proposal for $\omega_{CK2}$ is given by
the Grassmann integral
\begin{displaymath}
\omega_{CK2} \: = \: \int \prod d \lambda^{x} 
\, \exp( - {\cal A}_{CK2}) ,
\end{displaymath}
where    
\begin{displaymath}
{\cal A}_{CK2} \: = \: 
h_{x \overline{x}} s^x \overline{s}^{\overline{x}}
\: + \: 
h_{\gamma \overline{\gamma}} q^m 
\overline{q}^{\overline{m}} \tilde{E}_m^{\gamma} 
\tilde{\overline{E}}_{\overline{m}}^{ \overline{\gamma} }
\: + \:
\chi^{\overline{\imath}} \lambda^{x} 
\overline{D}_{\overline{\imath}} \overline{s}^{\overline{x}}\, h_{x \overline{x}} \: + \:
\chi^{\overline{m}} \lambda^{\gamma} \tilde{\overline{E}}_{\overline{m}}^{
\overline{\gamma}} \, h_{\gamma \overline{\gamma}}
 \: + \:
\chi^{\overline{\imath}} \lambda^{\gamma} \overline{q}^{\overline{m}}
\left( \overline{D}_{\overline{\imath}} \tilde{ \overline{E} }_{\overline{m}}^{
\overline{\gamma}} \right) h_{\gamma \overline{\gamma}}
\end{displaymath}
where
$x$ indexes local coordinates along the fibers of ${\cal G}$,
$m$ indexes local coordinates (denoted $q$) along the fibers of
${\cal F}_1$, $\gamma$ indexes local coordinates
along the fibers of ${\cal F}_2$, $i$ indexes local
coordinates on $M$, and $(m, i)$ index local coordinates on $X$.

Now, let us explain some aspects of $\omega_{CK2}$ in more detail.  In
${\cal A}_{CK2}$ above, every $\lambda^{x}$ is accompanied by a 
$\chi^{\overline{\imath}}$, so integrating out the $\lambda^{x}$'s
should result in ${\rm rk}\, {\cal G} = g$ $\chi^{\overline{\imath}}$'s,
hence a degree $g$ cohomology class.  
Furthermore, the Grassmann integral measure makes $\omega_{CK2}$ couple
to $\det {\cal G}^*$.  Thus, $\omega_{CK2}$ is 
a form of the type indicated in equation~(\ref{eq:MQ:cokernel:defn}).
We shall show it defines an element of hypercohomology momentarily.

Now, we can argue formally that the analogue of a Mathai-Quillen form
defined above represents an element of hypercohomology.  
The argument is a simple
variation of that seen previously.  The central point
is that for
\begin{displaymath}
\overline{D} \: = \: \chi^{\overline{\imath}} \, \overline{\partial}_{\overline{
\imath}} \: + \: \chi^{\overline{m}} \, \overline{\partial}_{\overline{m}}
\end{displaymath}
we have
\begin{eqnarray*}
\left( \overline{D} \: + \: s^x 
\frac{\partial}{\partial \lambda^{x}} \right) {\cal A}_{CK2} 
& = &
\chi^{\overline{m}} h_{\gamma \overline{\gamma}} q^m
\tilde{E}_m^{\gamma} \tilde{ \overline{E} }_{\overline{m}}^{
\overline{\gamma}} \: + \:
\chi^{\overline{\imath}} q^m \overline{q}^{\overline{m}}
\tilde{E}_m^{\gamma} \overline{\partial}_{\overline{\imath}}\left(
h_{\gamma \overline{\gamma}} \tilde{ \overline{E} }_{\overline{m}}^{
\overline{\gamma}} \right) ,
\\
& = & - q^m \tilde{E}_m^{\gamma} \frac{\partial}{\partial \lambda^{\gamma} }
{\cal A}_{CK2} .
\end{eqnarray*}
Given this result, we see that
although $\omega_{CK2}$ is not $\overline{\partial}$-closed, it
does define an element of hypercohomology of the following
sequence on $X$:
\begin{displaymath}
\pi^* \det {\cal G}^* \otimes \left(
\cdots \: \longrightarrow \: \wedge^2 \pi^* {\cal F}_2^* \: 
\longrightarrow \: \pi^* {\cal F}_2^* \: \longrightarrow \:
{\cal O}_X \right)
\end{displaymath}
with maps given by inclusion along $q \tilde{E}$,
and the $\pi^* \det {\cal G}^*$ factor determined by the Grassmann integral,
as desired.

Next we will argue that the cohomology class of $\omega_{CK2}$ is
unchanged by antiholomorphic deformations of the section $s$.
As the details are somewhat more complicated than the argument in
section~\ref{sect:kernels}, we sketch the details here.
Consider the one-parameter family
\begin{eqnarray*}
{\cal A}_{CK2, \tau} & = &
h_{x \overline{x}} s^x \left( \overline{s}^{\overline{x}}
+ \tau \overline{t}^{\overline{x}} \right)
\: + \: 
h_{\gamma \overline{\gamma}} q^m 
\overline{q}^{\overline{m}} \tilde{E}_m^{\gamma} 
\tilde{\overline{E}}_{\overline{m}}^{ \overline{\gamma} }
\: + \:
\chi^{\overline{\imath}} \lambda^{x} 
\overline{D}_{\overline{\imath}} \left( \overline{s}^{\overline{x}}
+ \tau \overline{t}^{\overline{x}} \right) \, h_{x \overline{x}}
\\
& & \hspace*{1.75in}
 \: + \:
\chi^{\overline{m}} \lambda^{\gamma} \tilde{\overline{E}}_{\overline{m}}^{
\overline{\gamma}} \, h_{\gamma \overline{\gamma}}
 \: + \:
\chi^{\overline{\imath}} \lambda^{\gamma} \overline{q}^{\overline{m}}
\left( \overline{D}_{\overline{\imath}} \tilde{ \overline{E} }_{\overline{m}}^{
\overline{\gamma}} \right) h_{\gamma \overline{\gamma}} ,
\end{eqnarray*}
so that
\begin{eqnarray*}
\frac{d}{d \tau} \omega_{CK2, \tau} & = &
\frac{d}{d \tau} \int \prod d \lambda^{x} \exp\left( - {\cal A}_{CK2, \tau} 
\right) ,
\\
& = &
- \int \prod d \lambda^{x} \left( h_{x \overline{x}} s^x \overline{t}^{
\overline{x}} \: + \: \chi^{\overline{\imath}} \lambda^{x} 
\overline{D}_{\overline{\imath}}  \overline{t}^{\overline{x}} h_{x \overline{x}} 
\right) \exp\left( - {\cal A}_{CK2, \tau} \right) ,
\\
& = & 
\int \prod d \lambda^{x} \left( \overline{D} \: + \: s^x 
\frac{\partial}{\partial \lambda^x} \: + \:
q^m \tilde{E}^{\gamma}_m \frac{\partial}{\partial \lambda^{\gamma}}
\right) \left( - h_{x \overline{x}}
\lambda^x \overline{t}^{\overline{x}} \right)
\exp\left( - {\cal A}_{CK2, \tau} \right) ,
\\
& = &
\left( \overline{\partial} \: + \: q^m \tilde{E}^{\gamma}_m 
\frac{\partial}{\partial \lambda^{\gamma}} \right)
\int \prod d \lambda^{x}
\left( - h_{x \overline{x}}
\lambda^x \overline{t}^{\overline{x}} \right)
\exp\left( - {\cal A}_{CK2, \tau} \right) ,
\end{eqnarray*}
from which the result follows.

In the special case that ${\cal G}=0$, we should note that
it was shown in \cite{GuffinSharpe:A-twistedheterotic}[appendix A]
that
\begin{displaymath}
{\mathbb H}^{\bullet}\left(X, \cdots \: \longrightarrow \: \wedge^2 \pi^* {\cal F}_2^* \: 
\longrightarrow \: \pi^* {\cal F}_2^* \: \longrightarrow \:
{\cal O}_X \right)
\: = \:
H^{\bullet}\left(M, \wedge^{\bullet} {\cal E}'^* \right) ,
\end{displaymath}
where ${\cal E}'$ is the cokernel, exactly as expected for the
hypercohomology described in this section to be related to ordinary
sheaf cohomology.

As we will see in section~\ref{sect:A2-cokernels},
this analogue of a Mathai-Quillen form appears in the A/2 model
pseudo-topological field theories \cite{GuffinSharpe:A-twistedheterotic},
just as the analogues in the last section.

\subsection{Cohomologies of short complexes}

In this section, we will consider two constructions that will relate
sheaf cohomology on
\begin{displaymath}
Y \: \equiv \: \{s=0\} \: \subset \: M
\end{displaymath}
to sheaf cohomology on the total space of a bundle over $M$,
where the coefficients in question are given as the cohomology of
a short complex.

\subsubsection{First short complex construction}

Suppose we want to compute
integrals of sheaf cohomology classes
\begin{displaymath}
{\cal O} \: \in \: H^{\bullet}(Y, \wedge^{\bullet} {\cal E}'^* )
\end{displaymath}
where ${\cal E}'$ is a holomorphic vector bundle on $Y$.
Suppose that
\begin{displaymath}
{\cal E}' \: = \: \frac{ \ker \tilde{F}|_Y }{ {\rm im}\, \tilde{E}|_Y } ,
\end{displaymath}
where $\tilde{E}: {\cal F}_1 \rightarrow {\cal F}_2$ is an injective
map between two holomorphic vector bundles on $M$, and
$\tilde{F}: {\cal F}_2 \rightarrow {\cal F}_3$ is a surjective map between
two holomorphic vector bundles on $M$, where $\tilde{E}$ is holomorphic
on all of $M$ but $\tilde{F}$ is only holomorphic along $Y \subset M$,
and whose restrictions to $Y$ form a complex:
\begin{displaymath}
0 \: \longrightarrow \: {\cal F}_1 |_Y \: 
\stackrel{ \tilde{E} }{\longrightarrow}
\: {\cal F}_2 |_Y \: \stackrel{ \tilde{F} }{\longrightarrow} \: 
{\cal F}_3 |_Y
\: \longrightarrow \: 0.
\end{displaymath}
(The composition $\tilde{F} \circ \tilde{E}$ vanishes everywhere\footnote{
Experts will note that this is not the most general possibility allowed
by physics.  We leave more general cases for future work.
}
on $M$.)

Then, we propose that, for those ${\cal O}$ such that
lifts $\tilde{\cal O}$ exist, 
\begin{displaymath}
\int_Y {\cal O}_1 \wedge \cdots \wedge {\cal O}_n \: \propto \:
\int_X \tilde{\cal O}_1 \wedge \cdots \wedge \tilde{\cal O}_n \wedge \omega_{MON1} ,
\end{displaymath}
where
\begin{displaymath}
X \: \equiv \: {\rm Tot}\, \left( {\cal F}_1 \:
\stackrel{\pi}{\longrightarrow} \: M \right),
\end{displaymath}
the lifts $\tilde{\cal O}$ are elements of hypercohomology
\begin{displaymath}
{\mathbb H}^{\bullet}\left(X, \cdots \longrightarrow \:
\wedge^2 \pi^* {\cal F}_2^* \: \longrightarrow \:
\pi^* {\cal F}_2^* \: \longrightarrow \: {\cal O}_X \right)
\end{displaymath}
with maps given by inclusion with $q^m \tilde{E}^{\gamma}_m$,
and
\begin{equation}   \label{eq:mon1:gp}
\omega_{MON1} \: \in \: {\mathbb H}^{g + f_3}
\left(X, \pi^* \det {\cal G}^* \otimes
\pi^* \det {\cal F}_3 \otimes \left(
\cdots \longrightarrow \:
\wedge^2 \pi^* {\cal F}_2^* \: \longrightarrow \:
\pi^* {\cal F}_2^* \: \longrightarrow \: {\cal O}_X \right)
\right)
\end{equation}
($g = {\rm rk}\, {\cal G}$, $f_i = {\rm rk}\, {\cal F}_i$)
given by
\begin{displaymath}
\omega_{MON1} \: = \: \int \prod d \lambda^{\overline{x}} d \chi^r
\exp\left( - {\cal A}_{MON1} \right),
\end{displaymath}
where
\begin{eqnarray*}
{\cal A}_{MON1} & = &
h^{x \overline{x}} s_x \overline{s}_{\overline{x}} \: + \:
\chi^{\overline{\imath}} \lambda^{\overline{x}} \,
\overline{D}_{\overline{\imath}} \overline{s}_{\overline{x}} \: + \:
\chi^{r} \lambda^{\gamma} 
\tilde{F}_{r \gamma} \: + \: 
F_{\overline{\imath} r \overline{x} \gamma } 
\chi^{\overline{\imath}} \chi^{r} \lambda^{\overline{x}}
\lambda^{\gamma}
\\
& & \hspace*{0.25in}
\: + \: 
h_{\gamma \overline{\gamma}} q^m \overline{q}^{\overline{m}} \, 
\tilde{ \overline{E} }_{\overline{m}}^{\overline{\gamma}} 
\tilde{ E }_{m}^{
\gamma} \: + \:
\chi^{\overline{m}} \lambda^{\gamma} \tilde{ \overline{E} }_{\overline{m}}^{
\overline{\gamma}} h_{
\gamma \overline{\gamma}} \: + \:
\chi^{\overline{\imath}} \lambda^{\gamma} \, \overline{q}^{\overline{m}} \,
\overline{D}_{\overline{\imath}} \tilde{ \overline{E} }_{\overline{m}}^{\overline{\gamma}}
h_{\gamma \overline{\gamma}} .
\end{eqnarray*}
In the expression above, 
$x$ indexes local coordinates along the fibers of ${\cal G}$,
$m$ indexes local coordinates along the fibers of ${\cal F}_1$,
$\gamma$ indexes local coordinates along the fibers of ${\cal F}_2$,
$r$ indexes local coordinates along the fibers of ${\cal F}_3^*$,
and $i$ indexes local coordinates on $M$.
The curvature term
\begin{displaymath}
F_{\overline{\imath} r \overline{x} \gamma } 
\chi^{\overline{\imath}} \chi^{r} \lambda^{\overline{x}}
\lambda^{\gamma}
\end{displaymath}
represents the pullback of an element of
\begin{displaymath}
H^1\left(M, {\cal F}_2^* \otimes {\cal F}_3 \otimes {\cal G}^* \right)
\end{displaymath}
which is defined by the condition
\begin{equation}   
\overline{\partial}_{\overline{\imath}} \tilde{F}_{r \gamma} \: = \:
h^{x \overline{x}} s_x F_{\overline{\imath} r \gamma \overline{x}} \: = \:
- h^{x \overline{x}} s_x F_{\overline{\imath} 
r \overline{x} \gamma} 
\end{equation}
(much as in the earlier discussion of kernels), 
and in addition, we assume that the curvature defined by $F$ annihilates
the image of $\tilde{E}$:
\begin{displaymath}
q^m \tilde{E}^{\gamma}_m F_{\overline{\imath} r \overline{x} \gamma}
\: = \: 0.
\end{displaymath}
Finally, there is an isomorphism
\begin{displaymath}
K_X \: \cong \: \pi^* \det {\cal G}^* \otimes \pi^* \det {\cal F}_2^* \otimes
\pi^* \det {\cal F}_3
\end{displaymath}
which restricts to the isomorphism needed to define the integrals of
sheaf cohomology classes on $Y$.

As before, we will check some elementary properties of $\omega_{MON1}$.

First, let us explain the relationship between 
\begin{displaymath}
{\cal O} \: \in \: H^{\bullet}\left(Y, \wedge^{\bullet} {\cal E}'^* \right)
\mbox{ and }
\tilde{\cal O} \: \in \:
{\mathbb H}^{\bullet}\left(X, \cdots \longrightarrow \:
\wedge^2 \pi^* {\cal F}_2^* \: \longrightarrow \:
\pi^* {\cal F}_2^* \: \longrightarrow \: {\cal O}_X \right).
\end{displaymath}
Define $S$ to be the cokernel
\begin{displaymath}
0 \: \longrightarrow \: {\cal F}_1 \: \stackrel{ \tilde{E} }{\longrightarrow}
\: {\cal F}_2 \: \longrightarrow \: S \: \longrightarrow \: 0
\end{displaymath}
We will use the isomorphism \cite{GuffinSharpe:A-twistedheterotic}[appendix A]
\begin{displaymath}
{\mathbb H}^{\bullet}\left(X, \cdots \longrightarrow \:
\wedge^2 \pi^* {\cal F}_2^* \: \longrightarrow \:
\pi^* {\cal F}_2^* \: \longrightarrow \: {\cal O}_X \right)
\: \cong \: H^{\bullet}\left(M, \wedge^{\bullet} S^* \right).
\end{displaymath}
Let $i: Y \hookrightarrow M$ denote the inclusion, and note that the
injective map
\begin{displaymath}
{\cal E}' \: = \: \frac{ \ker \tilde{F}|_Y }{ {\rm im}\, \tilde{E}|_Y } \:
\longrightarrow \: \frac{ {\cal F}_2 |_Y }{ {\rm im}\, \tilde{E}|_Y } \: = \:
i^* S
\end{displaymath}
defines a map
\begin{displaymath}
j_*: \: H^{\bullet}\left(Y, \wedge^{\bullet} {\cal E}'^* \right) \: 
\longrightarrow \: H^{\bullet}\left(Y, \wedge^{\bullet} i^* S^* \right).
\end{displaymath}
Then, using the isomorphism above, a pair ${\cal O}$, $\tilde{\cal O}$
(when it exists) is related by
\begin{displaymath}
j_* {\cal O} \: = \: i^* \tilde{\cal O} .
\end{displaymath}

Next, we will argue that $\omega_{MON1}$ above defines an element of
hypercohomology~(\ref{eq:mon1:gp}).  The argument is a variation of that
repeated previously.  As before, the central point is that
for
\begin{displaymath}
\overline{D} \: = \: \chi^{\overline{\imath}} \, \overline{\partial}_{
\overline{\imath}} \: + \: \chi^{\overline{m}} \, \overline{\partial}_{
\overline{m}}
\end{displaymath}
we have
\begin{eqnarray*}
\lefteqn{
\left( \overline{D} \: + \: h^{x \overline{x}} s_x \frac{\partial}{
\partial \lambda^{\overline{x}} } \: + \:
q^m \tilde{E}^{\gamma}_m \frac{\partial}{\partial \lambda^{\gamma}}
\right) {\cal A}_{MON1}
} \\
& \hspace*{1in}
 = &
- \chi^r q^m \tilde{E}^{\gamma}_m \tilde{F}_{r \gamma} \: + \:
\chi^{\overline{\imath}} \chi^r \lambda^{\gamma} \left(
\overline{\partial}_{\overline{\imath}} \tilde{F}_{r \gamma} \: + \:
h^{x \overline{x}} s_x 
F_{\overline{\imath} r \overline{x} \gamma} \right)
\\
& & \hspace*{0.5in}
\: - \: q^m \tilde{E}^{\gamma}_m F_{\overline{\imath} r \overline{x}
\gamma} \chi^{\overline{\imath}} \chi^r \lambda^{\overline{x}},
\\
& \hspace*{1in} = & 0,
\end{eqnarray*}
using the conditions discussed above.
Combined with the fact that the Grassmann integration measure couples to
\begin{displaymath}
\pi^* {\cal G}^* \otimes \pi^* {\cal F}_3
\end{displaymath}
and the fact that integrating over $\lambda^{\overline{x}}$ brings down
$g = {\rm rk}\, {\cal G}$ factors of $\chi^{\overline{\imath}}$,
$\chi^r$ brings down $f_3 = {\rm rk}\, {\cal F}_3$ factors of
$\lambda^{\gamma}$, we see that $\omega_{MON1}$ can be interpreted
as an element of the hypercohomology group~(\ref{eq:mon1:gp}).

Next, we demonstrate that the cohomology class of $\omega_{MON1}$ is
independent of antiholomorphic deformations of $s$.  Consider the
one-parameter family
\begin{eqnarray*}
{\cal A}_{MON1, \tau} & = &
h^{x \overline{x}} s_x \left( \overline{s}_{\overline{x}}
\: + \: \tau \overline{t}_{\overline{x}} \right) \: + \:
\chi^{\overline{\imath}} \lambda^{\overline{x}} 
\left( \overline{D}_{\overline{\imath}} \overline{s}_{\overline{x}}
\: + \: \overline{D}_{\overline{\imath}} \overline{t}_{\overline{x}} \right)  \: + \:
\chi^{r} \lambda^{\gamma} 
\tilde{F}_{r \gamma} \: + \: 
F_{\overline{\imath} r \overline{x} \gamma } 
\chi^{\overline{\imath}} \chi^{r} \lambda^{\overline{x}}
\lambda^{\gamma}
\\
& & \hspace*{0.25in}
\: + \: 
h_{\gamma \overline{\gamma}} q^m \overline{q}^{\overline{m}} \,
\tilde{ \overline{E} }_{\overline{m}}^{\overline{\gamma}} 
\tilde{ E }_{m}^{
\gamma} \: + \:
\chi^{\overline{m}} \lambda^{\gamma} \tilde{ \overline{E} }_{\overline{m}}^{
\overline{\gamma}} h_{
\gamma \overline{\gamma}} \: + \:
\chi^{\overline{\imath}} \lambda^{\gamma} \, \overline{q}^{\overline{m}} \, 
\overline{D}_{\overline{\imath}} \tilde{ \overline{E} }_{\overline{m}}^{\overline{\gamma}}
h_{\gamma \overline{\gamma}} ,
\end{eqnarray*}
so that
\begin{eqnarray*}
\frac{d}{d \tau} \omega_{MON1, \tau} & = &
\frac{d}{d \tau} \int \prod d \lambda^{\overline{x}} d \chi^r
\exp\left( - {\cal A}_{MON1, \tau} \right) ,
\\
& = &
- \int \prod d \lambda^{\overline{x}} d \chi^r
\left(  h^{x \overline{x}} s_x \overline{t}_{\overline{x}}
\: + \: \chi^{\overline{\imath}} \lambda^{\overline{x}} \, \overline{D}_{\overline{\imath}}
\overline{t}_{\overline{x}} \right)
\exp\left( - {\cal A}_{MON1, \tau} \right) ,
\\
& = &
\int \prod d \lambda^{\overline{x}} d \chi^r
\left( \overline{D} \: + \: h^{x \overline{x}} s_x \frac{\partial}{
\partial \lambda^{\overline{x}} } \: + \:
q^m \tilde{E}^{\gamma}_m \frac{\partial}{\partial \lambda^{\gamma}}
\right)
\left( - \lambda^{\overline{x}} \, \overline{t}_{\overline{x}} \right)
\exp\left( - {\cal A}_{MON1, \tau} \right) ,
\\
& = &
\left( \overline{\partial} \: + \:
q^m \tilde{E}^{\gamma}_m \frac{\partial}{\partial \lambda^{\gamma}}
\right)
\int \prod d \lambda^{\overline{x}} d \chi^r
\left( - \lambda^{\overline{x}} \, \overline{t}_{\overline{x}} \right)
\exp\left( - {\cal A}_{MON1, \tau} \right) ,
\end{eqnarray*}
from which the result follows.

The relevance of this construction to physics will be discussed
in section~\ref{sect:mon1:phys}.

\subsubsection{Second short complex construction}

Suppose we want to compute integrals of sheaf cohomology classes
\begin{displaymath}
{\cal O} \: \in \: H^{\bullet}\left(Y, \wedge^{\bullet} {\cal E}' \right) ,
\end{displaymath}
where ${\cal E}'$ is a holomorphic vector bundle on $Y$, of the form
\begin{displaymath}
{\cal E}' \: = \: \frac{\ker \tilde{F}|_Y}{{\rm im}\, \tilde{E}|_Y } ,
\end{displaymath}
where $\tilde{E}: {\cal F}_1 \rightarrow {\cal F}_2$ is an injective map
between two holomorphic vector bundles on $M$, and $\tilde{F}:
{\cal F}_2 \rightarrow {\cal F}_3$ is a surjective map between two holomorphic
vector bundles on $M$, where $\tilde{F}$ is holomorphic on all of $M$ but
$\tilde{E}$ need only be holomorphic along $Y \subset M$, and whose
restrictions form a complex
\begin{displaymath}
0 \: \longrightarrow \: {\cal F}_1 |_Y \: \stackrel{ \tilde{E} }{
\longrightarrow} \: {\cal F}_2 |_Y \: \stackrel{ \tilde{F} }{\longrightarrow}
\: {\cal F}_3 |_Y \: \longrightarrow \: 0.
\end{displaymath}
The composition $\tilde{F} \circ \tilde{E}$ vanishes everywhere\footnote{
More general cases are left for future work.
} on $M$.

Then, we propose that, for those ${\cal O}$ such that lifts $\tilde{\cal O}$
exist,
\begin{displaymath}
\int_Y {\cal O}_1 \wedge \cdots \wedge {\cal O}_n \: \propto \:
\int_X \tilde{\cal O}_1 \wedge \cdots \wedge \tilde{\cal O}_n \wedge \omega_{MON2} ,
\end{displaymath}
where
\begin{displaymath}
X \: \equiv \: {\rm Tot}\left( {\cal F}_3^* \: \stackrel{\pi}{\longrightarrow}
\: M \right)
\end{displaymath}
the lifts $\tilde{\cal O}$ are elements of hypercohomology
\begin{displaymath}
{\mathbb H}^{\bullet}\left(X, 
\cdots \: \longrightarrow \: \wedge^2 \pi^* {\cal F}_2 \:
\longrightarrow \: \pi^* {\cal F}_2 \: \longrightarrow \:
{\cal O}_X \right)
\end{displaymath}
with maps given by inclusion with $p^r \tilde{F}_{r \gamma}$, and
\begin{equation}   \label{eq:mon2:gp}
\omega_{MON2} \: \in \: {\mathbb H}^{g + f_1}\left(X,
\pi^* \det {\cal G}^* \otimes \pi^* \det {\cal F}_1^* \otimes \left(
\cdots \: \longrightarrow \: \wedge^2 \pi^* {\cal F}_2 \:
\longrightarrow \: \pi^* {\cal F}_2 \: \longrightarrow \:
{\cal O}_X \right)
\right)
\end{equation}
($g = {\rm rk}\, {\cal G}$, $f_i = {\rm rk}\, {\cal F}_i$)
given by
\begin{displaymath}
\omega_{MON2} \: = \: \int \prod d \lambda^x d \chi^m
\exp\left( - {\cal A}_{MON2} \right) ,
\end{displaymath}
where
\begin{eqnarray*}
{\cal A}_{MON2} & = &
h_{x \overline{x}} s^x \overline{s}^{\overline{x}} \: + \:
h^{\gamma \overline{\gamma}} p^r \overline{p}^{\overline{r}}
\tilde{F}_{r \gamma} \tilde{\overline{F}}_{\overline{r} \overline{\gamma}}
 \: + \:
\chi^{\overline{r}} \theta_{\gamma} h^{\gamma \overline{\gamma}} 
\tilde{\overline{F}}_{
\overline{r} \overline{\gamma}} \: + \:
\chi^{\overline{\imath}} \theta_{\gamma} h^{\gamma \overline{\gamma}}
\overline{p}^{\overline{r}} \, \overline{D}_{\overline{\imath}} \tilde{\overline{F}}_{
\overline{r} \overline{\gamma}}
\\
& & \hspace*{0.5in}
\: + \:
\chi^{\overline{\imath}} \lambda^{x} \overline{D}_{\overline{\imath}} \overline{s}^{
\overline{x}} \, h_{x \overline{x}} \: + \:
\chi^{m} \theta_{\gamma} 
 \tilde{E}_{
m}^{ \gamma}  \: + \:
F_{\overline{\imath} m x \overline{\gamma}} 
\chi^{\overline{\imath}} \chi^{m} \lambda^{x} 
\theta_{\gamma} h^{\gamma \overline{\gamma}} .
\end{eqnarray*}
In the expression above,
$x$ indexes local coordinates along the fibers of ${\cal G}$,
$m$ indexes local coordinates along the fibers of ${\cal F}_1$,
$\gamma$ indexes local coordinates along the fibers of ${\cal F}_2$,
$r$ indexes local coordinates along the fibers of ${\cal F}_3^*$,
and $i$ indexes local coordinates on $M$.
The curvature term
\begin{displaymath}
F_{\overline{\imath} m x \overline{\gamma}} 
\chi^{\overline{\imath}} \chi^{m} \lambda^{x} 
\theta_{\gamma} h^{\gamma \overline{\gamma}}
\end{displaymath}
represents the pullback of an element of
\begin{displaymath}
H^1\left(M, {\cal F}_1^* \otimes {\cal F}_2 \otimes {\cal G}^*  \right)
\end{displaymath}
and solves the equation
\begin{equation}
h_{\gamma \overline{\gamma}}
\overline{D}_{\overline{\imath}} \tilde{E}_m^{\gamma}
\: = \:
s^x F_{\overline{\imath} m  \overline{\gamma} x} 
\: = \: - s^x F_{\overline{\imath} m x \overline{\gamma}} .
\end{equation}
Its existence and properties were discussed in the earlier cokernels
section.  In addition, we assume the curvature defined by $F$
is in the kernel of $\tilde{F}$:
\begin{displaymath}
p^r \tilde{F}_{r \gamma} h^{\gamma \overline{\gamma}}
F_{\overline{\imath} m x \overline{\gamma}} \: = \: 0 .
\end{displaymath}

Finally, there is an isomorphism
\begin{displaymath}
K_X \: \cong \: 
\pi^* \det {\cal G}^* \otimes \pi^* \det {\cal F}_1^* \otimes 
\pi^* \det {\cal F}_2
\end{displaymath}
that restricts to the isomorphism that makes the integrals on $Y$
well-defined.

As before, we will check some elementary properties of $\omega_{MON2}$.

First, let us explain the relationship between
\begin{displaymath}
{\cal O} \: \in \: H^{\bullet}\left(Y, \wedge^{\bullet} {\cal E}' \right)
\mbox{ and }
\tilde{\cal O} \: \in \: 
{\mathbb H}^{\bullet}\left(X, 
\cdots \: \longrightarrow \: \wedge^2 \pi^* {\cal F}_2 \:
\longrightarrow \: \pi^* {\cal F}_2 \: \longrightarrow \:
{\cal O}_X \right) .
\end{displaymath}
Define $S$ to be the kernel
\begin{displaymath}
0 \: \longrightarrow \: S \: \longrightarrow \: {\cal F}_2 \:
\stackrel{ \tilde{F} }{\longrightarrow} \: {\cal F}_3 \: \longrightarrow
\: 0
\end{displaymath}
on $M$, and let $i: Y \hookrightarrow M$ denote the inclusion.
We will use the isomorphism \cite{GuffinSharpe:A-twistedheterotic}[appendix A]
\begin{displaymath}
{\mathbb H}^{\bullet}\left(X, 
\cdots \: \longrightarrow \: \wedge^2 \pi^* {\cal F}_2 \:
\longrightarrow \: \pi^* {\cal F}_2 \: \longrightarrow \:
{\cal O}_X \right) 
\: \cong \:
H^{\bullet}\left(M, \wedge^{\bullet} S \right).
\end{displaymath}
The map
\begin{displaymath}
i^* S \: = \: \ker \tilde{F}|_Y \: \longrightarrow \:
\frac{ \ker \tilde{F}|_Y }{ {\rm im}\, \tilde{E} |_Y } \: = \: {\cal E}'
\end{displaymath}
defines a map
\begin{displaymath}
j_*: \: H^{\bullet}\left(Y, \wedge^{\bullet} i^* S \right) \: 
\longrightarrow \: H^{\bullet}\left( Y, \wedge^{\bullet} {\cal E}' \right).
\end{displaymath}
Then, the pair ${\cal O}$, $\tilde{\cal O}$, when it exists, is related by
\begin{displaymath}
{\cal O} \: = \: j_* i^* \tilde{\cal O}.
\end{displaymath}

Next, we show that $\omega_{MON2}$ above defines an element
of the hypercohomology group~(\ref{eq:mon2:gp}).  
The argument is a variation of that
repeated several times already.  As before, the central point is that for
\begin{displaymath}
\overline{D} \: = \: \chi^{\overline{\imath}} \, \overline{\partial}_{
\overline{\imath}} \: + \: \chi^{\overline{r}} \, \overline{\partial}_{
\overline{r}}
\end{displaymath}
we have
\begin{eqnarray*}
\lefteqn{
\left( \overline{D} \: + \: s^x \frac{\partial}{\partial \lambda^x}
\: + \: p^r \tilde{F}_{r \gamma} 
\frac{\partial}{\partial \theta_{\gamma}  }
\right) {\cal A}_{MON2}
} \\
& \hspace*{1in} = &
- \chi^m p^r \tilde{E}^{\gamma}_m \tilde{F}_{r \gamma}
\: + \:
\chi^{\overline{\imath}} \chi^m \theta_{\gamma}  \left(
 \overline{D}_{\overline{\imath}} \tilde{E}^{\gamma}_m
\: + \:
s^x F_{\overline{\imath} m x \overline{\gamma}} h^{\gamma \overline{\gamma}}
\right)
\\
& \hspace*{1in} & \hspace*{0.5in}
- p^r \tilde{F}_{r \gamma} h^{\gamma \overline{\gamma}} F_{
\overline{\imath} m x \overline{\gamma}} \chi^{\overline{\imath}}
\chi^m \lambda^x,
\\
& \hspace*{1in} = & 0,
\end{eqnarray*}
using the conditions discussed above.  The Grassmann integration measure
contributes a factor of $\pi^* \det {\cal G}^* \otimes \pi^* \det 
{\cal F}_1^*$ to the coefficients.  Each $\lambda^x$ is accompanied by
a $\chi^{\overline{\imath}}$, and each $\chi^m$ is accompanied by a 
$\theta_{\gamma}$, so the Grassmann integration yields a result of
degree $g + f_1$ in $\chi^{\overline{\imath}}$, $\theta_{\gamma}$,
determining the degree.

Next, we will demonstrate that the cohomology class of 
${\cal A}_{MON2}$ is independent of antiholomorphic deformations of $s$.
Consider the one-parameter family
\begin{eqnarray*}
{\cal A}_{MON2, \tau} & = &
h_{x \overline{x}} s^x \left( \overline{s}^{\overline{x}}
\: + \: \tau \overline{t}^{\overline{x}} \right) \: + \:
h^{\gamma \overline{\gamma}} p^r \overline{p}^{\overline{r}}
\tilde{F}_{r \gamma} \tilde{\overline{F}}_{\overline{r} \overline{\gamma}}
 \: + \:
\chi^{\overline{r}} \theta_{\gamma} h^{\gamma \overline{\gamma}} \,
 \tilde{\overline{F}}_{
\overline{r} \overline{\gamma}} \: + \:
\chi^{\overline{\imath}} \theta_{\gamma} h^{\gamma \overline{\gamma}} \,
\overline{p}^{\overline{r}} \, \overline{D}_{\overline{\imath}} \tilde{\overline{F}}_{
\overline{r} \overline{\gamma}}
\\
& & \hspace*{0.5in}
\: + \:
\chi^{\overline{\imath}} \lambda^{x} \left( \overline{D}_{\overline{\imath}} \overline{s}^{
\overline{x}} \: + \: \tau \overline{D}_{\overline{\imath}} \overline{t}^{
\overline{x}} \right)\, h_{x \overline{x}} \: + \:
\chi^{m} \theta_{\gamma}  \tilde{E}_{
m}^{ \gamma}  \: + \:
F_{\overline{\imath} m x \overline{\gamma}} 
\chi^{\overline{\imath}} \chi^{m} \lambda^{x} 
\theta_{\gamma} h^{\gamma \overline{\gamma}} ,
\end{eqnarray*}
so that
\begin{eqnarray*}
\frac{d}{d \tau} \omega_{MON2, \tau} & = &
\frac{d}{d \tau} \int \prod d \lambda^x d \chi^m
\exp\left( - {\cal A}_{MON2, \tau} \right) ,
\\
& = &
- \int \prod d \lambda^x d \chi^m
\left( h_{x \overline{x}} s^x \overline{t}^{\overline{x}}
\: + \: \chi^{\overline{\imath}} \lambda^x \overline{D}_{\overline{\imath}}
\overline{t}^{\overline{x}} h_{x \overline{x}}   \right)
\exp\left( - {\cal A}_{MON2, \tau} \right) ,
\\
& = &
\int \prod d \lambda^x d \chi^m
\left( \overline{D} \: + \: s^x \frac{\partial}{\partial \lambda^x}
\: + \: p^r \tilde{F}_{r \gamma} 
\frac{\partial}{\partial \theta_{\gamma}  }
\right)
\left( - h_{x \overline{x}} \lambda^x \overline{t}^{\overline{x}} 
\right)
\exp\left( - {\cal A}_{MON2, \tau} \right) ,
\\
& = & \left( \overline{\partial} \: + \:
p^r \tilde{F}_{r \gamma} 
\frac{\partial}{\partial \theta_{\gamma}  }
\right)
\int \prod d \lambda^x d \chi^m
\left( - h_{x \overline{x}} \lambda^x \overline{t}^{\overline{x}} 
\right)
\exp\left( - {\cal A}_{MON2, \tau} \right) ,
\end{eqnarray*}
from which the result follows.

The relevance of this construction to physics will be discussed 
in section~\ref{sect:monad2:phys}.

\section{Applications in topological field theory}
\label{sect:apps:tft}

The original Mathai-Quillen form \cite{MathaiQuillen:Superconnections}
has appeared in topological
field theories in several ways.  One of its original uses was as
a route to define topological field theories 
(see {\it e.g.} 
\cite{Kalkman:BRST,Blau:The-Mathai-Quillen,Wu:On-the-Mathai-Quillen,CordesMooreRamgoolam:Lectures,Wu:Mathai-Quillen}),
but we are more concerned with a more recent application
to A-twisted Landau-Ginzburg models \cite{GuffinSharpe:A-twisted},
and heterotic generalizations thereof \cite{GuffinSharpe:A-twistedheterotic}.

In comparing expressions from heterotic strings and mathematics, we will
have to perform a convention switch.  Standard heterotic string conventions
result in $\partial$-closed forms, whereas standard mathematics conventions
involve $\overline{\partial}$-closed forms.  Rather than use nonstandard
conventions for either, we will simply complex conjugate whenever we wish
to compare heterotic string results to mathematics.

\subsection{Ordinary A-twisted Landau-Ginzburg models}

\subsubsection{(2,2) locus}

The paper \cite{GuffinSharpe:A-twisted} studied examples of 
A-twisted Landau-Ginzburg models which RG flow to nonlinear sigma models.
A prototypical example is the Landau-Ginzburg model on 
\begin{displaymath}
X =
{\rm Tot}\left(
\pi: {\cal G}^* \: \longrightarrow \: M \right),
\end{displaymath}
with superpotential
\begin{displaymath}
W \: = \: p \pi^* s,
\end{displaymath}
where $p$ is a fiber coordinate and $s$ a 
section of ${\cal G}$.
This model RG flows to a nonlinear sigma model on $Y \equiv \{s = 0 \}
\subset M$.

In A-twists of this Landau-Ginzburg model, the structure of a 
Mathai-Quillen form naturally arises, whose effect is to give a 
mathematical understanding of the effect of the renormalization group
in this case.  In other words, correlation functions in the A-twisted
Landau-Ginzburg theory look like wedge products of differential forms
on $M$,
but with an insertion of
the Mathai-Quillen form, which makes them equivalent to computations
on $Y \subset M$.

Briefly, the action for the Landau-Ginzburg theory is of the form
\begin{multline}
\label{S^(2,2)}
S^{(2,2)} 
   = 2t \int_{\Sigma} d^2 z
     \left[
            \frac{1}{2}
            \left( 
                    g_{\mu \nu}
                  + i B_{\mu \nu}
            \right)
            \partial_{z} \phi^{\mu}
            \overline{\partial}_{\overline{z}} \phi^{\nu}                      
          + i g_{\overline{a} a} 
            \psi^{\overline{a}}_+ 
            \overline{D}_{\overline{z}} \psi^a_+
          + i g_{b \overline{b}}
            \psi^b_- 
            D_z \psi^{\overline{b}}_-
     \right.
\\
     \bigl.  
          + R_{a \overline{a} b \overline{b}}
            \psi^a_+ 
            \psi^{\overline{a}}_+ 
            \psi^b_- 
            \psi^{\overline{b}}_-                      
          + g^{a\overline{b}} \partial_a W 
            \overline{\partial}_{\overline{b}} \overline{W}
          + \psi^a_+ \psi^b_- D_a \partial_b W 
          + \psi^{\overline{a}}_+ \psi^{\overline{b}}_- \overline{D}_{\overline{a}} 
            \overline{\partial}_{\overline{b}} \overline{W}          
    \biggr],
\end{multline}
where $ \phi^a = (p, \phi^i)$, $p$ a fiber coordinate, $\phi^i$ coordinates
on $M$, and in the A-twisted theory,
\begin{equation*}
\begin{aligned}
\psi^i_+ 
  &\equiv \chi^i \!
  &\in \
  &\Gamma
   \left(  
          \phi^* \!
          \left(
                 T^{1,0} M
          \right)       
  \right),        
&\psi^i_- 
  &\equiv \psi^i_{\overline{z}} \!
  &\in \
  &\Gamma
   \left( 
          \overline{K}_{\Sigma} \otimes
          \left( 
                 \phi^* \!
                 \left( 
                        T^{0,1} M
                 \right)       
          \right)\!^{*}
  \right),       
\\
\psi^{\overline{\imath}}_+ 
  &\equiv \psi^{\overline{\imath}}_z \!
  &\in \
  &\Gamma
   \left(
          K_{\Sigma} \otimes
          \left( 
                 \phi^* \!
                 \left(
                        T^{1,0} M
                 \right)       
          \right)\!^{*}
  \right),    
&\psi^{\overline{\imath}}_- 
  &\equiv \chi^{\overline{\imath}} \!
  &\in \
  &\Gamma
   \left( 
          \phi^* \!
          \left(
                 T^{0,1} M
          \right)        
   \right),
\\
\psi^p_+ 
  &\equiv \psi^p_z \!
  &\in \ 
  &\Gamma
   \left(
          K_{\Sigma} 
          \otimes 
          \phi^* \,
          T^{1,0}_{\pi}
   \right),
&\psi^p_- 
  &\equiv \chi^p \!
  &\in \ 
  &\Gamma
   \left(
          \left(
                 \phi^* \,
                 T^{0,1}_{\pi}
          \right)\!^{*}         
   \right),
\\
\psi^{\overline{p}}_+ 
  &\equiv \chi^{\overline{p}} \!
  &\in \ 
  &\Gamma
   \left(
          \left(
                 \phi^* \,
                 T^{1,0}_{\pi}
          \right)\!^{*} 
   \right),
&\psi^{\overline{p}}_- 
  &\equiv \psi^{\overline{p}}_{\overline{z}} \!
  &\in \ 
  &\Gamma
   \left(
          \overline{K}_{\Sigma} 
          \otimes 
          \phi^* \,
          T^{0,1}_{\pi}               
   \right).                                           
\end{aligned}
\end{equation*}
%%%%%%%%%%%%%%%%%
where $ K_{\Sigma} $ is the canonical bundle on $ \Sigma $ and
$ T_{\pi} $ is the relative tangent bundle of the projection
$\pi: {\cal G}^* \rightarrow M$.
To make sense of the A-twist of this theory, it was also necessary
to twist some of the bosons, specifically,
\begin{equation*} 
p \equiv p_z 
   \in \Gamma
        \left(
               K_{\Sigma} 
               \otimes
               \phi^* \,
               T^{1,0}_{\pi}
        \right),
\qquad
\overline{p} \equiv \overline{p}_{\overline{z}} 
   \in \Gamma
        \left(
               \overline{K}_{\Sigma}
               \otimes
               \phi^* \,
               T^{0,1}_{\pi}
        \right),        
\end{equation*}
and the $ \phi^i $ remain untwisted.

Although $\chi^i$, $\chi^{\overline{\imath}}$, $\chi^p$, and
$\chi^{\overline{p}}$ are all scalars, it can be shown
(following \cite{GuffinSharpe:A-twisted}) 
that only $\chi^i$, $\chi^{\overline{\imath}}$
are BRST-invariant.  Furthermore, as the $p$ fields are twisted, scalar
zero modes lie along $\{ p = 0 \} = M$, and so correlation functions take the
form of integrals over $M$ of wedge products of observables of the form
\begin{displaymath}
f(\phi^i) \chi^{i_1} \cdots \chi^{i_n}
\chi^{\overline{\imath}_1} \cdots \chi^{\overline{\imath}_m}
\end{displaymath}
with an insertion of an exponential of zero mode interactions we shall
discuss momentarily.

We can see the relevance of pullbacks of Mathai-Quillen forms as follows.
If, for example, we restrict to degree zero maps on a genus
zero worldsheet, then from restricting to zero modes we recover
the following interactions on the zero modes:
\begin{equation*}
     g^{p\overline{p}}
     s_p
     \overline{s}_{\overline{p}} 
   + \chi^i 
     \chi^p
     D_i s_p
   + \chi^{\overline{p}} 
     \chi^{\overline{\imath}} \,
     \overline{D}_{\overline{\imath}} 
     \overline{s}_{\overline{p}}
   + R_{i\overline{p}p\overline{\imath}} 
     \chi^i 
     \chi^{\overline{p}} 
     \chi^p 
     \chi^{\overline{\imath}} \, .
\end{equation*}
%%%%%%%%%%%%%%%%
If we now complex conjugate this expression to relate it to standard mathematics conventions, we obtain
%%%%%%%%%%%%%%%%
\begin{align}
\label{S^{(2,2)}_0}
     g^{p\overline{p}} &
     s_p
     \overline{s}_{\overline{p}}
   + \chi^{\overline{\imath}}
     \chi^{\overline{p}} \,
     \overline{D}_{\overline{\imath}} \overline{s}_{\overline{p}}      
   + \chi^p
     \chi^i
     D_i s_p
   + R_{\overline{\imath}p\overline{p}i} 
     \chi^{\overline{\imath}}
     \chi^p 
     \chi^{\overline{p}}  
     \chi^i ,  
\nonumber
\\[1ex]
  &= g^{p\overline{p}}
     s_p
     \overline{s}_{\overline{p}} 
   + \rho^{p}
     D s_p
   + \overline{D} \overline{s}_{\overline{p}}
     \rho^{\overline{p}}
   + \rho^{\overline{p}}
     \mathcal{R}_{\overline{p}p} 
     \rho^p ,
\nonumber
\\[1ex]
  &= \left( 
            s_p 
            e^p,
            \overline{s}_{\overline{p}} 
            e^{\overline{p}}  
     \right)_{\cal G}
   + \left\langle
            \rho^{p'} f_{p'},
            D s_p e^p
     \right\rangle_{\cal G}
   + \left\langle
            \overline{D} \overline{s}_{\overline{p}} e^{\overline{p}},   
            \rho^{\overline{p}'} f_{\overline{p}'} 
     \right\rangle_{\cal G}    
   + \left(
            \rho^{\overline{p}'} f_{\overline{p}'},
            f^{\overline{p}}
            \left(
                   \mathcal{R}_{\overline{p}p} f^p,
                   \rho^p f_p
            \right)_{{\cal G}^{*}}
     \right)_{{\cal G}^{*}} ,
\nonumber
\\[1ex]
  &= \frac{1}{2}
     \left(
            \mathfrak{s},
            \mathfrak{s}
     \right)_{\cal G}
   + \left\langle
            \nabla \mathfrak{s},
            \rho 
     \right\rangle_{\cal G}
   + \frac{1}{2}
     \left(
            \rho,
            \mathcal{R} \rho
     \right)_{{\cal G}^{*}} ,
\nonumber
\\[1ex]
  &= \cal{A} \, ,            
\end{align}
%%%%%%%%%%%%%%%%%
where
%%%%%%%%%%%%%%%%%
\begin{equation}
\begin{aligned}
d &= \partial 
   + \overline{\partial}
   = d \phi^i
     \partial_i
   + d \phi^{\overline{\imath}} \,
     \overline{\partial}_{\overline{\imath}} 
   = \chi^i
     \partial_i
   + \chi^{\overline{\imath}} \,
     \overline{\partial}_{\overline{\imath}} \, ,
\\[1ex]
\nabla 
  &= D 
   + \overline{D}
   = d \phi^i D_i
   + d \phi^{\overline{\imath}} \, 
     \overline{D}_{\overline{\imath}}
   = \chi^i D_i
   + \chi^{\overline{\imath}} \, 
     \overline{D}_{\overline{\imath}}   \, ,
\\[1ex]
\mathfrak{s} 
  &= s_p 
     e^p 
   + \overline{s}_{\overline{p}}
     e^{\overline{p}} \, ,
\\[1ex]
\rho 
  &= \rho^p f_p
   + \rho^{\overline{p}} f_{\overline{p}} 
   = \chi^p f_p
   + \chi^{\overline{p}} f_{\overline{p}} \, .  
\end{aligned}
\end{equation}
%%%%%%%%%%%%%%%%%
Thus, (\ref{S^{(2,2)}_0}) is minus the exponential of the
pullback of a Mathai-Quillen form, 
giving a mathematical understanding of the behavior of RG flow in this model.

In this language, the fact that A model correlators in nonlinear
sigma models are independent of the complex structure is a consequence
of the fact that the $ d $-cohomology class of the pullback of the 
Mathai-Quillen form by $ \mathfrak{s}$ is independent of $\mathfrak{s}$.

\subsubsection{A/2 deformation}

Now, let us turn to the A/2 model for a deformation of the model above,
describing a deformation of the tangent bundle.
Mathematically, the tangent bundle to $Y \equiv \{s=0\}$,
$s = (s_p)$, is defined by the
kernel in the short exact sequence
\begin{displaymath}
0 \: \longrightarrow \: 
T Y \: \longrightarrow \: TM |_Y \: \stackrel{(D_i s_p)}{
\longrightarrow} \:
{\cal G}|_Y \: \longrightarrow \: 0 .
\end{displaymath}
A deformation of the tangent bundle above is defined by
\begin{displaymath}
0 \: \longrightarrow \: 
{\mathcal E'} \: \longrightarrow \:
 TM |_Y \: \stackrel{(D_i s_p + (\delta s)_{i p})}{
\longrightarrow} \:
{\cal G}|_Y \: \longrightarrow \: 0 ,
\end{displaymath}
where the $(\delta s)_{i p}$ define the deformation.

A commonly-discussed special case of this involves deformations of
tangent bundles of hypersurfaces in projective spaces.  In such cases,
in homogeneous coordinates $z^i$, the $(\delta s)_{i p}$ are required to obey
\begin{displaymath}
z^i (\delta s)_{i p} \: = \: 0.
\end{displaymath}
In affine coordinates, this instead becomes the statement that, across
coordinate patches, there are several different $(\delta s)_{i p}$, 
but on any one given
coordinate patch, one is determined by the others.

The action of the A/2 twist of the heterotic Landau-Ginzburg model
that RG flows to a nonlinear sigma model with tangent bundle deformation above
is given by
\cite{GuffinSharpe:A-twistedheterotic}
%%%%%%%%%%%%%%%%%
\begin{align}
\label{S^(0,2)}
S^{(0,2)} 
  &= 2 t \int_{\Sigma} d^2 z
     \left[
            \frac{1}{2}
            \left(
                   g_{\mu\nu} 
                 + i B_{\mu\nu}
            \right)
            \partial_z \phi^{\mu}
            \overline{\partial}_{\overline{z}} \phi^{\nu}
          + i g_{\overline{a}a} \psi^{\overline{a}}_+ \overline{D}_{\overline{z}} \psi^a_+
          + i g_{b\overline{b}} \lambda^b_- D_z \lambda^{\overline{b}}_-
     \right.
\nonumber
\\
  &\phantom{= 2 t \int_{\Sigma} d^2 z \left[ \right.}
     \biggl.
          + R_{a\overline{a}b\overline{b}}
            \psi^a_+
            \psi^{\overline{a}}_+
            \lambda^b_-
            \lambda^{\overline{b}}_- 
          + g^{a\overline{a}}
            F_a 
            \overline{F}_{\overline{a}}
          + \psi^a_+ 
            \lambda^b_- 
            D_a F_b
          + \psi^{\overline{a}}_+ 
            \lambda^{\overline{b}}_- 
            \overline{D}_{\overline{a}} \overline{F}_{\overline{b}}        
     \biggr],
\end{align}
%%%%%%%%%%%%%%%%%
with target space 
\begin{displaymath}
X \: = \: {\rm Tot} \left( {\cal G}^* \: \stackrel{\pi}{\longrightarrow} \:
M \right)
\end{displaymath}
and gauge bundle ${\cal E} = TX$,
%%%%%%%%%%%%%%%%%
where
%%%%%%%%%%%%%%%%%
\begin{equation*}
F_a 
   = (F_p, F_i)
   = \left(
            s_p, p (D_i s_p + (\delta s)_{i p})
     \right) \, ,
\qquad       
\overline{F}_{\overline{a}} 
   = \left(
            \overline{F}_{\overline{p}}, \overline{F}_{\overline{\imath}}
     \right)
   = \left(
            \overline{s}_{\overline{p}}, 
            \overline{p} 
            \left( 
                   \overline{D}_{\overline{\imath}} \overline{s}_{\overline{p}} 
                 + (\delta \overline{s})_{\overline{\imath} \overline{p}}
            \right)
     \right) \, ,
\end{equation*}
%%%%%%%%%%%%%%%%%
\begin{equation*}
\begin{aligned}
D_a F_b 
  &= \partial_a F_b 
   - \Gamma^c_{ab} 
     F_c \, ,
\quad
&\overline{D}_{\overline{a}}  \overline{F}_{\overline{b}}
  &= \overline{\partial}_{\overline{a}} \overline{F}_{\overline{b}}
   - \Gamma^{\overline{c}}_{\overline{a}\overline{b}}
     \overline{F}_{\overline{c}} \, ,    
\\
\overline{D}_{\overline{z}} \psi^a_+ 
   &= \overline{\partial}_{\overline{z}} \psi^a_+ 
   - \overline{\partial}_{\overline{z}} \phi^b \Gamma^a_{bc} \psi^c_+ \, ,
\quad
&D_z \lambda^{\overline{b}}_-
  &= \partial_z \lambda^{\overline{b}}_- 
   - \partial_z \phi^{\overline{a}} 
     \Gamma^{\overline{b}}_{\overline{a}\overline{c}} \lambda^{\overline{c}}_- \, ,          
\end{aligned}
\end{equation*}
%%%%%%%%%%%%%%%%%
and
%%%%%%%%%%%%%%%%%
\begin{equation*}
\begin{aligned}
\psi^i_+ 
  &\equiv \chi^i \!
  &\in \
  &\Gamma
   \left(  
          \phi^* \!
          \left(
                 T^{1,0} M
          \right)       
  \right),
\quad          
&\lambda^i_-
  &\equiv \lambda^i_{\overline{z}} \!
  &\in \
  &\Gamma
   \left( 
          \overline{K}_{\Sigma}
          \otimes
          \left( 
                 \phi^* \!
                 \left(
                        T^{0,1} M
                 \right)       
         \right)\!^{*}
  \right),       
\\
\psi^{\overline{\imath}}_+
  &\equiv \psi^{\overline{\imath}}_z \!
  &\in \
  &\Gamma
   \left(
          K_{\Sigma}
          \otimes
          \left( 
                 \phi^* \!
                 \left(
                        T^{1,0} M
                 \right)       
          \right)\!^{*}
  \right),
\quad     
&\lambda^{\overline{\imath}}_-
  &\equiv \lambda^{\overline{\imath}} \!
  &\in \
  &\Gamma
   \left( 
          \phi^* \!
          \left(
                 T^{0,1} M
          \right)       
   \right),
\\
\psi^p_+
  &\equiv \psi^p_z \!
  &\in \ 
  &\Gamma
   \left(
          K_{\Sigma}
          \otimes
          \phi^* \,
          T^{1,0}_{\pi}
   \right),
\quad
&\lambda^p_-
  &\equiv \lambda^p \!
  &\in \ 
  &\Gamma
   \left(
          \left(
                 \phi^* \,
                 T^{0,1}_{\pi}
          \right)\!^{*}         
   \right),
\\
\psi^{\overline{p}}_+
  &\equiv \chi^{\overline{p}} \!
  &\in \ 
  &\Gamma
   \left(
          \left(
                 \phi^* \,
                 T^{1,0}_{\pi}
          \right)\!^{*} 
   \right),
\quad   
&\lambda^{\overline{p}}_-
  &\equiv \lambda^{\overline{p}}_{\overline{z}} \!
  &\in \ 
  &\Gamma
   \left(
          \overline{K}_{\Sigma}
          \otimes
          \phi^* \,
          T^{0,1}_{\pi}               
   \right),
\\
p 
  &\equiv p_z \!
  &\in \
  &\Gamma
   \left(
          K_{\Sigma} 
          \otimes 
          \phi^* \,
          T^{1,0}_{\pi}
   \right),
\quad
&\overline{p} 
  &\equiv \overline{p}_{\overline{z}} \! 
  &\in \
  &\Gamma
   \left(
          \overline{K}_{\Sigma} 
          \otimes 
          \phi^* \,
          T^{0,1}_{\pi}
   \right).                                              
\end{aligned}
\end{equation*}
%%%%%%%%%%%%%%%%%

Proceeding as in the last example, to illustrate the relevance
of the deformed object
$\omega_{\delta s}$ given by
(\ref{omega-delta-s}), if for example
we restrict
to zero modes on a genus zero worldsheet, in the degree zero sector
we find the following
interactions among zero modes:
%%%%%%%%%%%%%%%%%
\begin{align} 
     g^{\overline{p}p} &
     \overline{F}_{\overline{p}}
     F_p
   + \chi^i
     \lambda^p
     D_i F_p
   + \chi^{\overline{p}}
     \lambda^{\overline{\imath}} \,
     \overline{D}_{\overline{p}}
     \overline{F}_{\overline{\imath}}
   + R_{i\overline{p}p\overline{\imath}}
     \chi^i
     \chi^{\overline{p}}
     \lambda^p
     \lambda^{\overline{\imath}} \, ,      
\nonumber
\\[1ex]     
  &= g^{\overline{p}p}
     \overline{s}_{\overline{p}}
    s_p 
   + \chi^i
     \lambda^p
     D_i s_p           
   + \chi^{\overline{p}}
     \lambda^{\overline{\imath}}
     \left( 
            \overline{D}_{\overline{\imath}} \overline{s}_{\overline{p}} 
          + (\delta \overline{s})_{\overline{\imath} \overline{p}}
     \right)       
   + R_{i\overline{p}p\overline{\imath}}
     \chi^i
     \chi^{\overline{p}} 
     \lambda^p
     \lambda^{\overline{\imath}}\, .
\nonumber
\end{align}
If we now complex conjugate so as to relate the heterotic expression above to
standard mathematics conventions, we find
\begin{align}
     g^{p\overline{p}} &
     s_p
     \overline{s}_{\overline{p}}
   + \chi^{\overline{\imath}}
     \lambda^{\overline{p}} \,
     \overline{D}_{\overline{\imath}} \overline{s}_{\overline{p}}                        
   + \chi^{p}
     \lambda^i
     \left( 
            D_{i} s_p 
          + (\delta s)_{i p}
     \right)       
   + R_{\overline{\imath}p\overline{p}i}
     \chi^{\overline{\imath}}
     \chi^{p} 
     \lambda^{\overline{p}}
     \lambda^{i} \, ,
\nonumber
\\[1ex]
  &= g^{p\overline{p}}
     s_p
     \overline{s}_{\overline{p}}
   + \rho^p D s_p
   + \overline{D} \overline{s}_{\overline{p}} \, \rho^{\overline{p}}        
   + \rho^{\overline{p}}
     \mathcal{R}_{\overline{p}p}
     \rho^p
   + \rho^{p}
     d \phi^{i} \, 
     (\delta s)_{i p} \, ,
\nonumber
\\[1ex]
  &= \left(
            s_p e^p,
            \overline{s}_{\overline{p}} e^{\overline{p}}
     \right)_{\cal G}
   + \left\langle
            \rho^{p'} f_{p'},
            D s_p e^p
     \right\rangle_{\cal G}
   + \left\langle
            \overline{D} \overline{s}_{\overline{p}} e^{\overline{p}},
            \rho^{\overline{p}'} f_{\overline{p}'}
     \right\rangle_{\cal G}
\nonumber
\\
  &\phantom{=}
   + \left(
            \rho^{\overline{p}'} f_{\overline{p}'},
            f^{\overline{p}}
            \left(
                   \mathcal{R}_{\overline{p}p} f^p, 
                   \rho^p f_p
            \right)_{{\cal G}^*}
     \right)_{{\cal G}^*}
   + \left\langle
            \rho^{p'} f_{p'},
            d \phi^{i} \,
            (\delta s)_{i p}
            e^{p} 
     \right\rangle_{\cal G} \, ,
\nonumber
\\[1ex]
  &= {\cal A} 
   + \left\langle
            \rho^{p'} f_{p'},
            d \phi^{i} \,
            (\delta s)_{i p}
            e^{p} 
     \right\rangle_{\cal G} \, ,
\nonumber
\\[1ex]
  &= {\cal A}_{\delta s} \, ,                                         
\end{align}
%%%%%%%%%%%%%%%%%
which is minus the exponent of (\ref{omega-delta-s}).

\subsection{Kernels}

\subsubsection{A/2 model realization of first kernel construction}
\label{sect:A2-kernels}

We can write down an A/2 model describing a kernel as follows.
Suppose we wish to build a (0,2) Landau-Ginzburg model that
RG flows to a nonlinear sigma model on 
\begin{displaymath}
Y \: \equiv \: \{ s = 0 \} \: \subset \: M
\end{displaymath}
with gauge bundle ${\cal E}'$ defined by the kernel of the restriction of
a surjective map
\begin{displaymath}
\tilde{F}: \: {\cal F}_1 \: \longrightarrow \: {\cal F}_2
\end{displaymath}
to $\{s = 0 \}$.  The map $\tilde{F}$ is surjective everywhere
on $M$.  The restriction of $\tilde{F}$ to $Y$ is holomorphic;
however, over the rest of $M$, $\tilde{F}$ need be merely smooth.

Then, 
we consider a Landau-Ginzburg model on 
\begin{displaymath}
X \: = \: {\rm Tot} \left( {\cal F}_2^* \: \stackrel{\pi}{\longrightarrow}
\: M \right)
\end{displaymath}
with gauge bundle ${\cal E}$ given by an extension
\begin{equation}    \label{eq:kernel-ext}
0 \: \longrightarrow \: \pi^* {\cal G}^* \: \longrightarrow \:
{\cal E} \: \longrightarrow \: \pi^* {\cal F}_1
\: \longrightarrow \: 0 .
\end{equation}
To specify the physical theory, we need to specify both the extension
${\cal E}$ and a holomorphic
section of ${\cal E}^*$.  The details of the extension class are,
except for certain special cases, largely not relevant to this paper, so
let us give the holomorphic section first, and then we shall outline pertinent
facts on the extension class.
Dualizing the extension above to
\begin{displaymath}
0 \: \longrightarrow \: \pi^* {\cal F}_1^* \: \longrightarrow \:
{\cal E}^* \: \longrightarrow \: \pi^* {\cal G} \: \longrightarrow \: 0 ,
\end{displaymath}
it is straightforward to see that a holomorphic section of ${\cal E}^*$ uniquely
determines a holomorphic section of $\pi^* {\cal G}$, 
call it $\pi^* s$.
Furthermore, a holomorphic section of ${\cal E}^*$ 
noncanonically\footnote{The smooth section depends upon a choice of
splitting of the smooth bundle ${\cal E}^*$, forgetting the holomorphic
structure.
}
determines a smooth section of $\pi^* {\cal F}_1^*$ which is holomorphic
over $\{s = 0 \}$.  (Alternatively, we could work with holomorphic
sections in local trivializations, but in this paper it will be more
convenient to work with a global smooth section.)
To define the physical theory, we will pick a holomorphic
section of ${\cal E}^*$ determined by the pullback of $s \in \Gamma({\cal G})$,
determining $Y$, and we will take the smooth section of $\pi^* {\cal F}_1^*$
to be given by $p \tilde{F}$, where $p$ denotes fiber coordinates on $X$.

As an aside, the four-fermi term in ${\cal A}_{K1}$ in the
Mathai-Quillen analogue associated to this theory, defined by an element of
\begin{displaymath}
H^1\left(M, {\cal F}_1^* \otimes {\cal G}^* \otimes {\cal F}_2 \right)
\end{displaymath}
is related to the extension class of (\ref{eq:kernel-ext})
as follows.  One computes
\begin{eqnarray*}
{\rm Ext}^1_X(\pi^* {\cal F}_1, \pi^* {\cal G}^*) 
& = &
H^1(X, \pi^* {\cal F}_1^* \otimes \pi^* {\cal G}^*) , \\
& = & 
H^1(M, \pi_* \pi^* ( {\cal F}_1^* \otimes {\cal G}^* ) ) , \\
& = &
H^1(M, {\cal F}_1^* \otimes {\cal G}^* \otimes {\rm Sym}^{\bullet} 
{\cal F}_2) ,
\end{eqnarray*}
since $\pi$ is affine, by Leray.
The four-fermi term can be understood as living in one of the sheaf
cohomology groups above. 

In the special case that ${\cal E}'$ is the restriction of a bundle
on $M$ to $Y$ ({\it i.e.} the map $\tilde{F}$ is globally holomorphic, not
just smooth), the extension will be trivial:
${\cal E} = \pi^* {\cal G}^* \otimes \pi^* {\cal F}_1$.
Readers familiar with Distler-Kachru models can derive the same result 
physically by
thinking about the bundle defined by the fermi superfields.
In general, however, the extension need not be trivial.  For example,
if ${\cal E}' = TY$, as will be the case for ordinary Mathai-Quillen
forms, then the extension will be nontrivial.  In this special case,
${\cal E} = TX$ (as appropriate for a (2,2) supersymmetric theory),
which is realized via the nontrivial extension
\begin{displaymath}
0 \: \longrightarrow \: \pi^* {\cal G}^* \: \longrightarrow \:
{\cal E} \: \longrightarrow \: \pi^* TM \: \longrightarrow \: 0,
\end{displaymath}
where ${\cal F}_1 = TM$.
More generally, for
readers familiar with Distler-Kachru models, if one integrates out
fermionic shift symmetries, the extension will be nontrivial.

The action of the A/2 twisted Landau-Ginzburg model
that RG flows to the A/2 twist of the nonlinear
sigma model above is given in local coordinates 
by \cite{GuffinSharpe:A-twistedheterotic}
\begin{eqnarray*}
S & = & 2t \int_{\Sigma} d^2 z \Biggl[
\frac{1}{2} ( g_{\mu \nu} + i B_{\mu \nu}) \partial_z \phi^{\mu} 
\overline{\partial}_{\overline{z}} \phi^{\nu} \: + \:
i g_{\overline{a} a} \psi_+^{\overline{a}} \overline{D}_{\overline{z}} \psi_+^a
\: + \: i g_{\alpha \overline{\alpha}} \lambda_-^{\alpha}
D_z \lambda_-^{\overline{\alpha}}
\\
& & \hspace*{1in}
\: + \:
F_{a \overline{a} \alpha \overline{\alpha}} \psi_+^a \psi_+^{\overline{a}}
\lambda_-^{\alpha} \lambda_-^{\overline{\alpha}}
\: + \:
h^{x \overline{x}} s_x \overline{s}_{\overline{x}}
\: + \:
h^{\gamma \overline{\gamma}} p^r \overline{p}^{\overline{r}}
\tilde{F}_{r \gamma} 
\tilde{\overline{F}}_{\overline{r} \overline{\gamma}}
\\
& & \hspace*{1in}
\: + \:
\psi_+^i \lambda_-^x D_i s_x \: + \:
\psi_+^r \lambda_-^{\gamma} \tilde{F}_{r \gamma} \: + \:
\psi_+^i \lambda_-^{\gamma} p^r D_i \tilde{F}_{r \gamma}
\\
& & \hspace*{1in}
 \: + \:
\psi_+^{\overline{\imath}} \lambda_-^{\overline{x}} \overline{D}_{\overline{\imath}}
\overline{s}_{\overline{x}} \: + \:
\psi_+^{\overline{r}} \lambda_-^{\overline{\gamma}} \tilde{\overline{F}}_{
\overline{r} \overline{\gamma}} \: + \:
\psi_+^{\overline{\imath}} \lambda_-^{\overline{\gamma}} 
\overline{p}^{\overline{r}} \,
\overline{D}_{\overline{\imath}} \tilde{\overline{F}}_{\overline{r} \overline{\gamma}}
\Biggr] ,
\end{eqnarray*}
where $x$ indexes local coordinates along the fibers of ${\cal G}$,
$\gamma$ indexes local coordinates along the fibers of 
${\cal F}_1$, $r$ indexes local coordinates (denoted $p$) along
along the fibers of ${\cal F}_2^*$, $i$ indexes local
coordinates on $M$, $a \sim(r, i)$ indexes local coordinates on $X$,
and $\alpha \sim (x, \gamma)$ indexes local coordinates along the fibers
of ${\cal E}$.  In the notation of
\cite{GuffinSharpe:A-twistedheterotic}, in local coordinates,
$(F_{\alpha}) = (s_x, p^r \tilde{F}_{r \gamma})$.

Note that the space of vacua is given by $\{s=0\} \cap \{p=0\}$:
the first condition follows from the potential term $|s|^2$, the second
from the potential term $|p \tilde{F}|^2$ and the fact that $\tilde{F}$
is surjective everywhere on $M$.  (In a (2,2) theory, if the space
becomes singular so that $p$ gets a vev, the result seems to have
an interpretation in terms of cotangent complexes \cite{dbzpriv}, with
${\mathbb C}^{\times}$ rotations of the fibers of $X$ providing a grading.
In (0,2) theories, 
if surjectivity of $\tilde{F}$ breaks down and $p$ gets a vev, then
sometimes, under certain circumstances that are not well-understood,
the theory will still be well-behaved, but typically it will be singular
\cite{dgm}.)

The fermions and bosons are twisted as follows:
\begin{displaymath}
\begin{aligned}
\psi_+^i & \equiv \chi^i &\in \ &\Gamma( \phi^*(TM) ),
\quad
& \psi_+^{\overline{\imath}} & \equiv \psi_z^{\overline{\imath}}
& \in \ & \Gamma(K_{\Sigma} \otimes \phi^* T^*M ), 
\\
\psi_+^r & \equiv \psi_z^r & \in \
& \Gamma(K_{\Sigma} \otimes \phi^* T_{\pi}^{1,0} ),
\quad
& \psi_+^{\overline{r}} & \equiv \chi^{\overline{r}} & \in \
& \Gamma( (\phi^* T_{\pi}^{1,0})^* ), 
\\
\lambda_-^x & \equiv \lambda^x & \in \ & 
\Gamma( (\phi^* T_{\cal G}^{0,1} )^* ), 
\quad
& \lambda_-^{\overline{x}} & \equiv \lambda_{\overline{z}}^{\overline{x}}
& \in \ & \Gamma(\overline{K}_{\Sigma} \otimes \phi^* T_{\cal G}^{0,1} ),
\\
\lambda_-^{\gamma} & \equiv \lambda_{\overline{z}}^{\gamma} & \in \
& \Gamma( \overline{K}_{\Sigma} \otimes ( \phi^* T_{ {\cal F}_1 }^{0,1} )^* ),
\quad 
& \lambda_-^{\overline{\gamma}} & \equiv \lambda^{\overline{\gamma}}
& \in \ & \Gamma( \phi^* T_{ {\cal F}_1 }^{0,1} ),
\\
p & \equiv p_z & \in \ & \Gamma(K_{\Sigma} \otimes \phi^* T_{\pi}^{1,0} ),
\quad
& \overline{p} & \equiv \overline{p}_{\overline{z}} & \in \ &
\Gamma(\overline{K}_{\Sigma} \otimes \phi^* T_{\pi}^{0,1} ).
\end{aligned}
\end{displaymath}

Anomalies constrain the theory above.  Specifically, one must require that
\begin{displaymath}
\det {\cal G}^* \otimes \det {\cal F}_1^* \: \cong \:
\det {\cal F}_2^* \otimes K_M, \: \: \:
{\rm ch}_2({\cal E}) = {\rm ch}_2(TX) .
\end{displaymath}
(Note that the first condition is slightly different than merely
$\det {\cal E}^* \cong K_X$, ultimately because the twist acts
differently on various contributions to ${\cal E}$ and $TX$.)
One can show that anomaly-freedom in the UV implies anomaly-freedom in
the IR.

The effective interactions can be obtained by truncating to fermi
zero modes.  In the degree zero sector, they are
\begin{displaymath}
h^{x \overline{x}} s_x \overline{s}_{\overline{x}}
\: + \: \chi^i \lambda^x D_i s_x \: + \:
\chi^{\overline{r}} \lambda^{\overline{\gamma}} \, \tilde{\overline{F}}_{
\overline{r} \overline{\gamma}} \: + \:
F_{i \overline{r} x \overline{\gamma}} \chi^i \chi^{\overline{r}}
\lambda^x \lambda^{\overline{\gamma}} .
\end{displaymath}

The reader should recognize the expression above as the complex
conjugate of equation~(\ref{eq:MQ:kernels}).  The complex conjugation
is necessary because of a difference in standard conventions:  standard
heterotic string conventions yield $\partial$-closed operators, whereas
standard mathematics conventions in this context involve
$\overline{\partial}$-closed operators.  The two are related by a simple
complex conjugation.

Although $\chi^i$, $\chi^{\overline{r}}$, $\lambda^x$, and
$\lambda^{\overline{\gamma}}$ are all scalars,
it can be shown following \cite{GuffinSharpe:A-twistedheterotic}
that only $\chi^i$, $\lambda^{\overline{\gamma}}$ are
BRST-invariant.
Furthermore, since the $p$ fields are twisted, the bosonic zero modes lie
along $M$.  Restricting to zero modes, observables are then of the form
\begin{displaymath}
f(\phi^i) \chi^{i_1} \cdots \chi^{i_n} 
\lambda^{\overline{\gamma}_1} \cdots \lambda^{\overline{\gamma}_m} ,
\end{displaymath}
which after a complex conjugation are naively interpreted in terms of
\begin{displaymath}
H^{\bullet}\left(M, \wedge^{\bullet} {\cal F}_1^* \right) .
\end{displaymath}
(The more nearly correct interpretation of the chiral ring is in terms of
a restriction to $\{s=0\}$ of the cohomology above, but this is
not essential for this discussion.)
Correlation functions then are of the form of integrals over
$M$ of observables times the exponential of the zero mode interactions.

Since this theory flows under RG to an A/2-twisted nonlinear sigma model,
correlation functions of the observables above should coincide with
correlation functions in the nonlinear sigma model, and this is the
root of the claims in this paper regarding analogues of Mathai-Quillen forms.

Ordinary (2,2) Landau-Ginzburg models, and deformations thereof,
are special cases of this construction.  In a (2,2) Landau-Ginzburg
model on $X$, ${\cal E} = TX$ and the $(F_a)$ are given by derivatives
of a superpotential $W$.  By taking ${\cal F}_1 = TM$, ${\cal F}_2 = 
{\cal G}$, we can take the extension
\begin{displaymath}
0 \: \longrightarrow \: \pi^* {\cal G}^* \: \longrightarrow \:
{\cal E} \: \longrightarrow \: \pi^* {\cal F}_1 \: \longrightarrow \: 0
\end{displaymath}
to coincide with the tangent bundle $TX$,
\begin{displaymath}
0 \: \longrightarrow \: \pi^* {\cal G}^* \: \longrightarrow \:
TX \: \longrightarrow \: \pi^* TM \: \longrightarrow \: 0 ,
\end{displaymath}
where $X$ is the total space of $\pi: {\cal G}^* \rightarrow M$.
The (2,2) superpotential $W = p s$, $s$ here the pullback of a section of
${\cal G}$, so we take $(F_a) = (s, p D_i s)$.
(A deformation of the (2,2) locus would be described by the same
bundles but with $(F_a) = (s, p( D_i s + (\delta s)_i) )$.
It is straightforward to check that the Lagrangian in this special case
corresponds with the Lagrangian given earlier for the (2,2) theory.

In the special case of a (2,2) theory,
the curvature term appearing in the zero mode interactions, now an element
of
\begin{displaymath}
H^1\left(M, \Omega^1_M \otimes {\cal G} \otimes {\cal G}^* \right)
\end{displaymath}
can be shown to coincide with the Atiyah class of ${\cal G}$, from the
fact that the extension defined is $TX$.  Specifically\footnote{
We would like to thank T.~Pantev for a discussion of this point.
}, from earlier
work one has
\begin{displaymath}
{\rm Ext}^1_X( \pi^* TM, \pi^* {\cal G}^*) 
\: = \:
H^1\left(M, \Omega_M^1 \otimes {\cal G}^* \otimes {\rm Sym}^{\bullet}
{\cal G} \right) .
\end{displaymath} 
Now, there is a natural ${\mathbb C}^{\times}$ scaling action on the fibers
of $\pi: X \rightarrow M$, which induces an action on tangent bundles.
The original extension class has weight 0 under this ${\mathbb C}^{\times}$,
$\Omega_M^1$ has weight 0, ${\cal G}^*$ has weight 1, and
${\rm Sym}^{\bullet} {\cal G}$ has weight $- \bullet$, so the original extension
in $H^1(X, \pi^* (\Omega_M^1 \otimes {\cal G}^*)$ is an element of
$H^1(M, \Omega_M^1 \otimes {\cal G}^* \otimes {\cal G})$, and one can check
in local trivializations that this element is the Atiyah class of
${\cal G}$. 

In any event, the fact that the curvature term so determined coincides
with the Atiyah class of ${\cal G}$, means that on the (2,2) locus we
exactly reproduce the curvature term appearing in standard Mathai-Quillen
forms.

\subsubsection{B/2 model realization of second kernel construction}
\label{sect:b2-kernels}

We can also write down a B/2 model describing a kernel as follows.
Suppose we wish to build a (0,2) Landau-Ginzburg model that RG flows to
a nonlinear sigma model on 
\begin{displaymath}
Y \: \equiv \: \{s=0\} \: \subset \: M
\end{displaymath}
with gauge bundle defined by the kernel of the restriction of a
surjective holomorphic map
\begin{displaymath}
\tilde{F}: \: {\cal F}_1 \: \longrightarrow \: {\cal F}_2
\end{displaymath}
to $\{s=0\}$.
As before, $\tilde{F}$ is surjective everywhere on $M$, and in addition
we also impose the condition that $\tilde{F}$ be holomorphic everywhere
on $M$, not just the restriction to $Y$.

Then, we consider a Landau-Ginzburg model on 
\begin{displaymath}
X \: = \: {\rm Tot}\, \left( {\cal F}_2^* \: \stackrel{\pi}{\longrightarrow}
\: M \right)
\end{displaymath}
with gauge bundle ${\cal E} = \pi^* {\cal G}^* \oplus \pi^* {\cal F}_1$.

The action of the B/2 twisted Landau-Ginzburg model that RG flows to the
B/2 twist of the nonlinear sigma model above is of the same form as
that discussed previously for the A/2 twist:
\begin{eqnarray*}
S & = & 2t \int_{\Sigma} d^2 z \Biggl[
\frac{1}{2} ( g_{\mu \nu} + i B_{\mu \nu}) \partial_z \phi^{\mu} 
\overline{\partial}_{\overline{z}} \phi^{\nu} \: + \:
i g_{\overline{a} a} \psi_+^{\overline{a}} \overline{D}_{\overline{z}} \psi_+^a
\: + \: i g_{\alpha \overline{\alpha}} \lambda_-^{\alpha}
D_z \lambda_-^{\overline{\alpha}}
\\
& & \hspace*{1in}
\: + \:
F_{a \overline{a} \alpha \overline{\alpha}} \psi_+^a \psi_+^{\overline{a}}
\lambda_-^{\alpha} \lambda_-^{\overline{\alpha}}
\: + \:
h^{x \overline{x}} s_x \overline{s}_{\overline{x}}
\: + \:
h^{\gamma \overline{\gamma}} p^r \overline{p}^{\overline{r}}
\tilde{F}_{r \gamma} 
\tilde{\overline{F}}_{\overline{r} \overline{\gamma}}
\\
& & \hspace*{1in}
\: + \:
\psi_+^i \lambda_-^x D_i s_x \: + \:
\psi_+^r \lambda_-^{\gamma} \tilde{F}_{r \gamma} \: + \:
\psi_+^i \lambda_-^{\gamma} p^r D_i \tilde{F}_{r \gamma}
\\
& & \hspace*{1in}
 \: + \:
\psi_+^{\overline{\imath}} \lambda_-^{\overline{x}} \overline{D}_{\overline{\imath}}
\overline{s}_{\overline{x}} \: + \:
\psi_+^{\overline{r}} \lambda_-^{\overline{\gamma}} \tilde{\overline{F}}_{
\overline{r} \overline{\gamma}} \: + \:
\psi_+^{\overline{\imath}} \lambda_-^{\overline{\gamma}} 
\overline{p}^{\overline{r}} \,
\overline{D}_{\overline{\imath}} \tilde{\overline{F}}_{\overline{r} \overline{\gamma}}
\Biggr] ,
\end{eqnarray*}
where $x$ indexes local coordinates along the fibers of ${\cal G}$,
$\gamma$ indexes local coordinates along the fibers of
${\cal F}_1$, $r$ indexes local coordinates (denoted $p$) along
along the fibers of ${\cal F}_2^*$, $i$ indexes local
coordinates on $M$, $a \sim(r, i)$ indexes local coordinates on $X$,
and $\alpha \sim (x, \gamma)$ indexes local coordinates along the fibers
of ${\cal E}$.  In the notation of
\cite{GuffinSharpe:A-twistedheterotic}, in local coordinates,
$(F_{\alpha}) = (s_x, p^r \tilde{F}_{r \gamma})$.

The fermions are twisted as follows:
\begin{displaymath}
\begin{aligned}
\psi_+^i & \equiv \chi^i &\in \ &\Gamma( \phi^*(TM) ),
\quad
& \psi_+^{\overline{\imath}} & \equiv \psi_z^{\overline{\imath}}
& \in \ & \Gamma(K_{\Sigma} \otimes \phi^* T^*M ), 
\\
\psi_+^r & \equiv \chi^r & \in \
& \Gamma( \phi^* T_{\pi}^{1,0} ),
\quad
& \psi_+^{\overline{r}} & \equiv \psi_z^{\overline{r}} & \in \
& \Gamma(K_{\Sigma} \otimes (\phi^* T_{\pi}^{1,0})^* ), 
\\
\lambda_-^x & \equiv \lambda^x & \in \ & 
\Gamma( (\phi^* T_{\cal G}^{0,1} )^* ), 
\quad
& \lambda_-^{\overline{x}} & \equiv \lambda_{\overline{z}}^{\overline{x}}
& \in \ & \Gamma(\overline{K}_{\Sigma} \otimes \phi^* T_{\cal G}^{0,1} ),
\\
\lambda_-^{\gamma} & \equiv \lambda^{\gamma} & \in \
& \Gamma(  ( \phi^* T_{ {\cal F}_1 }^{0,1} )^* ),
\quad 
& \lambda_-^{\overline{\gamma}} & \equiv 
\lambda_{\overline{z}}^{\overline{\gamma}}
& \in \ & \Gamma( \overline{K}_{\Sigma} \otimes
\phi^* T_{ {\cal F}_1 }^{0,1} ).
\end{aligned}
\end{displaymath}
In the B/2 twisted theory, no bosons need be twisted.

Anomalies constrain the theory as follows:
\begin{displaymath}
K_X \: \cong \: \pi^* \det {\cal G}^* \otimes \pi^* \det {\cal F}_1, 
\: \: \:
{\rm ch}_2({\cal E}) \: = \: {\rm ch}_2(TX) .
\end{displaymath}
In the IR, the first condition becomes 
$\det {\cal E}' \cong K_Y$, as needed to define the B/2 twist of the
nonlinear sigma model.

The effective interactions can be obtained by truncating to fermi zero modes.
In the degree zero sector, they are
\begin{displaymath}
h^{x \overline{x}} s_x \overline{s}_{\overline{x}}
\: + \:
h^{\gamma \overline{\gamma}} p^r \overline{p}^{\overline{r}}
\tilde{F}_{r \gamma} 
\tilde{\overline{F}}_{\overline{r} \overline{\gamma}}
\: + \:
\chi^i \lambda^x D_i s_x \: + \:
\chi^r \lambda^{\gamma} \tilde{F}_{r \gamma} \: + \:
\chi^i \lambda^{\gamma} p^r D_i \tilde{F}_{r \gamma}.
\end{displaymath}

Of the scalars $\chi^i$, $\chi^r$, $\lambda^x$, $\lambda^{\gamma}$,
$\chi^i$, $\chi^r$ are BRST invariant, and if we define
\begin{displaymath}
\theta_{\overline{x}} \: \equiv \: h_{x \overline{x}} \lambda^x, \: \: \:
\theta_{\overline{\gamma}} \: \equiv \: h_{\gamma \overline{\gamma}}
\lambda^{\gamma},
\end{displaymath}
then it can be shown
\begin{displaymath}
Q \cdot \theta_{\overline{x}} \: \propto \: \overline{s}_{\overline{x}},
\: \: \:
Q \cdot \theta_{\overline{\gamma}} \: \propto \: \overline{p}^{
\overline{r}} \, \tilde{ \overline{F} }_{\overline{r} \overline{\gamma} }.
\end{displaymath}

Observables built from $\chi^i$, $\chi^r$, $\theta_{\overline{\gamma}}$
are then elements of hypercohomology
\begin{displaymath}
{\mathbb H}^{\bullet}\left(X, \cdots \: \longrightarrow \:
\wedge^2 {\cal F}_1 \: \longrightarrow \: {\cal F}_1 \: \longrightarrow
\: {\cal O}_X \right)
\end{displaymath}
(with maps given by contraction with $p \tilde{F}$).
Under RG flow these become observables in a B/2-twisted nonlinear sigma
model, and the zero-mode interactions define an analogue of a
Mathai-Quillen form.

\subsection{Cokernels}

\subsubsection{B/2 model realization of first cokernel construction}
\label{sect:b2-cokernels}

In this section we will describe a B/2 twisted (0,2) Landau-Ginzburg
model for a cokernel.  This will yield another analogue of a Mathai-Quillen
form.

Suppose we wish to build a (0,2) Landau-Ginzburg model that RG flows to a
nonlinear sigma model on 
\begin{displaymath}
Y \: \equiv \: \{s = 0 \} \: \subset \: M
\end{displaymath}
(where $s \in \Gamma({\cal G})$), and with gauge bundle
given by the cokernel
of the restriction of an injective map
\begin{displaymath}
\tilde{E}: \: {\cal F}_1 \: \longrightarrow \: {\cal F}_2
\end{displaymath}
to $Y$.  The map $\tilde{E}$ is injective everywhere on $M$.
The restriction of $\tilde{E}$ to $Y$ is holomorphic; however, over the
rest of $M$, $\tilde{E}$ need be merely smooth.

Then, we consider a Landau-Ginzburg model
on
\begin{displaymath}
X \: = \: {\rm Tot}\left( {\cal F}_1 \: \stackrel{\pi}{\longrightarrow} \:
M \right),
\end{displaymath}
with gauge bundle ${\cal E} \rightarrow X$ given by an extension
\begin{equation}   \label{eq:cokernel-ext}
0 \: \longrightarrow \: \pi^* {\cal F}_2 \: \longrightarrow \:
{\cal E} \: \longrightarrow \: \pi^* {\cal G} \:
\longrightarrow \: 0 .
\end{equation}

To uniquely determine the physics, we must specify which extension,
and also a holomorphic section of ${\cal E}$, such that the resulting
(0,2) Landau-Ginzburg theory will renormalization-group flow to the (0,2)
nonlinear sigma model above.

Let us begin with the choice of holomorphic section of ${\cal E}$.
This determines a holomorphic section of $\pi^* {\cal G}$ together with
a smooth section of $\pi^* {\cal F}_2$ that becomes holomorphic over
the vanishing locus of the section of $\pi^* {\cal G}$.
We will take the holomorphic section of $\pi^* {\cal G}$ to be
the pullback of $s$ (whose vanishing locus is $Y$), and the smooth
section of $\pi^* {\cal F}_2$ to be $q \tilde{E}$, where $q$ is a
fiber coordinate on $X$.

The choice of extension class is largely not relevant to the purpose of
this paper, so we shall not describe it in detail.  Suffice it to say,
in general, the extension will be nontrivial, with certain exceptions.
In the special case that ${\cal E}'$ is the restriction of a bundle
on $M$ to $Y$ ({\it i.e.} the map $\tilde{E}$ is globally holomorphic),
the extension will be trivial:  ${\cal E} = 
\pi^* {\cal G} \otimes \pi^* {\cal F}_2$.

The action of the B/2 twisted Landau-Ginzburg model that RG flows to the
B/2 twist of the nonlinear sigma model above is given in local coordinates
by \cite{GuffinSharpe:A-twistedheterotic}
\begin{eqnarray*}
S & = & 2t \int_{\Sigma} d^2 z \Biggl[
\frac{1}{2} ( g_{\mu \nu} + i B_{\mu \nu}) \partial_z \phi^{\mu} 
\overline{\partial}_{\overline{z}} \phi^{\nu} \: + \:
i g_{\overline{a} a} \psi_+^{\overline{a}} \overline{D}_{\overline{z}} \psi_+^a
\: + \: i g_{\alpha \overline{\alpha}} \lambda_-^{\alpha}
D_z \lambda_-^{\overline{\alpha}}
\\
& & \hspace*{1in}
\: + \:
F_{a \overline{a} \alpha \overline{\alpha}} \psi_+^a \psi_+^{\overline{a}}
\lambda_-^{\alpha} \lambda_-^{\overline{\alpha}}
\: + \:
h_{x \overline{x}} s^x \overline{s}^{\overline{x}}
\: + \:
h_{\gamma \overline{\gamma}} q^m 
\overline{q}^{\overline{m}} \tilde{E}_m^{\gamma} 
\tilde{\overline{E}}_{\overline{m}}^{ \overline{\gamma} }
\\
& & \hspace*{1in}
\: + \:
\psi_+^i \lambda_-^{\overline{x}} D_i s^x \, h_{x \overline{x}} \: + \:
\psi_+^{m} \lambda_-^{\overline{\gamma}} \tilde{E}_m^{\gamma}
\, h_{\gamma \overline{\gamma}} \: + \:
\psi_+^i \lambda_-^{\overline{\gamma}} q^m \left(D_i \tilde{E}_m^{\gamma}
\right)  h_{\gamma \overline{\gamma}}
\\
& & \hspace*{1in}
 \: + \:
\psi_+^{\overline{\imath}} \lambda_-^{x} \overline{D}_{\overline{\imath}}
\overline{s}^{\overline{x}} \, h_{x \overline{x}} \: + \:
\psi_+^{\overline{m}} \lambda_-^{\gamma} \tilde{\overline{E}}_{
\overline{m}}^{ \overline{\gamma}} \, h_{\gamma \overline{\gamma}} \: + \:
\psi_+^{\overline{\imath}} \lambda_-^{\gamma} 
\overline{q}^{\overline{m}} \left(
\overline{D}_{\overline{\imath}} \tilde{\overline{E}}_{\overline{m}}^{ \overline{\gamma}}
\right) h_{\gamma \overline{\gamma}}
\Biggr] ,
\end{eqnarray*}
where, much as in the last section,
$x$ indexes local coordinates along the fibers of ${\cal G}$,
$m$ indexes local coordinates (denoted $q$) along the fibers of
${\cal F}_1$, $\gamma$ indexes local coordinates
along the fibers of ${\cal F}_2$, $i$ indexes local
coordinates on $M$, $a \sim(m, i)$ indexes local coordinates on $X$,
and $\alpha \sim (x, \gamma)$ indexes local coordinates along the fibers
of ${\cal E}$, and $s^x$ now denotes a component of a holomorphic
section of ${\cal G}$ (rather than $s_x$ as was used elsewhere,
because of the different way ${\cal G}$ appears in ${\cal E}$).  
In the notation of
\cite{GuffinSharpe:A-twistedheterotic}, in local coordinates,
$(E^{\alpha}) = (s^x, q^m \tilde{E}_m^{\gamma})$.

The space of vacua is of the form $\{s=0\} \cap \{q=0\}$.  The first
condition is a result of the bosonic potential $|s|^2$, and the second
is a result of the bosonic potential $|q \tilde{E}|^2$ plus the fact that
$\tilde{E}$ is injective everywhere.

The fermions and bosons are twisted as follows:
\begin{displaymath}
\begin{aligned}
\psi_+^i & \equiv \chi^i &\in \ &\Gamma( \phi^*(TM) ),
\quad
& \psi_+^{\overline{\imath}} & \equiv \psi_z^{\overline{\imath}}
& \in \ & \Gamma(K_{\Sigma} \otimes \phi^* T^*M ), 
\\
\psi_+^{m} & \equiv \psi_z^{m} & \in \ 
& \Gamma( K_{\Sigma} \otimes \phi^* T_{\pi}^{1,0} ),
\quad
& \psi_+^{\overline{m}} & \equiv \chi^{\overline{m}} & \in \
& \Gamma( (\phi^* T_{\pi}^{1,0})^* ), 
\\
\lambda_-^x & \equiv \lambda_{\overline{z}}^x & \in \ & 
\Gamma( \overline{K}_{\Sigma} \otimes (\phi^* T_{\cal G}^{0,1} )^* ), 
\quad
& \lambda_-^{\overline{x}} & \equiv \lambda^{\overline{x}}
& \in \ & \Gamma( \phi^* T_{\cal G}^{0,1} ),
\\
\lambda_-^{\gamma} & \equiv \lambda^{\gamma} & \in \
& \Gamma( ( \phi^* T_{ {\cal F}_2 }^{0,1} )^* ),
\quad 
& \lambda_-^{\overline{\gamma}} & \equiv \lambda_{\overline{z}}^{
\overline{\gamma}}
& \in \ & \Gamma( \overline{K}_{\Sigma} \otimes \phi^* T_{ {\cal F}_2 }^{0,1} ),
\\
q & \equiv q_z & \in \ & \Gamma(K_{\Sigma} \otimes \phi^* T_{\pi}^{1,0}),
\quad
& \overline{q} & \equiv \overline{q}_{\overline{z}} & \in \ &
\Gamma( \overline{K}_{\Sigma} \otimes \phi^* T_{\pi}^{0,1}).
\end{aligned}
\end{displaymath}

Anomalies constrain the theory as follows:
\begin{displaymath}
K_M \: \cong \: \det {\cal G}^* \otimes \det {\cal F}_1^* \otimes
\det {\cal F}_2, \: \: \:
{\rm ch}_2({\cal E}) \: = \: {\rm ch}_2(TX) .
\end{displaymath}

The effective interactions can be obtained by truncating to fermi zero modes.
In the degree zero sector, they are
\begin{displaymath}
h_{x \overline{x}} s^x \overline{s}^{\overline{x}} \: + \:
\chi^i \lambda^{\overline{x}} D_i s^x \, h_{x \overline{x}} \: + \:
\chi^{\overline{m}} \lambda^{\gamma} \tilde{ \overline{E} }_{
\overline{m}}^{\overline{\gamma}} h_{\gamma \overline{\gamma}} \: + \:
F_{i \overline{m} \overline{x} \gamma } \chi^i \chi^{\overline{m}} 
\lambda^{\overline{x}} \lambda^{\gamma} .
\end{displaymath}
It will be useful to define $\theta_{\overline{\gamma}} \equiv h_{
\overline{\gamma} \gamma} \lambda^{\gamma}$, so that the effective interactions
become
\begin{displaymath}
h_{x \overline{x}} s^x \overline{s}^{\overline{x}} \: + \:
\chi^i \lambda^{\overline{x}} D_i s^x \, h_{x \overline{x}} \: + \:
\chi^{\overline{m}} \theta_{\overline{\gamma}} \tilde{ \overline{E} }_{
\overline{m}}^{\overline{\gamma}}  \: + \:
F_{i \overline{m} \overline{x} \gamma } \chi^i \chi^{\overline{m}} 
\lambda^{\overline{x}} \theta_{\overline{\gamma}} 
h^{\gamma \overline{\gamma}} .
\end{displaymath}

It can be shown that the scalars
$\chi^i$ and $\theta_{\overline{\gamma}}$ are
BRST invariant, and the other scalars $\chi^{\overline{m}}$,
$\lambda^{\overline{x}}$ are not.  Furthermore, since the $q$ fields are
twisted, the bosonic zero modes lie along $M$.  Restricting to zero modes,
observables are then of the form
\begin{displaymath}
f(\phi^i) \chi^{i_1} \cdots \chi^{i_n} \theta_{\overline{\gamma}_1}
\cdots \theta_{\overline{\gamma}_m}
\end{displaymath}
which after a complex conjugation are naively interpreted in terms of
\begin{displaymath}
H^{\bullet}\left(M, \wedge^{\bullet} {\cal F}_2 \right).
\end{displaymath}
(The more nearly correct interpretation of this chiral ring is in
terms of a restriction to $\{s=0\}$ of the cohomology above, but this is
not essential for this discussion.)

Correlation functions then are of the form of integrals over $M$ of 
observables times the exponential of the zero mode interactions.

As an aside, the four-fermi term in the zero mode interactions above,
defined by an element of
\begin{displaymath}
H^1\left(M, {\cal F}_1^* \otimes {\cal G}^* \otimes {\cal F}_2 \right)
\end{displaymath}
is related to the extension class of (\ref{eq:cokernel-ext}) as follows.
One computes
\begin{eqnarray*}
{\rm Ext}^1_X\left( \pi^* {\cal G}, \pi^* {\cal F}_2 \right)
& = &
H^1\left(X, \pi^* {\cal G}^* \otimes \pi^* {\cal F}_2 \right), \\
& = &
H^1\left(M, \pi_* \pi^* \left( {\cal G}^* \otimes {\cal F}_2 \right)
\right), \\
& = &
H^1\left(M, {\cal G}^* \otimes {\cal F}_2 \otimes {\rm Sym}^{\bullet}
{\cal F}_1^* \right),
\end{eqnarray*}
as previously, as the four-fermi term lives in one of the sheaf cohomology
groups above.

\subsubsection{A/2 model realization of second cokernel construction}
\label{sect:A2-cokernels}

We can also write down an A/2 model describing a cokernel.
Our description will be dual to the description of kernels
in section~\ref{sect:b2-kernels}, but for
completeness, we give the details here.
Suppose we wish to build a (0,2) Landau-Ginzburg model that RG flows to a 
nonlinear sigma model on 
\begin{displaymath}
Y \: \equiv \: \{s = 0 \} \: \subset \: M
\end{displaymath}
(where $s \in \Gamma({\cal G})$), and with gauge bundle
given by the cokernel
of the restriction of an injective map
\begin{displaymath}
\tilde{E}: \: {\cal F}_1 \: \longrightarrow \: {\cal F}_2
\end{displaymath}
to $Y$, where $\tilde{E}$ is both injective and holomorphic everywhere on $M$.

Then, we consider a Landau-Ginzburg model
on
\begin{displaymath}
X \: = \: {\rm Tot}\left( {\cal F}_1 \: \stackrel{\pi}{\longrightarrow} \:
M \right),
\end{displaymath}
with gauge bundle ${\cal E} \rightarrow X$ given as the sum
\begin{displaymath}
{\cal E} \: = \: \pi^* {\cal F}_2 \oplus \pi^* {\cal G} .
\end{displaymath}

To uniquely determine the physics, we must specify which extension,
and also a holomorphic section of ${\cal E}$, such that the resulting
(0,2) Landau-Ginzburg theory will renormalization-group flow to the (0,2)
nonlinear sigma model above.

Let us begin with the choice of holomorphic section of ${\cal E}$.
This is determined by holomorphic sections of each of ${\cal F}_2$,
${\cal G}$.  
We will take the holomorphic section of $\pi^* {\cal G}$ to be
the pullback of $s$ (whose vanishing locus is $Y$), and the holomorphic
section of $\pi^* {\cal F}_2$ to be $q \tilde{E}$, where $q$ is a 
fiber coordinate on $X$.

The action of the A/2 twisted Landau-Ginzburg model that RG flows to the
A/2 twist of the nonlinear sigma model above is given in local coordinates
by \cite{GuffinSharpe:A-twistedheterotic}
\begin{eqnarray*}
S & = & 2t \int_{\Sigma} d^2 z \Biggl[
\frac{1}{2} ( g_{\mu \nu} + i B_{\mu \nu}) \partial_z \phi^{\mu} 
\overline{\partial}_{\overline{z}} \phi^{\nu} \: + \:
i g_{\overline{a} a} \psi_+^{\overline{a}} \overline{D}_{\overline{z}} \psi_+^a
\: + \: i g_{\alpha \overline{\alpha}} \lambda_-^{\alpha}
D_z \lambda_-^{\overline{\alpha}}
\\
& & \hspace*{1in}
\: + \:
F_{a \overline{a} \alpha \overline{\alpha}} \psi_+^a \psi_+^{\overline{a}}
\lambda_-^{\alpha} \lambda_-^{\overline{\alpha}}
\: + \:
h_{x \overline{x}} s^x \overline{s}^{\overline{x}}
\: + \:
h_{\gamma \overline{\gamma}} q^m 
\overline{q}^{\overline{m}} \tilde{E}_m^{\gamma} 
\tilde{\overline{E}}_{\overline{m}}^{ \overline{\gamma} }
\\
& & \hspace*{1in}
\: + \:
\psi_+^i \lambda_-^{\overline{x}} D_i s^x\, h_{x \overline{x}} \: + \:
\psi_+^{m} \lambda_-^{\overline{\gamma}} \tilde{E}_m^{\gamma}
\, h_{\gamma \overline{\gamma}} \: + \:
\psi_+^i \lambda_-^{\overline{\gamma}} q^m \left(D_i \tilde{E}_m^{\gamma}
\right)  h_{\gamma \overline{\gamma}}
\\
& & \hspace*{1in}
 \: + \:
\psi_+^{\overline{\imath}} \lambda_-^{x} \overline{D}_{\overline{\imath}}
\overline{s}^{\overline{x}}\, h_{x \overline{x}} \: + \:
\psi_+^{\overline{m}} \lambda_-^{\gamma} \tilde{\overline{E}}_{
\overline{m}}^{ \overline{\gamma}} \, h_{\gamma \overline{\gamma}} \: + \:
\psi_+^{\overline{\imath}} \lambda_-^{\gamma} 
\overline{q}^{\overline{m}} \left(
\overline{D}_{\overline{\imath}} \tilde{\overline{E}}_{\overline{m}}^{ \overline{\gamma}}
\right) h_{\gamma \overline{\gamma}}
\Biggr] ,
\end{eqnarray*}
where, much as in the last section,
$x$ indexes local coordinates along the fibers of ${\cal G}$,
$m$ indexes local coordinates (denoted $q$) along the fibers of
${\cal F}_1$, $\gamma$ indexes local coordinates 
along the fibers of ${\cal F}_2$, $i$ indexes local
coordinates on $M$, $a \sim(m, i)$ indexes local coordinates on $X$,
and $\alpha \sim (x, \gamma)$ indexes local coordinates along the fibers
of ${\cal E}$.  In the notation of
\cite{GuffinSharpe:A-twistedheterotic}, in local coordinates,
$(E^{\alpha}) = (s^x, q^m \tilde{E}_m^{\gamma})$.

The fermions are twisted as follows:
\begin{displaymath}
\begin{aligned}
\psi_+^i & \equiv \chi^i &\in \ &\Gamma( \phi^*(TM) ),
\quad
& \psi_+^{\overline{\imath}} & \equiv \psi_z^{\overline{\imath}}
& \in \ & \Gamma(K_{\Sigma} \otimes \phi^* T^*M ), 
\\
\psi_+^{m} & \equiv \chi^{m} & \in \ 
& \Gamma( \phi^* T_{\pi}^{1,0} ),
\quad
& \psi_+^{\overline{m}} & \equiv \psi_z^{\overline{m}} & \in \
& \Gamma( K_{\Sigma} \otimes (\phi^* T_{\pi}^{1,0})^* ), 
\\
\lambda_-^x & \equiv \lambda_{\overline{z}}^x & \in \ & 
\Gamma( \overline{K}_{\Sigma} \otimes (\phi^* T_{\cal G}^{0,1} )^* ), 
\quad
& \lambda_-^{\overline{x}} & \equiv \lambda^{\overline{x}}
& \in \ & \Gamma( \phi^* T_{\cal G}^{0,1} ),
\\
\lambda_-^{\gamma} & \equiv \lambda_{\overline{z}}^{\gamma} & \in \
& \Gamma( \overline{K}_{\Sigma} \otimes ( \phi^* T_{ {\cal F}_2 }^{0,1} )^* ),
\quad 
& \lambda_-^{\overline{\gamma}} & \equiv \lambda^{
\overline{\gamma}}
& \in \ & \Gamma( \phi^* T_{ {\cal F}_2 }^{0,1} ).
\end{aligned}
\end{displaymath}
Unlike the A/2 kernels theory, in the A/2 cokernels case there is no need to
twist bosons.

Anomalies constrain the theory, as follows:
\begin{displaymath}
K_M \: \cong \: \det {\cal G}^* \otimes \det {\cal F}_1 \otimes
\det {\cal F}_2^*, \: \: \:
{\rm ch}_2({\cal E}) \: = \: {\rm ch}_2(TX).
\end{displaymath}
For later use, the first condition implies
\begin{displaymath}
K_X \: \cong \: \pi^* \det {\cal G}^* \otimes \pi^* \det {\cal F}_2^*.
\end{displaymath}
One can show that
anomaly-freedom in the UV implies anomaly-freedom in the IR.

The effective interactions can be obtained by truncating to fermi
zero modes.  In the degree zero sector, they are
\begin{displaymath}
h^{x \overline{x}} s_x \overline{s}_{\overline{x}}
\: + \: 
h_{\gamma \overline{\gamma}} q^m 
\overline{q}^{\overline{m}} \tilde{E}_m^{\gamma} 
\tilde{\overline{E}}_{\overline{m}}^{ \overline{\gamma} }
\: + \:
\chi^i \lambda^{\overline{x}} D_i s^x\, h_{x \overline{x}} \: + \:
\chi^{m} \lambda^{\overline{\gamma}} \tilde{E}_{
m}^{ \gamma}\, h_{\gamma \overline{\gamma}} \: + \:
\chi^i \lambda^{\overline{\gamma}} q^m \left( D_i
\tilde{E}_m^{\gamma} \right) h_{\gamma \overline{\gamma}},
\end{displaymath}
which descend to define an analogue of a Mathai-Quillen form.

Although $\chi^i$, $\chi^{m}$, $\lambda^{\overline{x}}$, and
$\lambda^{\overline{\gamma}}$ are all scalars,    
it can be shown following \cite{GuffinSharpe:A-twistedheterotic}
that only $\chi^i$, $\chi^m$ are genuinely BRST-invariant, and 
the others merely nearly BRST invariant:
\begin{displaymath}
Q \cdot \lambda^{\overline{x}} \: \propto \: \overline{s}^{\overline{x}},
\: \: \:
Q \cdot \lambda^{\overline{\gamma}} \: \propto \: \overline{q}^{
\overline{m}} \,
\tilde{\overline{E}}_{\overline{m}}^{\overline{\gamma}}.
\end{displaymath}
Furthermore, since the $q$ fields are not twisted, the bosonic zero modes lie
along $X$.

Restricting to zero modes, observables are then of the form
\begin{displaymath}
f(\phi^i, \phi^m) \chi^{i_1} \cdots \chi^{m_n}
\lambda^{\overline{\gamma}_1} \cdots \lambda^{\overline{\gamma}_{n'}}
\end{displaymath}
which after a complex conjugation are naively interpreted in terms of
the hypercohomology on $X$
of a complex of the form
\begin{displaymath}
\cdots \: \longrightarrow \: \wedge^2 \pi^* {\cal F}_2^* \: 
\longrightarrow \: \pi^* {\cal F}_2^* \: \longrightarrow \:
{\cal O}_X
\end{displaymath}
with maps given by inclusion along $q \tilde{E}$.
Note that in order for this interpretation to hold,
$\tilde{E}$ must be holomorphic everywhere on $M$, which is the reason
we restricted to that case in this section.

It was shown in \cite{GuffinSharpe:A-twistedheterotic}[section 4.1,
appendix A] that
the hypercohomology on $X$ of the sequence above is isomorphic to
\begin{displaymath}
H^{\bullet}\left(M, \wedge^{\bullet} {\cal E}' \right) ,
\end{displaymath}
where ${\cal E}'$ is the cokernel of the map $\tilde{E}:
{\cal F}_1 \rightarrow {\cal F}_2$.
Correlation functions then are of the
form of integrals over $X$ of observables times the exponential of
the zero mode interactions.

\subsection{Cohomologies of short complexes}

Suppose we want to build a Landau-Ginzburg model that flows to a nonlinear
sigma model on
\begin{displaymath}
Y \: \equiv \: \{s=0\} \: \subset \: M,
\end{displaymath}
as before,
with gauge bundle given by 
\begin{displaymath}
\frac{ \ker \tilde{F}|_Y }{ {\rm im}\, \tilde{E}|_Y } ,
\end{displaymath}
where $\tilde{E}: {\cal F}_1 \rightarrow {\cal F}_2$ is injective
everywhere on $M$, $\tilde{F}: {\cal F}_2 \rightarrow {\cal F}_3$ is
surjective everywhere on $M$.
\begin{displaymath}
0 \: \longrightarrow \: {\cal F}_1 \: 
\stackrel{ \tilde{E} }{\longrightarrow} \: {\cal F}_2 \:
\stackrel{ \tilde{F} }{\longrightarrow} \: {\cal F}_3
\: \longrightarrow \: 0
\end{displaymath}
The restrictions of both
$\tilde{E}$ and $\tilde{F}$ to $Y$ are holomorphic, but only one need
be holomorphic on all of $M$.  Furthermore, 
the composition $\tilde{F} \circ \tilde{E}$ should vanish everywhere
on $M$, making
the restriction into a complex:
\begin{displaymath}
0 \: \longrightarrow \: {\cal F}_1 |_Y \: \stackrel{ \tilde{E} }{
\longrightarrow} \: {\cal F}_2 |_Y \: \stackrel{ \tilde{F} }{\longrightarrow}
\: {\cal F}_3 |_Y \: \longrightarrow \: 0.
\end{displaymath}
(Experts will note that this is not the most general possibility
allowed by physics; we leave such more general cases for future work.)

Depending upon whether $\tilde{E}$ or $\tilde{F}$ is holomorphic on 
$M$, one gets two slightly different Landau-Ginzburg models that 
renormalization-group flow to the nonlinear sigma model described above.
We will describe each in turn.

\subsubsection{A/2 model realization of first short complex construction}
\label{sect:mon1:phys}

Suppose that $\tilde{E}$ is holomorphic on all of $M$,
where $\tilde{F}$ is only holomorphic after restriction to $Y \subset M$.
Corresponding to this nonlinear sigma model is a Landau-Ginzburg model
on
\begin{displaymath}
Z \: = \: {\rm Tot}\left( {\cal F}_1 \oplus {\cal F}_3^* \:
\stackrel{\tilde{\pi}}{\longrightarrow} \: M \right)
\end{displaymath}
with gauge bundle
\begin{displaymath}
0 \: \longrightarrow \: \tilde{\pi}^* {\cal G}^* \: \longrightarrow \:
{\cal E} \: \longrightarrow \: \tilde{\pi}^* {\cal F}_2 \: \longrightarrow
\: 0 .
\end{displaymath}

Physically, following \cite{GuffinSharpe:A-twistedheterotic}, to
specify this theory, we must specify a holomorphic section of ${\cal E}$
and a holomorphic section of ${\cal E}^*$, whose compositions vanish,
along with the precise extension above.

We will begin by specifying the sections.  Let us denote fiber coordinates
on ${\cal F}_1$, ${\cal F}_3^*$ by $q$, $p$, respectively.

A holomorphic section of ${\cal E}$ determines a holomorphic section of
$\tilde{\pi}^* {\cal F}_2$ and a smooth section of 
$\tilde{\pi}^* {\cal G}^*$,
holomorphic over the vanishing locus of the section of 
$\tilde{\pi}^* {\cal F}_2$.
We will take the holomorphic section of $\tilde{\pi}^* {\cal F}_2$ to be
$q \tilde{E}$, and the smooth section of $\tilde{\pi}^* {\cal G}^*$ to be 
identically zero.

Using the dual sequence
\begin{displaymath}
0 \: \longrightarrow \: \tilde{\pi}^* {\cal F}_2^* \: \longrightarrow \:
{\cal E}^* \: \longrightarrow \: \tilde{\pi}^* {\cal G} \: \longrightarrow \: 
0 ,
\end{displaymath}
a holomorphic section of ${\cal E}^*$ determines a holomorphic section of
$\tilde{\pi}^* {\cal G}$ and a smooth section of 
$\tilde{\pi}^* {\cal F}_2^*$ which is
holomorphic over the vanishing locus of the section of 
$\tilde{\pi}^* {\cal G}$.
We will take the holomorphic section of $\tilde{\pi}^* {\cal G}$ to be the
pullback of $s$
(whose vanishing locus is $Y$), and the smooth section of 
$\tilde{\pi}^* {\cal F}_2^*$ to be $p \tilde{F}$.

Consistency requires the composition of these sections to vanish,
and indeed, $\tilde{F} \tilde{E} = 0$ and $(s)(0) = 0$.

The data above -- a smooth not-necessarily holomorphic $\tilde{F}$
and a globally holomorphic $\tilde{E}$ --
are only compatible with the A/2 twist in general.

The action of the A/2 twisted Landau-Ginzburg model that RG flows
to the A/2 twist of the nonlinear sigma model above is given in local
coordinates by \cite{GuffinSharpe:A-twistedheterotic}
\begin{eqnarray*}
S & = & 2t \int_{\Sigma} d^2 z \Biggl[
\frac{1}{2} ( g_{\mu \nu} + i B_{\mu \nu}) \partial_z \phi^{\mu} 
\overline{\partial}_{\overline{z}} \phi^{\nu} \: + \:
i g_{\overline{a} a} \psi_+^{\overline{a}} \overline{D}_{\overline{z}} \psi_+^a
\: + \: i g_{\alpha \overline{\alpha}} \lambda_-^{\alpha}
D_z \lambda_-^{\overline{\alpha}}
\\
& & \hspace*{1in}
\: + \:
F_{a \overline{a} \alpha \overline{\alpha}} \psi_+^a \psi_+^{\overline{a}}
\lambda_-^{\alpha} \lambda_-^{\overline{\alpha}}
\: + \:
h^{x \overline{x}} s_x \overline{s}_{\overline{x}}
\: + \:
h^{\gamma \overline{\gamma}} p^r \overline{p}^{\overline{r}}
\tilde{F}_{r \gamma} 
\tilde{\overline{F}}_{\overline{r}
\overline{\gamma} }
\\
& & \hspace*{1in}
\: + \:
\psi_+^i \lambda_-^x D_i s_x \: + \:
\psi_+^r \lambda_-^{\gamma} \tilde{F}_{r \gamma} \: + \:
\psi_+^i \lambda_-^{\gamma} p^r D_i \tilde{F}_{r \gamma}
\\
& & \hspace*{1in}
\: + \:
\psi_+^{\overline{\imath}} \lambda_-^{\overline{x}} 
\overline{D}_{\overline{\imath}} \overline{s}_{\overline{x}} \: + \:
\psi_+^{\overline{r}} \lambda_-^{\overline{\gamma}} 
\tilde{\overline{F}}_{\overline{r} \overline{\gamma}} \: + \:
\psi_+^{\overline{\imath}} \lambda_-^{\overline{\gamma}} 
\overline{p}^{\overline{r}} \, \overline{D}_{\overline{\imath}}
\tilde{\overline{F}}_{\overline{r} \overline{\gamma}}
\\
& & \hspace*{1in}
\: + \:
h_{\gamma \overline{\gamma}} q^m \overline{q}^{\overline{m}}
\tilde{E}_m^{\gamma}
 \tilde{\overline{E}}_{\overline{m}}^{
\overline{\gamma}}
\\
& & \hspace*{1in}
\: + \: \psi_+^m \lambda_-^{\overline{\gamma}} \tilde{E}_m^{\gamma}
h_{\gamma \overline{\gamma}} 
\: + \:
\psi_+^i \lambda_-^{\overline{\gamma}} q^m D_i \tilde{E}_m^{\gamma}
h_{\gamma \overline{\gamma}}
\\
& & \hspace*{1in}
\: + \:
\psi_+^{\overline{m}} \lambda_-^{\gamma} \tilde{\overline{E}}_{
\overline{m}}^{\overline{\gamma}} h_{\gamma \overline{\gamma}}
\: + \:
\psi_+^{\overline{\imath}} \lambda_-^{\gamma} \overline{q}^{\overline{m}} \,
\overline{D}_{\overline{\imath}} \tilde{\overline{E}}_{\overline{m}}^{\overline{
\gamma}} h_{\gamma \overline{\gamma} }
\Biggr] ,
\end{eqnarray*}
where $x$ indexes local coordinates along the fibers of ${\cal G}$,
$m$ indexes local coordinates along the fibers of ${\cal F}_1$,
$\gamma$ indexes local coordinates along the fibers of ${\cal F}_2$,
$r$ indexes local coordinates along the fibers of ${\cal F}_3^*$,
$i$ indexes local coordinates on $M$,
$a \sim (m, r, i)$ indexes local coordinates on $X$,
and $\alpha \sim (x, \gamma)$ indexes local coordinates along the
fibers of ${\cal E}$.

The fermions and bosons are twisted as follows:
\begin{displaymath}
\begin{aligned}
\psi_+^i & \equiv \chi^i &\in \ &\Gamma( \phi^*(TM) ),
\quad
& \psi_+^{\overline{\imath}} & \equiv \psi_z^{\overline{\imath}}
& \in \ & \Gamma(K_{\Sigma} \otimes \phi^* T^*M ), 
\\
\psi_+^m & \equiv \psi^m & \in \ & \Gamma( \phi^* T_{{\cal F}_1}^{1,0}),
\quad
& \psi_+^{\overline{m}} & \equiv \psi_z^{\overline{m}}
& \in \ & \Gamma( K_{\Sigma} \otimes (\phi^* T_{{\cal F}_1}^{1,0})^*),
\\
\psi_+^r & \equiv \psi_z^r & \in \
& \Gamma(K_{\Sigma} \otimes \phi^* T_{\pi}^{1,0} ),
\quad
& \psi_+^{\overline{r}} & \equiv \chi^{\overline{r}} & \in \
& \Gamma( (\phi^* T_{\pi}^{1,0})^* ), 
\\
\lambda_-^x & \equiv \lambda^x & \in \ & 
\Gamma( (\phi^* T_{\cal G}^{0,1} )^* ), 
\quad
& \lambda_-^{\overline{x}} & \equiv \lambda_{\overline{z}}^{\overline{x}}
& \in \ & \Gamma(\overline{K}_{\Sigma} \otimes \phi^* T_{\cal G}^{0,1} ),
\\
\lambda_-^{\gamma} & \equiv \lambda_{\overline{z}}^{\gamma} & \in \
& \Gamma( \overline{K}_{\Sigma} \otimes ( \phi^* T_{ {\cal F}_1 }^{0,1} )^* ),
\quad 
& \lambda_-^{\overline{\gamma}} & \equiv \lambda^{\overline{\gamma}}
& \in \ & \Gamma( \phi^* T_{ {\cal F}_1 }^{0,1} ),
\\
p & \equiv p_z & \in \ & \Gamma(K_{\Sigma} \otimes \phi^* T_{\pi}^{1,0} ),
\quad
& \overline{p} & \equiv \overline{p}_{\overline{z}} & \in \ &
\Gamma(\overline{K}_{\Sigma} \otimes \phi^* T_{\pi}^{0,1} ).
\end{aligned}
\end{displaymath}
The bosons $q$, $\overline{q}$ are untwisted.

Anomalies constrain this theory as follows:
\begin{displaymath}
K_M \: \cong \: \det {\cal G}^* \otimes \det {\cal F}_1 \otimes
\det {\cal F}_2^* \otimes \det {\cal F}_3, \: \: \:
{\rm ch}_2(Z) \: = \: {\rm ch}_2({\cal E}).
\end{displaymath}

The effective interactions can be obtained by truncating to fermi zero modes.
In the degree zero sector, they are
\begin{eqnarray*}
\lefteqn{
h^{x \overline{x}} s_x \overline{s}_{\overline{x}} \: + \:
\chi^i \lambda^x D_i s_x \: + \:
\chi^{\overline{r}} \lambda^{\overline{\gamma}} \,
\tilde{\overline{F}}_{\overline{r} \overline{\gamma}} \: + \: 
F_{i \overline{r} x \overline{\gamma} } \chi^i \chi^{\overline{r}} \lambda^x
\lambda^{\overline{\gamma}}
} \\
& & \hspace*{0.25in}
\: + \: 
h_{\gamma \overline{\gamma}} q^m  \overline{q}^{\overline{m}}
\tilde{E}_m^{\gamma} 
 \tilde{ \overline{E} }_{\overline{m}}^{\overline{
\gamma}} \: + \:
\chi^m \lambda^{\overline{\gamma}} \tilde{E}_m^{\gamma} h_{
\gamma \overline{\gamma}} \: + \:
\chi^i \lambda^{\overline{\gamma}} q^m D_i \tilde{E}_m^{\gamma}
h_{\gamma \overline{\gamma}}
\: + \:
F_{m \overline{r} x \overline{\gamma} }
\chi^m  \chi^{\overline{r}} \lambda^x
\lambda^{\overline{\gamma}} .
\end{eqnarray*}

The curvature term
\begin{displaymath}
F_{m \overline{r} x \overline{\gamma} }
\chi^m  \chi^{\overline{r}} \lambda^x 
\lambda^{\overline{\gamma}}
\end{displaymath}
will always vanish, so we omit it from further discussion.

Since the $q$'s are untwisted, the resulting analogue of a
Mathai-Quillen form will live on a 
bundle over $M$.
Specifically, they live on 
\begin{displaymath}
X \: \equiv \: {\rm Tot}\, \left( {\cal F}_1 \:
\stackrel{\pi}{\longrightarrow} \: M \right) .
\end{displaymath}

The scalars $\psi^i$, $\psi^m$ are BRST-invariant.
Similarly,
\begin{displaymath}
Q \cdot \lambda^{\overline{\gamma}} \: \propto \:
\overline{q}^{\overline{m}} \, \tilde{ \overline{E} }^{\overline{\gamma}}_{
\overline{m}}
\end{displaymath}
so, after complex conjugation,
we interpret observables built from $\chi^i$, $\chi^m$, 
$\lambda^{\overline{\gamma}}$ in terms of hypercohomology
\begin{displaymath}
{\mathbb H}^{\bullet}\left(X, \cdots \longrightarrow \:
\wedge^2 \pi^* {\cal F}_2^* \: \longrightarrow \:
\pi^* {\cal F}_2^* \: \longrightarrow \: {\cal O}_X \right)
\end{displaymath}
with maps given by inclusion with $q^m \tilde{E}^{\gamma}_m$.

\subsubsection{B/2 model realization of second short complex construction}
\label{sect:monad2:phys}

Suppose that $\tilde{E}$ is holomorphic only after restriction to $Y \subset M$,
whereas $\tilde{F}$ is holomorphic everywhere on $M$.
Corresponding to this nonlinear sigma model is a Landau-Ginzburg model on
\begin{displaymath}
Z \: = \: {\rm Tot}\left( {\cal F}_1 \oplus {\cal F}_3^* \:
\stackrel{\tilde{\pi}}{\longrightarrow} \: M \right)
\end{displaymath}
with gauge bundle
\begin{displaymath}
0 \: \longrightarrow \: \tilde{\pi}^* {\cal F}_2 \: \longrightarrow \:
{\cal E} \: \longrightarrow \: \tilde{\pi}^* {\cal G} \: \longrightarrow \: 0 .
\end{displaymath}

Physically, following \cite{GuffinSharpe:A-twistedheterotic}, to
specify this theory, we must specify a holomorphic section of ${\cal E}$
and a holomorphic section of ${\cal E}^*$, whose compositions vanish,
along with the precise extension above.

We will begin by specifying the sections.  Let us denote fiber coordinates
on ${\cal F}_1$, ${\cal F}_3^*$ by $q$, $p$, respectively.

A holomorphic section of ${\cal E}$ determines a holomorphic section of
$\tilde{\pi}^* {\cal G}$ and a smooth section of $\tilde{\pi}^* {\cal F}_2$, 
holomorphic
over the vanishing locus of the section of $\tilde{\pi}^* {\cal G}$.
We will take the holomorphic section of $\tilde{\pi}^* {\cal G}$ to be the
pullback of $s$ (whose vanishing locus is $Y$), and the smooth section of
$\tilde{\pi}^* {\cal F}_2$ to be $p \tilde{E}$.

Using the dual sequence
\begin{displaymath}
0 \: \longrightarrow \: \tilde{\pi}^* {\cal G}^* \: \longrightarrow \:
{\cal E}^* \: \longrightarrow \: \tilde{\pi}^* {\cal F}_2^* 
\: \longrightarrow \:
0 ,
\end{displaymath}
a holomorphic section of ${\cal E}^*$ determines a holomorphic section of
$\tilde{\pi}^* {\cal F}_2^*$ and a smooth section of 
$\tilde{\pi}^* {\cal G}^*$ which is
holomorphic over the vanishing locus of the section of 
$\tilde{\pi}^* {\cal F}_2^*$.
We will take the holomorphic section of $\tilde{\pi}^* {\cal F}_2^*$ to be 
$q \tilde{F}$, and the smooth section of $\tilde{\pi}^* {\cal G}^*$ 
to be identically
zero.

Consistency requires the composition of these sections to vanish,
and indeed, $\tilde{F} \tilde{E} = 0$, $(0)(s) = 0$.

The data above -- a smooth not-necessarily holomorphic $\tilde{E}$ and
a globally holomorphic $\tilde{F}$ -- are only compatible with the
B/2 twist in general.

The action of the B/2 twisted Landau-Ginzburg model that RG flows to the
B/2 twist of the nonlinear sigma model above is given in local coordinates
by \cite{GuffinSharpe:A-twistedheterotic}
\begin{eqnarray*}
S & = & 2t \int_{\Sigma} d^2 z \Biggl[
\frac{1}{2} ( g_{\mu \nu} + i B_{\mu \nu}) \partial_z \phi^{\mu} 
\overline{\partial}_{\overline{z}} \phi^{\nu} \: + \:
i g_{\overline{a} a} \psi_+^{\overline{a}} \overline{D}_{\overline{z}} \psi_+^a
\: + \: i g_{\alpha \overline{\alpha}} \lambda_-^{\alpha}
D_z \lambda_-^{\overline{\alpha}}
\\
& & \hspace*{1in}
\: + \:
F_{a \overline{a} \alpha \overline{\alpha}} \psi_+^a \psi_+^{\overline{a}}
\lambda_-^{\alpha} \lambda_-^{\overline{\alpha}}
\: + \:
h^{\gamma \overline{\gamma}} p^r \overline{p}^{\overline{r}}
\tilde{F}_{r \gamma} 
\tilde{\overline{F}}_{\overline{r} \overline{\gamma}}
\\
& & \hspace*{1in}
\: + \:
\psi_+^r \lambda_-^{\gamma} \tilde{F}_{r \gamma} \: + \:
\psi_+^i \lambda_-^{\gamma} p^r D_i \tilde{F}_{r \gamma}
\\
& & \hspace*{1in}
 \: + \:
\psi_+^{\overline{r}} \lambda_-^{\overline{\gamma}} \tilde{\overline{F}}_{
\overline{r} \overline{\gamma}} \: + \:
\psi_+^{\overline{\imath}} \lambda_-^{\overline{\gamma}} 
\overline{p}^{\overline{r}}
\overline{D}_{\overline{\imath}} \tilde{\overline{F}}_{\overline{r} \overline{\gamma}}
\\
& & \hspace*{1in}
\: + \:
h_{x \overline{x}} s^x \overline{s}^{\overline{x}} \: + \:
h_{\gamma \overline{\gamma}} q^m 
\overline{q}^{\overline{m}} \tilde{E}_m^{\gamma} 
\tilde{\overline{E}}_{\overline{m}}^{ \overline{\gamma} }
\\
& & \hspace*{1in}
\: + \:
\psi_+^i \lambda_-^{\overline{x}} D_i s^x \, h_{x \overline{x}} \: + \:
\psi_+^{m} \lambda_-^{\overline{\gamma}} \tilde{E}_m^{\gamma}
\, h_{\gamma \overline{\gamma}} \: + \:
\psi_+^i \lambda_-^{\overline{\gamma}} q^m \left(D_i \tilde{E}_m^{\gamma}
\right)  h_{\gamma \overline{\gamma}}
\\
& & \hspace*{1in}
 \: + \:
\psi_+^{\overline{\imath}} \lambda_-^{x} \overline{D}_{\overline{\imath}}
\overline{s}^{\overline{x}} \, h_{x \overline{x}} \: + \:
\psi_+^{\overline{m}} \lambda_-^{\gamma} \tilde{\overline{E}}_{
\overline{m}}^{ \overline{\gamma}} \, h_{\gamma \overline{\gamma}} \: + \:
\psi_+^{\overline{\imath}} \lambda_-^{\gamma} 
\overline{q}^{\overline{m}} \left(
\overline{D}_{\overline{\imath}} \tilde{\overline{E}}_{\overline{m}}^{ \overline{\gamma}}
\right) h_{\gamma \overline{\gamma}}
\Biggr] ,
\end{eqnarray*}
where $x$ indexes local coordinates along the fibers of ${\cal G}$,
$m$ indexes local coordinates along the fibers of ${\cal F}_1$,
$\gamma$ indexes local coordinates along the fibers of ${\cal F}_2$,
$r$ indexes local coordinates along the fibers of ${\cal F}_3^*$,
$i$ indexes local coordinates on $M$,
$a \sim (m, r, i)$ indexes local coordinates on $X$,
and $\alpha \sim (x, \gamma)$ indexes local coordinates along the
fibers of ${\cal E}$.

The fermions and bosons are twisted as follows:
\begin{displaymath}
\begin{aligned}
\psi_+^i & \equiv \chi^i &\in \ &\Gamma( \phi^*(TM) ),
\quad
& \psi_+^{\overline{\imath}} & \equiv \psi_z^{\overline{\imath}}
& \in \ & \Gamma(K_{\Sigma} \otimes \phi^* T^*M ), 
\\
\psi_+^{m} & \equiv \psi_z^{m} & \in \
& \Gamma( K_{\Sigma} \otimes \phi^* T_{\pi}^{1,0} ),
\quad
& \psi_+^{\overline{m}} & \equiv \chi^{\overline{m}} & \in \
& \Gamma( (\phi^* T_{\pi}^{1,0})^* ), 
\\
\psi_+^r & \equiv \chi^r & \in \
& \Gamma( \phi^* T_{\pi}^{1,0} ),
\quad
& \psi_+^{\overline{r}} & \equiv \psi_z^{\overline{r}} & \in \
& \Gamma(K_{\Sigma} \otimes (\phi^* T_{\pi}^{1,0})^* ), 
\\
\lambda_-^x & \equiv \lambda_{\overline{z}}^x & \in \ & 
\Gamma( \overline{K}_{\Sigma} \otimes (\phi^* T_{\cal G}^{0,1} )^* ), 
\quad
& \lambda_-^{\overline{x}} & \equiv \lambda^{\overline{x}}
& \in \ & \Gamma( \phi^* T_{\cal G}^{0,1} ),
\\
\lambda_-^{\gamma} & \equiv \lambda^{\gamma} & \in \
& \Gamma(  ( \phi^* T_{ {\cal F}_1 }^{0,1} )^* ),
\quad 
& \lambda_-^{\overline{\gamma}} & \equiv 
\lambda_{\overline{z}}^{\overline{\gamma}}
& \in \ & \Gamma( \overline{K}_{\Sigma} \otimes
\phi^* T_{ {\cal F}_1 }^{0,1} ),
\\
q & \equiv q_z & \in \ & \Gamma(K_{\Sigma} \otimes \phi^* T_{\pi}^{1,0}),
\quad
& \overline{q} & \equiv \overline{q}_{\overline{z}} & \in \ &
\Gamma( \overline{K}_{\Sigma} \otimes \phi^* T_{\pi}^{0,1}).
\end{aligned}
\end{displaymath}
In the B/2 twisted theory, $p$, $\overline{p}$ are not twisted.

Anomalies constrain the theory as follows:
\begin{displaymath}
K_M \: \cong \: \det {\cal G}^* \otimes \det {\cal F}_1^* \otimes
\det {\cal F}_2 \otimes \det {\cal F}_3^*, \: \: \:
{\rm ch}_2(TZ) \: = \: {\rm ch}_2({\cal E}).
\end{displaymath}

The effective interactions can be obtained by truncating to fermi zero
modes.  In the degree zero sector, they are
\begin{eqnarray*}
\lefteqn{
h_{x \overline{x}} s^x \overline{s}^{\overline{x}} \: + \:
h^{\gamma \overline{\gamma}} p^r \overline{p}^{\overline{r}}
\tilde{F}_{r \gamma} 
\tilde{\overline{F}}_{\overline{r} \overline{\gamma}}
 \: + \:
\chi^r \lambda^{\gamma} \tilde{F}_{r \gamma} \: + \:
\chi^i \lambda^{\gamma} p^r D_i \tilde{F}_{r \gamma}
} \\
& & \hspace*{0.5in}
\: + \:
\chi^i \lambda^{\overline{x}} D_i s^x \, h_{x \overline{x}} \: + \:
\chi^{\overline{m}} \lambda^{\gamma} \tilde{\overline{E}}_{
\overline{m}}^{ \overline{\gamma}} \, h_{\gamma \overline{\gamma}} \: + \:
F_{i \overline{m} \overline{x} \gamma} 
\chi^i \chi^{\overline{m}} \lambda^{\overline{x}} \lambda^{\gamma} 
\: + \:
F_{r \overline{m} \overline{x} \gamma}
\chi^r \chi^{\overline{m}} \lambda^{\overline{x}} \lambda^{\gamma} ,
\end{eqnarray*}
and they descend to define an analogue of a Mathai-Quillen form.
The curvature term
\begin{displaymath}
F_{r \overline{m} \overline{x} \gamma}
\chi^r \chi^{\overline{m}} \lambda^{\overline{x}} \lambda^{\gamma}
\end{displaymath}
will always vanish, so we omit it from further discussion.

Since the $p$'s are untwisted, the observables live on
\begin{displaymath}
X \: \equiv \: {\rm Tot}\,\left(
{\cal F}_3^* \: \stackrel{\pi}{\longrightarrow} \: M \right) .
\end{displaymath}

The scalars $\chi^i$, $\chi^r$,   are BRST-invariant,
and if we define $\theta_{\overline{\gamma}} \equiv h_{\gamma 
\overline{\gamma}} \lambda^{\gamma}$, then
\begin{displaymath}
Q \cdot \lambda_-^{\overline{x}} \: \propto \: \overline{s}^{\overline{x}},
Q \cdot \theta_{\gamma} \: \propto \: \overline{p}^{\overline{r}} \,
\tilde{ \overline{F} }_{\overline{r} \overline{\gamma} }.
\end{displaymath}
Observables built from $\psi_+^i$, $\psi_+^r$, $\theta_{\overline{\gamma}}$
can then, after complex conjugation, be interpreted in terms of
hypercohomology
\begin{displaymath}
{\mathbb H}^{\bullet}\left(X, 
\cdots \: \longrightarrow \: \wedge^2 \pi^* {\cal F}_2 \:
\longrightarrow \: \pi^* {\cal F}_2 \: \longrightarrow \:
{\cal O}_X \right)
\end{displaymath}
with maps given by inclusion with $p^r \tilde{F}_{r \gamma}$.

\subsection{A note on anomalies}

In the A/2 and B/2 models, we have encountered two different anomalies,
one a condition on determinants of bundles, the second a condition on
second Chern characters.  The second condition is the standard
Green-Schwarz condition; the first is specific to the A/2 and B/2 models.

The condition on determinants of bundles is applied mathematically to give
well-defined integrals of products of sheaf cohomology classes.  However,
the first condition, the Green-Schwarz condition, did not appear in the
mathematical discussion.

Although it is possible that a more detailed examination of the
analogues of Mathai-Quillen forms we have proposed will require the
Green-Schwarz condition, it is our belief that they will not.  The reason
is the manner in which the Green-Schwarz condition appears in {\it e.g.}
quantum sheaf cohomology computations.  There, its role is to ensure that
in worldsheet instanton sectors, corresponding integrals of sheaf
cohomology classes over moduli spaces of instantons are well-defined.
In other words, its role is to help provide an analogue of the
determinants condition over moduli spaces of instantons.  As worldsheet
instantons are not discussed in this paper, as they do not arise in
our constructions of Mathai-Quillen analogues, it is our suspicion that
Green-Schwarz is not relevant to the mathematics of the constructions
presented here.

\section{Conclusions}

In this paper we have presented some sheaf-cohomological analogues of
Mathai-Quillen forms, which is to say, $\overline{\partial}$-closed
bundle-valued differential forms which generalize Mathai-Quillen forms.
We have shown that the cohomology classes of these forms are invariant
under certain deformations, and we have conjectured (based on their
physical origin relating UV and IR theories via renormalization group flow)
that these analogues have Thom-form-like properties, though we have not
given a mathematical argument to justify that assertion.

One of the original hopes of this work was to give a mathematical understanding
of some claims of Melnikov and McOrist \cite{McOristMelnikov:Summing}
regarding A/2 correlation functions and their dependence (or lack thereof)
on certain complex and bundle moduli.  Unfortunately, we were not able
in this work to explicitly confirm more than a part of their claims,
but neither have we disproven them, their verification remains an
open problem.

\section{Acknowledgements}

We would like to thank R.~Donagi, S.~Katz, V.~Mathai, and T.~Pantev 
for useful conversations.
R.G. thanks the Department of Mathematical and Statistical Sciences at 
the University of Alberta for hospitality during the initial stages of 
this project.  E.S. thanks the Aspen Center for Physics for hospitality
while this work was completed, under its NSF grant PHYS-1066293.
E.S. was partially supported by NSF grant PHY-1068725.

\end{document}